\pdfoutput=1
\documentclass[11pt]{article}
\usepackage{jheppub}
\usepackage[T1]{fontenc}
\usepackage[latin9]{inputenc}
\usepackage{xcolor}
\usepackage{amsmath}
\usepackage{amssymb}
\usepackage{graphicx}
\usepackage{float}
\usepackage{graphicx}
\usepackage{graphbox}
\usepackage{tikz}
\usepackage{tikz-cd} 
\usetikzlibrary{shapes.geometric, arrows}
\usepackage{mathtools}
\usepackage{dsfont}
\usepackage{empheq}
\usepackage{cancel}
\usepackage{tensor}
\usepackage{mathrsfs}

\def \nn {\nonumber}
\def \la {\langle}
\def \ra {\rangle}
\def \< {\langle}
\def \> {\rangle}

\def \be{\begin{equation}}
\def \ee{\end{equation}}
\def \bea{\begin{eqnarray}}
\def \eea{\end{eqnarray}}
\def \braket#1#2{\big\langle{#1}|{#2}\big\rangle}
\def \bra#1{\big\langle{#1}|}
\def \ket#1{|{#1}\big\rangle}
\def \dual{\vee}

\def \O {\mathcal{O}}
\def \B {\mathcal{B}}
\def \G {\mathcal{G}}
\def \C {\mathcal{C}}

\def \gl {{\text{gl}(1)}}

\def \embedd{\frac{\delta(Y_\perp^2 + Y^2)}{\gl}}

\def \CP {\mathbb{CP}}

\def \Cbb {\mathbb{C}}

\def \res {{\rm Res}}
\def \ker {{\rm ker}}

\def \alg {{\rm alg}}
\def \dr {{\rm dR}}
\def \drc {{\rm dR,c}}
\def \ibp {{\rm{IBP}}}

\def \tri{\text{tri}}

\def \bub {{\rm{bub}_{12}}}
\def \bubb {{\rm{bub}_{23}}}
\def \bubbb {{\rm{bub}_{13}}}

\def \tad {{\rm{tad}_1}}

\def \Space{\Cbb^n\setminus \{\mathcal{G}=0\} \cup \{D=0\}}
\def \Spacedual{\Cbb^n\setminus \{\mathcal{G}=0\} }
\def \div{\{D=0\}}

\def \dint{d_\text{int}}

\def \vphi {\varphi}

\def \vep {\varepsilon}

\def\eps{\varepsilon}
\def\th{\theta}
\def\disc{{\rm Disc}}

\def\be{\begin{equation}}
\def\ee{\end{equation}}
\def\ba{\begin{eqnarray}}
\def\ea{\end{eqnarray}}

\def\eps{\epsilon}

\newcommand{\lsim}{\mathrel{\hbox{\rlap{\lower.55ex \hbox{$\sim$}} \kern-.3em \raise.4ex \hbox{$<$}}}}
\newcommand{\gsim}{\mathrel{\hbox{\rlap{\lower.55ex \hbox{$\sim$}} \kern-.3em \raise.4ex \hbox{$>$}}}}

\newtheorem{theorem}{Theorem}

\newcommand{\bs}[1]{\boldsymbol{#1}} 
\newcommand{\mat}[1]{\underline{\boldsymbol{#1}}}

\newcommand{\ipaa}[2]{\left\langle #1 \vert #2 \right\rangle}
\newcommand{\ipsa}[2]{\left[ #1 \vert #2 \right\rangle}
\newcommand{\ipas}[2]{\left\langle #1 \vert #2 \right]}

\title{Duals of Feynman Integrals, I: Differential Equations}
\author[a]{Simon Caron-Huot,}
\author[a]{Andrzej Pokraka}
\affiliation[a]{
	Department of Physics, McGill University, 
	3600 Rue University, 
	Montr\'eal, QC Canada H3A 2T8
}
\emailAdd{schuot@physics.mcgill.ca}
\emailAdd{andrzej.pokraka@mail.mcgill.ca}

\abstract{
We elucidate the vector space (twisted relative cohomology) that is Poincar\'e dual to the vector space of Feynman integrals (twisted cohomology) in general
spacetime dimension.  The pairing between these spaces -- an algebraic invariant called the intersection number --
extracts integral coefficients for a minimal basis, bypassing the generation of integration-by-parts identities.
Dual forms turn out to be much simpler than their Feynman counterparts: they are supported on maximal cuts of various sub-topologies (boundaries).
Thus, they provide a systematic approach to generalized unitarity, the reconstruction of amplitudes
from on-shell data.
In this paper, we introduce the idea of dual forms and study their mathematical structures.
As an application, we derive compact differential equations satisfied by arbitrary one-loop integrals in non-integer spacetime dimension.
A second paper of this series will detail intersection pairings and their use to extract integral coefficients.
}

\begin{document}

\maketitle

\section{Introduction}

Generalized unitarity \cite{Bern:1994cg, Bern:1995db, Bern:2004ky, Britto:2004nc, Buchbinder:2005wp, Anastasiou:2006jv, Britto:2006fc, Ossola:2006us, Britto:2007tt, Forde:2007mi, Badger:2008cm, Bern:2008pv, Britto:2008vq, Cachazo:2008vp, Bern:2010tq, Bern:2010qa, Bourjaily:2017wjl, Bourjaily:2019iqr, Feng:2021spv} is a powerful technique for constructing loop-level scattering amplitudes. 
It is a generic procedure that applies to scattering amplitudes in any theory (including non-planar contributions).  Generalized unitarity has played a large part in obtaining state of the art predictions relevant for LHC phenomenology \cite{Berger:2008sj, KeithEllis:2009bu, Bevilacqua:2009zn, Mastrolia:2010nb, Badger:2013gxa, Abreu:2017xsl, Chicherin:2018yne, Badger:2019djh}; for illustration,
5-point amplitudes at NNLO accuracy in both QCD and  $\mathcal{N}=4$ sYM \cite{Carrasco:2011mn, Abreu:2018aqd, Chicherin:2018yne, Badger:2019djh} have been recently obtained.

Unitarity has a long history in quantum field theory since the optical theorem and Cutkosky rules \cite{Cutkosky:1960sp}. Unlike unitarity cuts, generalized unitarity places multiple internal lines on shell, which subdivides an amplitude into more than two pieces. Then, by taking many combinations of cuts and matching against a basis of known master integrals, the original amplitude can be expressed as a linear combination of a basis of so-called master integrals, with coefficients that are rational functions in the kinematic data.

The master integrals originate from the idea that Feynman integrals contributing to a given process
only need to be known modulo total derivatives, also called integration-by-parts identities.
Indeed, one important property of dimensional regularization is that total derivatives integrate to zero \cite{Chetyrkin:1981qh, Tkachov:1981wb}. This is widely used to simplify the integrand and project it onto a basis (which is always finite \cite{Smirnov:2010hn, Lee:2013hzt, Bitoun:2018afx}), expressing the amplitude for any given process and loop order as a sum:
\be
 \mathcal{A}^{(L)}(\{p_k\},\eps)  = \sum_i c_i(\{p_k\},\eps) F_i(\{p_k\},\eps)\,. \label{generalized unitarity}
\ee
The master integrals $F_i$ are complicated transcendental functions which depend on
the process but not on the theory. On the other hand, the coefficients $c_i$ depend on the theory but are algebraic functions of momenta (and polarization spinors etc.). The aim of this work
will be on the extraction of the coefficients $c_i$ of master integrals, starting from a given representation of the loop integrand, or more simply, of its cuts. We will also discuss differential equations satisfies by the $F_i$.

In principle, the coefficients $c_i$ may be computed by starting with some representative of the integrand, for example Feynman diagrams, and systematically simplifying it by generating IBP (integration-by-parts) identities following variants of the Laporta algorithm \cite{Laporta:2001dd, vonManteuffel:2012np, vonManteuffel:2014ixa, Maierhoefer:2017hyi, Smirnov:2019qkx}. While this is has been an extremely powerful and successful strategy, it is often computationally intensive and leaves one with little understand as to when or how cancellations occur in the final result. Therefore, it is important to contemplate more analytical approaches. Recently, an alternative mathematical perspective was proposed \cite{Mastrolia:2018uzb, Frellesvig:2019kgj, Frellesvig:2019uqt,  Mizera:2019vvs, Mizera:2020wdt, Frellesvig:2020qot}: working modulo integration-by-parts identities means that we are interested only in the cohomology class of the integrand (defined more precisely below). Algebraic geometry defines invariant pairings -- so called \emph{intersection numbers} -- whose value is the same on all representatives of a given cohomology class. Since the pairing is non-degenerate (by construction), it facilitates the extraction of generalized unitarity coefficients:
\be
 c_i = \langle f_i^\dual| \mathcal{I} \rangle  \label{pairing intro}
\ee
where the $f_i^\dual$ are dual to corresponding master integrals $F_i$, and $\mathcal{I}$ is the given integrand for the process at hand. Since the pairing is unchanged upon adding IBPs to the integrand, the need to generate IBP identities is circumvented altogether!

The main goal of this paper is to elucidate the space in which the ``dual'' integrands $f_i^\dual$ reside.
Since this space is dual to that of Feynman integrals (modulo IBPs), the two spaces have the same dimensions. Furthermore, any orthonormal bases of these spaces satisfy equivalent differential equations with respect to variations of external kinematic data. Thus, constructing a basis of dual Feynman integrals $f_i^\dual$ for a given process is equivalent to enumerating the transcendental functions $F_i$ which can appear in eq.~\eqref{generalized unitarity}. The pairing then gives the coefficients $c_i$.

Our main finding is that the dual integrands $f_i^\dual$ are supported on cuts: contrary to the original loop integrand, which runs over all $d$-dimensional momenta, the duals contain $\delta$-functions that force some propagators to be on-shell. On these cuts, dual integrands are essentially polynomials.
Yet, no information is lost, because the pairing is non-degenerate. This provides a mathematical basis for unitarity methods, and an alternative explanation for why scattering amplitudes $\mathcal{A}$ are determined by on-shell subprocesses.

Specifically, we consider generic Feynman integrals in $d$-dimensional momentum space, which takes the schematic form:
\be \label{eq:schematic Feynman integral}
 \int \frac{d^d\ell_1}{\pi^{d/2}}\cdots\frac{d^d\ell_L}{\pi^{d/2}}\ \mathcal{I}(\ell,p) \quad\mbox{with}\quad \mathcal{I}(\ell,p) = \frac{N(\ell,p)}{\prod_i D_i},
\ee
where the integrand $\mathcal{I}$ is manifestly a rational function of $\ell$ with some numerator $N$ (polynomial in $\ell$) and denominators quadratic in loop momenta ($D_i= (\sum_k a_{ik}\ell_k+\sum_j b_{ij} p_j)^2+m^2_k$ where $a_{ik}\in \{0,\pm 1\}$), and, where the $m^2_k$ are (possibly zero) constant mass parameters.

The key features of the integrand in eq.~\eqref{eq:schematic Feynman integral}  is the presence of \emph{poles} (where one of the $D_i=0$) and \emph{branch points}.  The latter are less obvious and hidden inside the symbol $d^d\ell$ when the spacetime dimension is not an integer, and, occur at the zeros of certain Gram determinants $\mathcal{G}$ (to be reviewed below).

As mentioned, working modulo IBPs means that the integrand is really a representative of a cohomology class
\be
  \mathcal{I} \hookrightarrow H^n(\Space;\nabla_{\vep \omega})
\ee
where the singular points have been removed and  the covariant derivative $\nabla_{\vep \omega}$ will account for multi-valuedness of the integration measure in non-integer dimensions \cite{bott1995, aomoto2011theory, Matsumoto:2018aa, Mastrolia:2018uzb, Frellesvig:2019kgj, Frellesvig:2019uqt,  Mizera:2019vvs, Mizera:2020wdt, Frellesvig:2020qot}. Moreover, since $\frac{N(\ell,p)}{\prod_i D_i}$ is algebraic, the reality properties of $\ell$ play no role and therefore, it is best to work with complexified momenta (where algebraic equations always have solutions).

The degree $n$ of the form deserves comment. Due to the peculiarities of dimensional regularization, a one-loop integral near four space-time dimensions is really a 5-form (see eq.~\eqref{eq:volform}) where the fifth variable $\ell_\perp^2$ accounts for extra-dimensional components. In this work, it is crucial that the dimension is not an integer $d=\dint\pm2\vep$. Since we will often have use for the integer part of $d$, we denote it by $\dint$.

Our main claim can now be stated concisely in this notation: dual Feynman integrands are representatives of a relative cohomology where the boundaries are the zero locus of the propagators $D_i$:
\be
  \mbox{dual-integrands:} \quad f_i^\dual 
  \in H^n(\CP^n\setminus\{\mathcal{G}=0\}, \{D=0\}; \nabla_{-\vep\omega}) \ .
\ee
These dual objects are effectively Feynman integrands, but with the sign of $\vep$ reversed (the dimensional regularization parameter), and where propagators $D_i$ define geometric boundaries instead of denominators \cite{Hwa:102287}.

If $f$ is a form representing the integrand $\mathcal{I}$, and $f^\vee_i$ a basis dual to the master integrals $F_i$, then the coefficients in eq.~\eqref{generalized unitarity} are obtained by the pairing\footnote{The factor $(-1)^\frac{n(n+1)}{2}$ simply transposes the left form, and saves many annoying minus signs.}
\be
	c_i= \braket{f_i^\dual}{f}
	\equiv \frac{ (-1)^{\frac{n(n+1)}{2}}}{(2\pi i)^n} 
	\int_{\Cbb^n} f_{i,c}^\dual \wedge f. 
	\label{eq:pairing}
\ee
The subscript $c$ indicates that $f_{i,c}^\vee$ is a compactly supported representative of $f^\vee_i$. 
For now, being ``compactly supported'' means that all $f_i^\dual$ vanish in a neighborhood of singularities of the integrand, $\{\mathcal{G}=0\}$ and $\{D=0\}$. This ensures that the integral \eqref{eq:pairing} is always well-defined  (absolutely convergent).  Note that we have a $2n$-dimensional integral ranging over all of $\Cbb^n$, so no contour choice is necessary.

Since the integral runs over all complexified momenta and the $f_{i,c}^\dual$ have compact support,
the pairing  (\ref{eq:pairing}) is manifestly invariant upon adding total derivatives to $f$.  
However, this sort of integral would seem rather impractical to compute.
What saves the day is that it turns out to be always possible to choose 
representatives $f_{i,c}^\dual$ for which all anti-holomorphic dependence 
is concentrated inside $\delta$-functions. The integral reduces to residues---around propagator poles, and possibly other locations.   The evaluation of eq.~\eqref{eq:pairing} is \emph{purely algebraic}!

While extracting generalized unitarity coefficients is our main motivation for this work, in this first paper of the series we focus on introducing the idea of dual forms and explaining their mathematical structure. 
Without introducing the machinery of multi-variate intersection numbers, the concept of dual forms proves to be very powerful in its own right. 
As an application, we give a basis of for uniform transcendental one-loop dual forms and derive the associated canonical differential equations using only ``naive'' integration-by-parts techniques.
In fact, the IBP relations for dual forms are much simpler than the analogous relations for Feynman integrals since IBP vectors are localized to cuts and squared propagators never appear (although it would be possible to construct IBP vectors that do not square propagators \cite{Kosower:2018obg}). 
Since we work in generic dimension, our results generalize the well known differential equations for one-loop Feynman integrals in integer dimensions \cite{Spradlin:2011wp,Arkani-Hamed:2017ahv, Bourjaily:2019exo}.

The structure of this paper is as follows. 
Section \ref{sec:fyn int and intersections} establishes the connection between Feynman integrals and intersection theory.
Relative twisted cohomology is introduced as the vector space dual to Feynman integrands.
This is exemplified by providing an explicit basis of one-loop dual forms.
In section~\ref{sec:one-loop deqs}, we study the differential equations satisfied by dual-integrals.
As a pedagogical application of the formalism, we derive the differential equations for generic one-loop dual-integrals near $2$-dimensions in section~\ref{sec:2dDEqs}.
To further illustrate the utility of dual forms, we derive differential equations for generic one-loop dual-integrals near any integer dimension  $\dint$ in section~\ref{sec:4dDEqs}. 
The duality with Feynman integrals is established in section~\ref{sec:FI and degenerate limits} by comparing the differential equations of dual-integrals with those of Feynman integrals in various degenerate limits.
Subtleties relating to the possible over-counting of dual forms (seen in section~\ref{sec:degenerate limits}) are explained in appendix \ref{app:equal mass dual tadpole}. 
We conclude in section~\ref{sec:conclusion}.
In a subsequent  paper \cite{schap}, we focus on the computation of multi-variate intersection numbers and the extraction of generalized unitarity coefficients. 

\section{Feynman integrals, forms and intersection theory \label{sec:fyn int and intersections}}

In this section, we introduce some basic mathematical concepts from algebraic geometry and explain their relation to Feynman integrals and dual forms. The key first step is to define the differential form $d^d\ell$ when $d$ is continuous (subsection~\ref{ssec:dim reg}). Then, we explain how to understand Feynman integrals as elements of a twisted cohomology group and the associated dual-cohomology group (subsection~\ref{ssec:duals}). We then review the relevant ideas from relative cohomology that are applicable to dual forms while subsection~\ref{sec:dual form localization} explains why dual forms are localized to cuts. Lastly, we provide a uniform transcendental basis of one-loop dual forms in subsection~\ref{sec:generalized unitarity}.

\subsection{Differential forms for $d$-dimensional integrals}
\label{ssec:dim reg}

A defining property of dimensionally-regulated integrals, originally from \cite{THOOFT1972189, tHooft:1973mfk}, is the product rule: a $d$-dimensional integral is the product of a $4$-dimensional one with a $(-2\eps)$-dimensional one. Correspondingly, all momenta can be split into $4$d and perpendicular components: 
\be
\ell^\mu = (\ell^0,\ell^1,\ell^2,\ell^3,\ell_\perp),  \qquad
p_k^\mu = (p_k^0, p_k^1,p_k^2, p_k^3,0_\perp).
\ee
Throughout this paper, we separate the dimension of spacetime $d=\dint-2\vep$ into its integer part $\dint$ and non-integer part $-2\vep$. Unless specified otherwise, we assume that all external momenta lie within the $(\dint=4)$-dimensional physical subspace; such integrals are referred to as being \emph{near} four dimensions.  (If the external momenta span a higher-dimensional subspace, then we would say that this integral is near a higher dimension, even if $d\to4$ in the end.) The variables to be integrated are then the 4-dimensional components of the loop momenta, plus dot products of the extra components:
\be
 f = f(\ell_k^0,\ell_k^1,\ell_k^2,\ell_k^3,\ell_{j\perp}{\cdot}\ell_{k\perp}), \qquad 1\leq j\leq k\leq L,
\ee
for a total of $n=d_{\rm int}L+\frac{L(L+1)}{2}$ variables.  More precisely, the integrand 
is a $n$-form in these variables.\footnote{
A step of symmetrical integration of the numerator,
with respect to perpendicular directions \emph{only}, may sometimes be necessary.
}

As a simple example, consider a one-loop integral. The radial integration 
measure in the extra dimensions is 
\be
 d^{-2\eps}\ell_\perp \propto (\ell_\perp^2)^{-1-\eps} d(\ell_\perp^2),
\ee
where the proportionality constant is simply half the volume of a unit sphere in $(-2\vep)$-dimensions. Thus, the one-loop integration measure around four-dimensions is really a \emph{five-form}:\footnote{We have assumed that a Wick rotation has already been performed.}
\be
  \boxed{
    \frac{d^{4-2\eps}\ell}{\pi^{(4-2\vep)/2}}
    = d\ell^0\wedge d\ell^1\wedge \cdots d\ell^3 
    \wedge \frac{d(\ell_\perp^2)}{\ell_\perp^2}\times \mathcal{C}_{1}(4-2\vep)\ u_1(\ell_\perp^2)\
   }
\ee
where 
\begin{align}
	u_1 \equiv \mathcal{G}_1(\ell_\perp^2)^{-\eps},
	\quad
	\G_1(\ell_\perp^2) \equiv \ell_\perp^2,
	\quad
	\C_1(m-2\vep) \equiv \frac{1}{\pi^{m/2}\ \Gamma(\frac{m}{2}-2-\vep)}.
\end{align}
Here, the twist $u_1$ is universal to all one-loop Feynman integrals and is the only multi-valued piece of the Feynman integrand. At higher loops, the integrand also depends on \emph{angles} or dot products between the extra-dimensional components. For example, there is the dot product $\ell_{1\perp}{\cdot}\ell_{2\perp}$ in addition to $\ell_{1\perp}^2$ and $\ell_{2\perp}^2$ at two-loops. Thus, a two-loop integrand near $d=4$
is really an \emph{11-form}. Specifically, the measure can be readily calculated (see \cite{Baikov:1996iu}):
\be \label{eq:volform}
 	\frac{d^{4-2\eps}\ell_{1}}{\pi^{(4-2\vep)/2}}
	\frac{d^{4-2\eps}\ell_{2}}{\pi^{(4-2\vep)/2}}
 	= \C_2(4-2\vep)\ u_2\
	\left(\bigwedge_{\mu=0}^3 d\ell_1^\mu\right)\left(\bigwedge_{\mu=0}^3 d\ell_2^\mu\right)
	\wedge d\ell_{1\perp}^2 \wedge d\ell_{2\perp}^2 \wedge d(\ell_{1\perp}{\cdot}\ell_{2\perp})
\ee
where the twist now involves a Gram determinant:
$u_2=\mathcal{G}_2^{-\tfrac12-\eps}$ with $\mathcal{G}_2=\ell_{1\perp}^2\ell_{2\perp}^2-(\ell_{1\perp}{\cdot}\ell_{2\perp})^2$.

Importantly, this measure does not depend on the labelling choice of internal momenta. For example, using a different labelling might replace $\ell_1\mapsto -\ell_1-\ell_2$, under which the formula stays unchanged thanks to properties of the Gram determinant.

Generalizing to $L$-loops, the degree of the integrand is given in table \ref{tab:formdegree}. Explicitly, 
\begin{align}
\label{L loops from Gram}
 	\frac{d^{4-2\eps}\ell_1}{\pi^{(4-2\vep)/2}} {\cdots} \frac{d^{4-2\eps}\ell_L}{\pi^{(4-2\vep)/2}}
 	&=  \frac{\C_L\ u_L}{\G_L^{\lfloor\frac{L+1}{2}\rfloor}} 
		\left(\prod_{i=1}^L\bigwedge_{\mu=0}^3 d\ell_i^\mu \right)
		\bigwedge_{i\leq j\leq L} d(\ell_{i\perp}{\cdot}\ell_{j\perp}), 
\end{align}
where
\begin{align}
	u_L(\ell_1,\dots,\ell_L) &= \mathcal{G}_L^{-\eps-\frac{L+1}{2} + \lfloor\frac{L+1}{2}\rfloor },
	\\
	\G_L (\ell_1,\dots,\ell_L) &= \det \ell_{i\perp} \cdot \ell_{j\perp},
	\\
	\C_L (\dint-2\vep) &= \frac{1}{(\pi^{\dint/2})^L} \prod_{a=0}^{L-1} \frac{1}{\Gamma(-\vep-\dint/2)}.
\end{align}

\begin{table}
	\centering
	\bgroup
	\def\arraystretch{1.5}
	\begin{tabular}{l|cccc}
		number of loops $L$ & 1&2&3&4 \\\hline
		form degree: $n=4L+\frac{L(L+1)}{2}$ & 5&11& 18& 26
	\end{tabular}
	\egroup 
	\caption{
	\label{tab:formdegree}
	The degree of the $L$-loop Feynman integrand near $d=4$.
	}
\end{table}

The standard integration contour for a Feynman integral is the $\mathbb{R}^n$ subspace consisting of real Minkowski momenta (with the usual Feynman $i0$ prescription), times the region over which the Gram matrix $\ell_{i\perp}{\cdot}\ell_{j\perp}$ is positive definite.  As stressed already, this contour is irrelevant for the integral reduction problem that is the focus of this paper: the intersection pairing involves a $(2n)$-dimensional integral over all of $\mathbb{C}^n$.

\subsection{Twisted cohomology and the duals of Feynman integrals}
\label{ssec:duals}

Due to the non-integer exponent in the twist $u_L = \mathcal{G}_L^{-\eps}$, the integrand in dimensional-regularization is multi-valued when viewed as function of complexified momenta. It is convenient to factor out this multi-valuedness:
\be
 d^{4-2\eps}\ell_1{\cdots}d^{4-2\eps}\ell_L\ f(\ell,p) \equiv u_L\ \vphi,
\ee
where $\vphi$ is a single-valued $n$-form. Integration-by-parts identities shift $u_L\ \vphi$ by $d(u_L\ \psi)$ where $\psi$ is a $(n{-}1)$-form, which amounts to shifting $\vphi$ by a \emph{covariant derivative}:
\be
 d\psi + \omega\wedge \psi \equiv \nabla_\omega \psi\quad\mbox{where}\quad \omega=d\log u_L\,.
\ee
Modding out by IBPs means that $\vphi\simeq \vphi + \nabla_\omega \psi$. The connection, $\omega$, is a 1-form with simple poles at the zeros of $u$ (equivalently the zeros of the Gram determinant $\mathcal{G}_L$) and linear in the dimensional regularization parameter $\vep \notin \mathbb{Z}$.
It is curvature-free and keeps track of the multi-valuedness of the original integrand, somewhat analogously to the gauge potential in the Aharonov-Bohm effect.

Feynman integrands, $\vphi$, are automatically closed on $X=\Space$ since they are represented by holomorphic top forms (they are holomorphic on $X$ because all singularities are outside of $X$). This means that Feynman integrals are part of the (twisted) de Rham cohomology group, defined formally as:
\be \vphi\in H^p_\dr (X;\nabla_\omega) 
	= \frac{ \text{ker}\,\nabla_\omega : \Omega^p_\dr (X) \to \Omega^{p+1}_\dr(X)}
	{\text{im}\,\nabla_\omega : \Omega^{p-1}_\dr(X) \to \Omega^{p}_\dr(X)},
\ee
where $\Omega^p_\dr(X)$ is the space of (smooth) $p$-dimensional forms on $X$ \cite{bott1995}. Intuitively, $H^p_\dr$ is the set of $p$-dimensional forms on $X$ that are closed (so their integral is unchanged under small contour deformations), modulo those that are exact (modulo IBPs -- addition of a total derivative).

To define an intersection pairing (inner product), we use Poincar\'e duality in real $(2n)$-dimensional space: $p$-forms are dual to $(2n-p)$-form, where we get a number by integrating their wedge product over the full space.
For this to make sense, the product must be single-valued, which requires the dual integrands to be $u^{-1}\vphi^\dual$ where
$\vphi^\dual$ is single-valued. This is why dual forms come with the opposite sign of $\vep$.
The intersection pairing is then \cite{aomoto2011theory, yoshida2013hypergeometric}:
\be
 	\braket{\vphi^\dual}{\vphi} 
 	\equiv \frac{(-1)^{ \frac{p(p+1)}{2}} }{(2\pi i)^n} \int_X \vphi^\dual_c \wedge \vphi
	= \frac{1}{(2\pi i)^n} \int_X (\vphi^\dual_c)^T \wedge \vphi . 
 \label{pairing}
\ee
where $T$ denotes the transpose of the wedge product.\footnote{This way, if we integrate one variable at a time, anti-holomorphic and holomorphic pairs of differentials are always adjacent: $\int d\bar{z}_n \wedge \cdots \wedge d\bar{z}_1 \wedge dz_1 \wedge \cdots \wedge dz_n \to \int d\bar{z}_n \wedge \cdots \wedge d\bar{z}_2 \wedge dz_2 \wedge \cdots \wedge dz_n \to \cdots \to \int d\bar{z}_n \wedge dz_n$.}
Since $X$ is non-compact, $\vphi^\dual_c$ must have compact support so that the integral is well-defined for any representative
$\vphi$ ({\it ie.} converge near poles and branch points of the Feynman integral).\footnote{Equivalently, one could make $\vphi$ have compact support instead. All that is required is that the \emph{product} $\vphi^\vee\wedge\vphi$ has compact support. However, modifying the Feynman integrals would not
work as well for physical applications in the context of generalized unitarity.}
The set of dual Feynman integrals is thus simply the set of compactly supported forms: $\vphi^\vee_c \in H_\drc^\bullet$.

Demanding \eqref{pairing} to be invariant under changes of representative $\vphi$, forces $\vphi^\dual$ to live in a cohomology space. Suppose that we shift $\vphi$, in \eqref{pairing}, by an IBP identity: $\vphi\mapsto \vphi +\nabla_\omega \psi$. Then, integrating by parts, we see that the pairing is unchanged if and only if the dual forms are closed: $\nabla_{-\omega}\vphi^\dual=0$.
Similarly, if $\vphi^\dual=\nabla_{-\omega}\psi^\dual$ for some $\psi^\dual$, we can integrate-by-parts and get zero since $\vphi$ is closed.

While this definition (almost tautologically) leads to a non-degenerate pairing between cohomology spaces,
the space $H_\drc^\bullet$ is not very convenient for algebraic manipulations.
Fortunately, it is isomorphic to a space of \emph{algebraic} ``relative'' cohomology \cite{Matsumoto:2018aa}.
To explain this, we must distinguish between two different kinds of boundaries: \emph{twisted} and \emph{relative} boundaries. 

Twisted boundaries occur when $u=0$ or $u=\infty$. In these neighborhoods, $\nabla_{\pm\omega}$ is locally invertible and we say that twisted boundaries are \emph{regulated} by $\vep$. For us, the Gram determinant $\{\mathcal{G}=0\}\cup\{\mathcal{G}=\infty\}$ defines the twisted boundaries (also
called twisted singularities).

Relative boundaries are \emph{unregulated} and occur when $\vphi$ has singularities that are not zeros of $u$.
Near these boundaries, $\nabla_{\pm\omega}$ is only invertible up to an integration constant.
These are the dangerous boundaries. For us, propagators $\{D=0\}$ define relative boundaries. 

Put simply, relative cohomology is a scheme to track integration constants at relative boundaries.
These are analogous to the surface terms produced by integration-by-parts on a manifold with boundaries.

Our main proposal is that the dual of Feynman integrals are represented by algebraic forms
relative to the $\{D=0\}$ boundaries: 
\begin{empheq}[box=\fbox]{align*} 
\mbox{Feynman integrands } &\in H^n_\text{alg} (\Space;\nabla_\omega) \quad &\mbox{(alg./holo. forms)},\\
\mbox{dual integrands } &\in H_\text{alg}^n(\Spacedual,\div;\nabla_{-\omega}) \quad &\mbox{(relative forms)}.
\end{empheq}
Equivalently, we can also think of Feynman integrands as smooth forms. Then, the dual forms are smooth relative forms, which are isomorphic to compactly supported forms:
\begin{empheq}[box=\fbox]{align*}
	\mbox{Feynman integrands } 
		&\in H^n_\dr(\Space;\nabla_\omega) \quad &\mbox{(smooth forms)},\\
	\mbox{dual integrands } 
		&\in H^n_\dr(\Spacedual,\div;\nabla_{-\omega}) \quad &\mbox{(smooth relative)},\\
		&\simeq H^n_\drc(\Space;\nabla_{-\omega}) \quad &\mbox{(compact)}.
\end{empheq}
The algebraic perspective will be the most useful one for enumerating dual forms and to derive differential equations.
On the other hand, the second perspective is better suited for computing intersection numbers and extracting master integral coefficients.
Our calculations will exploit the following chain of isomorphisms that connect these spaces:\footnote{If in any situation the definition of the other spaces
were to be ambiguous, the primary definition of the dual space should be taken to be $H_{\rm dR,c}$,
with the other spaces defined such that the isomorphisms holds.}
\begin{equation*}
\begin{tikzcd}
	H^p_\alg \arrow[r] 
	& H^p_\dr \arrow[r, "c"]       
	& H^p_\drc                   \,.
\end{tikzcd} 
\end{equation*}
Algebraic forms are, by definition, holomorphic. The $c$-map, illustrated in appendices \ref{app:delta} and 
\ref{app:equal mass dual tadpole} and further detailed in our subsequent paper \cite{schap},
produces representatives whose anti-holomorphic dependence enters \emph{exclusively} in a very simple manner:
through $\delta$-functions supported on products of small circles. Thus, intersection numbers are computed algebraically via residues.

\subsection{Review of relative cohomology \label{sec:rel cohom}}

We now review relative cohomology, which deals with integration-by-parts on manifolds with boundaries.
One could simply restrict to forms that vanish on all boundaries. Then there would never be boundary terms. 
However, such forms are cumbersome to work with and a better solution is to simply track boundary terms.

By Stoke's theorem, boundary terms are lower-dimensional forms supported on boundaries.
For this presentation, we will assume that some space $Y$ has codimension-1 boundaries defined by equations: $D^{(i)}=\{D_i=0\}$.
For simplicity these will be assumed to be normal crossing divisors\footnote{Normal crossing means that the intersection of all boundaries cross in transversal way. For example, this condition prohibits the intersection of 3 lines at a point as well as lines that that are tangent to a curve.},
so that all codimension-2 boundaries are intersections $D^{(i,j)}=D^{(i)}\cap D^{(j)}$, etc. This situation is realized in our setup of dimensionally regularized
Feynman integrals, with $Y=\Spacedual$.
Then the space of degree-$p$ forms is
\begin{align}
\Omega^p(Y,\{D=0\}) =&\  \Omega^p(Y)
\ \bigoplus_i\  \Omega^{p-1}(D^{(i)})
\  \bigoplus_{(i,j)}\  \Omega^{p-2}(D^{(i,j)})
\cdots
\end{align}
This direct sum is sometimes written in vector notation in the mathematical literature \cite{bott1995, pham2011singularities, Huber2017}. For example $(0,\psi_1,0,\ldots)$ would denote the $(p{-}1)$-form
$\psi_1$ supported on the first boundary component $D^{(1)}$.
Since we have multiple boundaries, we prefer to use the following more readable notation for elements of the direct sum:
\be
 	\th\ \psi, 
 	\quad \delta_{1}(\th\ \psi_1),
	\quad \delta_{2}(\th\ \psi_2),\ \ldots\ , 
 	\quad \delta_{1} \delta_{2} (\th\ \psi_{12}), 
	\quad\ldots 
	\label{delta notation}
\ee
where $\delta_i$ is a purely combinatorial device that anti-commutes with $\nabla$, other $\delta_j$'s and any $1$-form.
We also denote $\delta_I = \delta_{i_1, \cdots, i_{|I|}} = \delta_{i_1} \cdots \delta_{i_{|I|}}$ for any ordered set $I$. Here, $|I|$ is the cardinality of the set $I$.

The ``step functions'' $\theta$ serve two purposes. First, they remind us that we are discussing representative of relative cohomology,
not of cohomology without boundaries. Second, they keep track of the boundaries generated by 
the action of the exterior derivative 
\be
	d \theta  =  \sum_{j} \delta_j\ \theta 
	\implies 
	\nabla (\theta\ \psi) 
	= \theta\ \nabla \psi + \sum_{j} \delta_j \left( \theta\ \psi\big|_{j} \right)	
\label{boundary term}
\ee
where the slash denotes the restriction of the form $\psi$ to the $j$-th boundary $\{D_j=0\}$.
More generally, if $\psi_I$ is a form on the boundary $I$, these definitions give
\be
	\nabla \delta_I (\theta\ \psi_I) 
	= (-1)^{|I|} \delta_I \left(\theta\ \nabla\psi_I + \sum_{j \notin I} \delta_j \left( \theta\ \psi_I\big|_{I,j} \right) \right).
\ee 
The anti-commuting nature of $\delta_j$ ensures that $\nabla^2=0$.

Colloquially, commuting a derivative past a step function produces delta functions.
In fact, the symbols $\theta$ and $\delta$ become literally step and delta functions upon using the isomorphism
to the smooth-compact category (see appendix \ref{app:delta})!

Let us consider an example in order to understand the above seemingly abstract notation.
Suppose that we have two boundaries $D_1$ and $D_2$. Then, a general 2-form $\vphi$ is represented by 
\begin{align}
	\vphi 
	= \theta\ \psi 
	+ \delta_1( \theta\ \psi_1 ) 
	+ \delta_2( \theta\ \psi_2 ) 
	+ \delta_{1,2} ( \theta\ \psi_{12} )
\end{align}
where $\psi \in \Omega^2(Y)$, $\psi_i \in \Omega^1(D^{(i)})$ and $\psi_{12} \in \Omega^0(D^{(1,2)})$. To be an element of relative cohomology, $\vphi$ must be closed:
\begin{align} \label{dphi example 1}
	\nabla\vphi 
	= \theta\ d\psi
	{+} \delta_1\Big( \theta\ \big(\psi\big|_1 {-} \nabla\psi_1 \big) \Big)
	{+} \delta_2\Big( \theta\ \big(\psi\big|_2 {-} \nabla\psi_2 \big) \Big)
	{+} \delta_{1,2} \Big( \nabla\psi_{12} {-} \psi_1\big|_2 {+}\psi_2\big|_1 \Big) 
	= 0.
\end{align}
This puts constraints on the components $\psi, \psi_i, \psi_{12}$. For example, $\psi$ must be an exact 1-form when restricted to any boundary $\{D_i=0\}$:
$\psi\big|_1 = \nabla\psi_1$. 

As a second example, suppose that $\psi_1 \in H^q(D^{(1)},\{D_{\neq 1}=0\};\nabla)$ represents a relative class on a boundary,
where we use the notation $D_{\neq 1}$ to denote all boundaries other than $D_1$.
Then, it is simple to show that $\vphi = \delta_1(\psi_1)$ is a cohomology class in $H^{q+1}(Y, \{D=0\};\nabla)$:
\be
	\nabla \vphi
	= \nabla\delta_1 (\psi_1) 
	= - \delta_1 (\nabla \psi_1) 
	= - \delta_1(0)
	= 0.
\ee
This trivial example illustrates the powerful concept of \emph{Leray coboundary}:
we can generate (co)homology classes on a larger ambient space simply by embedding classes from a boundary sub-manifold.

The process can be repeated to map any codimension-$|I|$ boundary into the ambient space:
\be
 \delta_{I}: \ H^q\bigg(D^{(I)},\bigg\{\prod_{j\notin I}D_j=0\bigg\};\nabla\bigg) \longrightarrow H^{q+|I|}\big(Y,\{D=0\};\nabla\big)\,. \label{delta map}
\ee
This map plays a crucial role in our story. Recall that the \emph{duals} of Feynman integrals live in the relative cohomology $H^n(\Spacedual,\{D=0\};\nabla_{-\omega})$, where the boundaries are cut surfaces: sub-manifolds where some propagators are put on-shell, $D_i=0$. Eq.~\eqref{delta map} creates dual forms starting from lower-dimensional ones that live on cuts. We will find that \emph{all} dual forms are produced this way: all dual forms live on cuts! 

For the algebraic description of relative cohomology in eq.~\eqref{delta notation}, the $\delta_p$ were introduced as boundary labels.
However, when computing intersection pairings \eqref{eq:pairing},
it is necessary to construct an explicit realization of the $\delta$-map for smooth-compact forms on $Y\setminus\{D=0\}$.
A standard construction following Leray is described in appendix \ref{app:delta}. It realizes the image of $\delta_i$ as a distribution
proportional to $d\theta(|D_i|>\epsilon)$ which is supported on a small circle around $D^{(i)}$ (times $D^{(i)}$ itself). Since a representative is supported on a circle, the intersection number reduces to a residue.
As shown in appendix \ref{app:delta},
\be
 \bigg\la \delta_i(\psi^\dual) \bigg\vert \vphi\bigg\ra =
 \bigg\la \psi^\dual  \bigg\vert \res_{D_i=0}\left(\frac{u}{u|_i}\vphi\right)\bigg\ra\,.
\label{eq:residue1}
\ee
where the residue gives a $(p{-}1)$-form if $\vphi$ is a $p$-form.
The factor involving the twist $u$ can be interpreted as parallel transport between the cut $D_i=0$ and its neighborhood.

Eq.~\eqref{eq:residue1} will be important for us: it shows that dual forms, which live on cuts, lead to residue operations on Feynman integrals!

Someone working in the algebraic category could treat eq.~\eqref{eq:residue1} as a definition of the intersection pairing
between relative and algebraic cohomologies: Leray's coboundary $\delta$ is transpose to the residue map.
For our purposes, since our primary definition of the pairing was the integral eq.~\eqref{eq:pairing}, it is satisfying to find eq.~\eqref{eq:residue1} as a consequence
of the isomorphism between relative and smooth-compact forms just mentioned.

For multiple boundaries, the analogous relation is:
\be \label{eq:action of delta in intersection numbers}
\boxed{
	\bigg\la \delta_{1,\dots,p} ( \psi^\vee ) \bigg\vert \vphi \bigg\ra
	= \bigg\la \psi^\vee \bigg\vert \res_{1,\ldots,p} \left(\frac{u}{u\vert_{1,\dots,p}} \vphi \right) \bigg\ra
}
\ee
where $\res_{1,2,\ldots,p} = \res_{\{D_p=0\}} \cdots \res_{\{D_1=0\}}$. 
Note that we don't pick extra minus signs thanks to the transpose operation hidden in our definition of the intersection pairing (see eq.~\eqref{eq:pairing}).
The indices on $\res$ anti-commute since $\res_{i}$ is equivalent to integrating against $d\theta(D_i)$.

This formula makes it clear that the dual form $\delta_{1,\dots,p}(\psi^\vee)$ picks up the $D_1\cdots D_p$ cut of the Feynman integrand.
Furthermore, $\delta_{1,\dots,p}(\psi^\vee)$ is orthogonal to $\vphi$ if $\vphi$ does not have at least simple poles in each of $\{D_i\}_{i=1}^p$.
Thus, the contribution of a boundary component $I$ to the dual cohomology is trivial if there exists no master integrals for the topology $I$ (modulo sub-topologies).

These are the main advantages of working with relative cohomology as opposed to deforming integrals by twisting
propagator singularities as done in \cite{Mastrolia:2018uzb, Frellesvig:2019kgj, Frellesvig:2019uqt,  Mizera:2019vvs, Mizera:2020wdt, Frellesvig:2020qot}.  All integrals contributing to a given physical process combine into a single structure.

\subsection{All dual forms live on cuts \label{sec:dual form localization} }

We have seen that the Leray map can create dual Feynman integrals starting from forms that live on some cut $D_i=0$.
What could this miss?

This question was answered long ago by Leray's \emph{long exact sequence}:
\be \label{longexact}
\begin{array}{rclcl} \displaystyle
&&& \ldots \longrightarrow & H^{q-1}(Y,\{ D_{\neq i}=0\})\ \underset{\kappa_i}{\longrightarrow}\\\displaystyle
H^{q-1}(D^{(i)},\{D_{\neq i}=0\})  &
	\underset{\delta_i}{\longrightarrow}& H^q(Y,\{D=0\}) &
	\underset{\iota_i}{\longrightarrow} & H^q(Y,\{ D_{\neq i}=0\})\ \underset{\kappa_i'}{\longrightarrow} \\\displaystyle
	H^{q}(D^{(i)},\{D_{\neq i}=0\}) &\longrightarrow \ldots &&&
\end{array}\ee
Focusing on the middle line, we are interested in  knowing the central element.
The arrow entering it is the, by now familiar, $\delta$-map, while $\iota_i$ is the map which forgets about boundary terms at $D_i=0$.
Since the sequence is exact, the image of one arrow is the kernel of the next arrow.
Equivalently, all missing forms $H^q(Y,\{D=0\})/{\rm im}\ \delta_i$ are captured by the image of $\iota_i$ in the third factor.
Or, in words, dual forms that do not live on a cut with $D_i=0$
can be understood from simpler sub-topologies that don't involve the propagator $D_i$.
The propagator $D_i$ is either cut or forgotten about.

Analogous sequences for other boundaries $D_i$ can be used to recursively compute the left and right factors,
ultimately reducing to forms that are \emph{simultaneously}
in the image of $\delta_I$ for some multi-cut $I=\{i_1,i_2,\ldots,i_k\}$, \emph{and} in the image of $\iota_{\bar{I}}$ for its complement,
where one forgets about all the other cuts.  In the end, \emph{each} propagator is either cut or forgotten.

These cohomology groups, where a propagator is either cut or forgotten, are guaranteed to cover
all dual forms.  Leray's sequence tells us precisely when and how they can overcount.
This is captured by the groups shown on the first and third lines of eq.~\eqref{longexact}.

First, it could happen that there are relations between dual forms, meaning that forms from different
cuts map to the same element of $H^q(Y,\{D=0\})$.  This phenomenon is familiar from Cauchy's residue
theorem, which provides relations between residues, and is captured by the group in the top line of eq.~\eqref{longexact}.
A simple example is the Riemann sphere relative to three points:
\be
\begin{array}{cccccc} \displaystyle
	{}
	& {}
	& H^0_c(\CP^1,\{0,1,\infty\}) = 0 
	& \underset{\iota_{\{0,1,\infty\}} }{\longrightarrow}
	& H^{0}_c(\CP^1)
	& \underset{\kappa_{\{0,1,\infty\}}}{\longrightarrow}
\\ \displaystyle
	H^{0}_c(\{0,1,\infty\})  
	& \underset{\delta_{\{0,1,\infty\}}}{\longrightarrow}
	& H^1_c(\CP^1,\{0,1,\infty\}) 
	& \underset{\iota_{\{0,1,\infty\}}}{\longrightarrow} 
	& H^1_c(\CP^1)=0
	&
\end{array}
\ee
Notice that we have added the three boundaries in one go here instead of one-by-one as in eq.~\eqref{longexact}. Either way, the sequence works.
The fact that the residues around three points sum up to zero is related
to the existence of constant function on the sphere ($H^{0}_c(\CP^1)$):
\be
0\simeq \nabla(\theta\ 1) = \delta_0(1)+\delta_1(1)+\delta_{\infty}(1)\,.
\ee
Recall from eq.~\eqref{eq:residue1} that $\delta_x(1)$ is supported on a small circle around the point $x$, such
that the pairing computes residues: $\langle \delta_x(1)|\varphi\rangle = {\rm Res}_{z=x}\varphi$.
Like homology, dual forms capture residue theorems without having to discuss the forms they act on.

Second, it could happen that the group immediately after the central one is too big. 
This would mean that some forms which make sense after forgetting propagators,
are not actually in the image of $\iota_i$ and can't be uplifted to forms in the full space.
This obstruction is captured by $\kappa_i'$.
This map is simply the restriction to the boundary, and this obstruction
was stated already in eq.~\eqref{dphi example 1}: in order to be closed,
a form on a manifold with boundary must be exact when restricted to all boundaries.

Both phenomena are captured by ``wrong-dimension'' cohomology groups --- either lower or higher than middle-dimensional.
Based on the general principle that twisted cohomology is ``generically'' concentrated in middle dimension \cite{aomoto2011theory, Matsumoto:2018aa, aomoto1975, cho_matsumoto_1995, AOMOTO1997119, matsumoto1998} ---higher-than-middle dimensional forms vanish whenever holomorphic differentials are the full story--- we expect ${\rm preim}\ \kappa_i$ and ${\rm im}\ \kappa_i'$ to be empty in dimensional regularization,
at least for sufficiently generic external momenta and/or internal masses.
In this case, the middle line of eq.~\eqref{longexact} collapses to a short exact sequence:
\begin{multline}
 0\longrightarrow H^{q-1}(D^{(i)},\{D_{\neq i}=0\}) \longrightarrow_{\delta_i} H^q(Y,\{D=0\}) \longrightarrow_{\iota_i} H^q(Y,\{ D_{\neq i}=0\}) \longrightarrow 0
\\ \mbox{(generically expected in dimensional regularization)}\,.
\label{shortexact}
\end{multline}
The precise meaning of ``generic'' in the context of Feynman integrals remains to be clarified
(we will see some explicit non-generic examples).
When this holds, the space of dual Feynman integrals, modulo those which
come from $\delta_i$ (the $D_i=0$ cut), is exactly the elements of the third group: subtopologies which do not know about the $D_i$ propagator.
Applying this short-exact-sequence recursively to compute the third group, we conclude that the full space is spanned by forms living on boundaries:
\be \label{duals from cuts}
\boxed{ H^n(\Spacedual,\{D=0\};\nabla) = \mbox{span}_I \delta_I\bigg(H^{n-|I|}(\ldots;\nabla)\bigg)} \qquad\mbox{(expected generic case)},
\ee
where $I$ runs over all propagator subsets, {\it ie}. over sub-topologies where we cut all propagators in the set $I$ and forget about the existence of
all the other ones. 

Let us summarize: eq.~\eqref{duals from cuts}, together with the proposed duality (eq.'s~\eqref{eq:action of delta in intersection numbers} and \eqref{pairing intro}),
constitute the main results of this section.  
They show that Feynman integrals are paired one-to-one with dual forms, which can be enumerated from the cohomology of cut surfaces $\{x: D_i(x){=}0\ \forall\ i\in I\}$ \emph{where one forgets about the existence of all uncut propagators}.

In non-generic cases, it could happen that the true space is smaller than a direct sum would predict
(either because of relations or of obstructions), but it can \emph{never} be larger.

Even when the dimensions match, it is important to note that the decomposition \eqref{duals from cuts} is not a direct sum:
the inverse of the third arrow in eq.~\eqref{shortexact} is non-canonical.
The ``unforget'' map on $D_{(I)}$ is only defined modulo dual forms on higher cuts $D_{(IJ)}$.
This is dual to the fact that, on the (non-dual) Feynman integral side, we can add inverse propagators to the numerator of a Feynman integral
to shift it by sub-topologies, without changing its maximal cut.
In practice, we fix these ambiguities by picking particular representatives for forms on cuts;
we then ``unforget'' the remaining cuts by multiplying with the formal step function introduced in eq.~\eqref{delta notation}.
In this paper, we will focus on the one-loop case and discover that natural choices automatically
yield orthogonal pairings.

\subsection{A basis of one-loop dual forms \label{sec:generalized unitarity}}

At one-loop, all cut surfaces are of the form $\Cbb^m$ minus a quadratic form.
For example, if we cut a propagator $D_1=\ell^2+m_1^2=0$,
we can eliminate $\ell_\perp^2= -\sum_{i=1}^4 \ell_i^2-m_1^2$, and we are left with
\be
 \mathcal{S}^m \equiv \Cbb^m \setminus\{\mathcal{G}=0\}, \qquad \mathcal{G}=m^2_1-\sum_i z_i^2.
\ee
We will refer to this space as a (complexified) $m$-sphere.\footnote{The name is natural
if we think about the equation of a $m$-sphere in $\Cbb^{m+1}$: $r^2=\sum_{i=1}^{m+1} z_i^2$.
Projecting onto $\Cbb^{m}$ by eliminating $z_{m+1}$ will give the above space, the singularity $\G=0$
representing the equator.}

On the support of such a cut, all other propagators become linear, so cutting further propagators only involve linear equations:
the intersection of a sphere with a hyperplane is again a sphere.
Thus, the variety obtained by cutting $k$ propagators at one-loop is a $(n-k)$-dimensional sphere $\mathcal{S}^{n-k}$.
The cohomology of these spheres is one-dimensional if the sphere is non-degenerate, and trivial otherwise:\footnote{
More precisely, the cohomology is nontrivial if both Gram determinants in the numerator and denominator of eq.~\eqref{eq:radii}
are nonzero.
}
\be
 \mbox{Proposition:}\hspace{10mm} \dim H^m(\mathcal{S}^m) =\left\{\begin{array}{ll} 
 1, & \mbox{if } r\neq 0,\infty \\
 0, & \mbox{otherwise}\end{array}\right.
\ee
Natural generators for $H^m(\mathcal{S}^m)$ preserve rotational symmetry; including normalization:
\be \label{eq:sphereforms}
	d\Omega^{(m)} =
	c_{m-1}\ r^{\delta_{m,\text{odd}}}\ \frac{dz_1\wedge \cdots \wedge dz_m}{(r^2+\sum_i z_i^2)^{\lfloor\frac{m+1}{2}\rfloor}}
\ee
where 
\begin{align} \label{eq:ci's}
	c_{l} = (2\vep)^{\lfloor\frac{l+1}{2}\rfloor}\ (-\vep)_{\lfloor\frac{l}{2}\rfloor+1} 
\end{align}
$(\vep)_n = \Gamma(\vep+N)/\Gamma(\vep)$ denotes the Pochhammer symbol and $l = -1,0,\dots \dint-1$. The $c_{l}$ have been chosen such that the resulting differential equations for the dual forms is canonical \cite{Henn:2013pwa} and such that the Feynman integrals dual to \eqref{eq:ci's} have the standard transcendental weight (this will be verified in the following sections). It is easy to see that these forms are nontrivial, because they do not integrate to zero on the interior of the real disc $z^2\leq r^2$, which is a valid contour in $H_m(\mathcal{S}^m)$.

We can embed this form into the full loop-momentum-space via the $\delta$-map in \eqref{duals from cuts}, and we conclude (from the long exact sequence \eqref{longexact}) that $H$ is spanned by a sphere for each possible cut topology:
\be \label{eq:delta(sphere)}
\boxed{
H^n(X\setminus \B\cup D) =
\oplus_{I\supset \{D\}, r_I\neq 0,\infty}
\delta_{I} \left[d\Omega^{(n-k)}\right]}
\ee
where the sum runs over non-trivial scalar integral topologies. 

Since the $I$'th term in eq.~\eqref{duals from cuts} is dual to ordinary integrals with $|I|$ propagators in the denominator, it follows that the space is trivial when $I=\emptyset$: $\dim\ H^{n}(\Spacedual;\nabla)=0$. More generally, each contributing term in eq.~\eqref{duals from cuts} must have at least \emph{one cut propagator per loop}. (In a massive theory, there will be single-cut contributions dual to tadpole integrals. In a massless theory, we expect all dual forms to involve at least one Cutkosky cut.)

In \cite{schap}, it will be shown that this basis is dual to the standard scalar integral expansion:
for example, the form $\delta_{D_1,D_2,D_3} [d\Omega^{(2)}]$ extracts the coefficient of the scalar triangle 
with the three propagators $D_1,D_2,D_3$.\footnote{
While the mathematics of intersection numbers involving linear twist has been well studied, not much is known about quadric and higher degree twists (see \cite{aomoto2012, aomoto2015, Aomoto:2017npl} for some recent progress on hypersphere arrangements).
}
The above basis thus parallels the standard one of
scalar tadpoles, bubbles, triangles, boxes and pentagons
for dimensionally regulated integrals with 4-dimensional external momenta. 
While the basis is derived using integration-by-parts identities on the Feynman integral side, it is a consequence of the simple propositions above on the dual-integral side.

\begin{figure}[]
\centering
\includegraphics[align=c,width=.25\textwidth,]{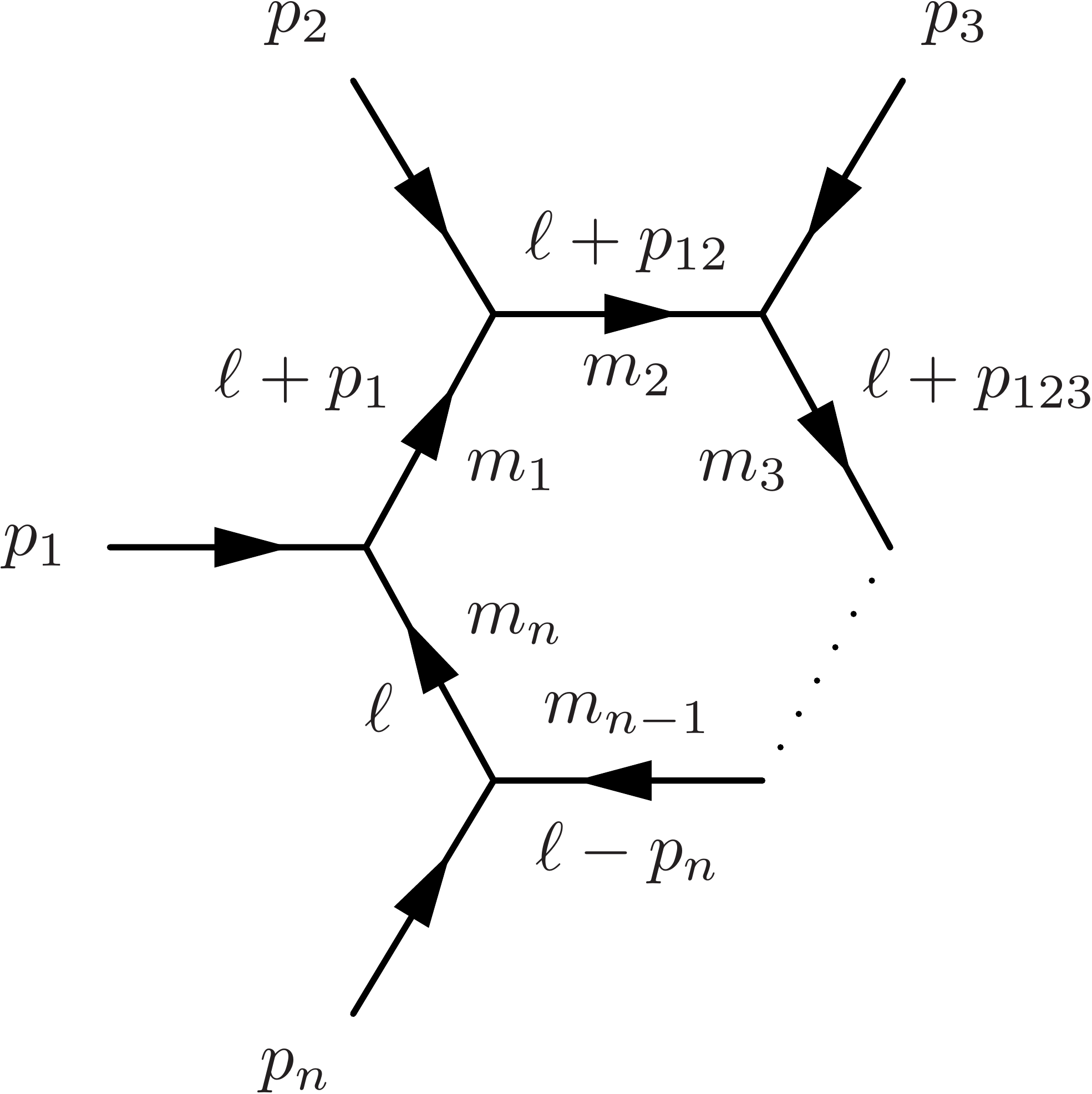}
\qquad
\includegraphics[align=c,width=.25\textwidth]{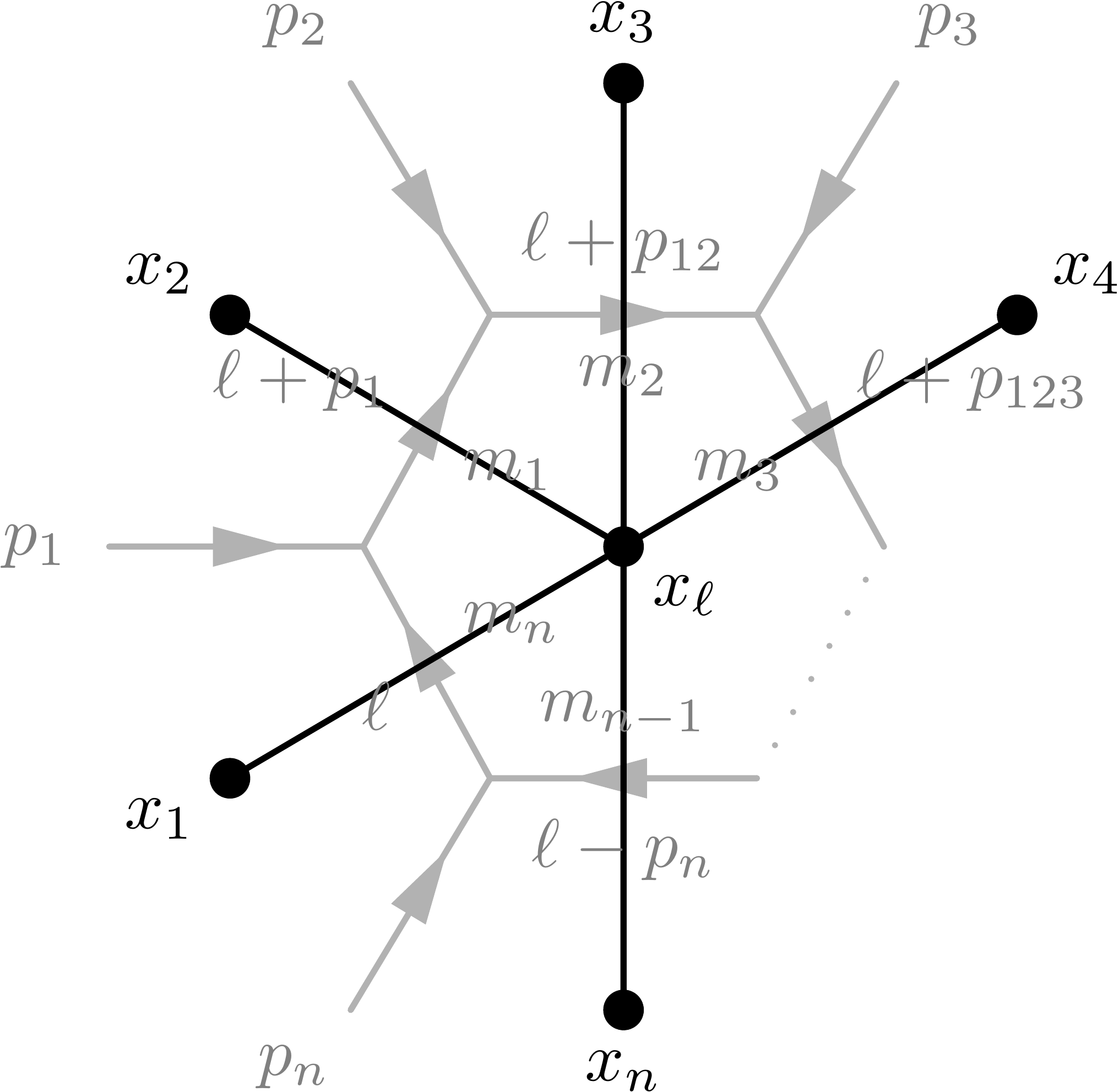}
\caption{
\label{fig:region momenta}
Conventions for one-loop propagators and region momenta.
}
\end{figure}

The radius of the sphere obtained by cutting the set of propagators $I=\{i,j,k,\ldots\}$
can be computed from the minors of the Gram determinant:
\be \label{eq:gram}
 G_{00} = 0, \qquad 
 G_{0i} = G_{i0} = 1, \qquad
 G_{ij} = -\frac{x_{ij}^2+m_i^2+m_j^2}{2}
\ee
where the $x_{i+1}-x_{i} = p_i$ are region momenta (see Fig.~\ref{fig:region momenta}) and $x_{ij}=x_i-x_j$.\footnote{This convention for $G$ differs from \cite{Bourjaily:2019exo} by a relative sign. This was chosen so that $G_{ij} = (X_iX_j)$ where $X_i$ are embedding space vectors and $\eta_\text{embedd}$ is mostly plus.}
Writing $(I)\cdot(J)$ ($(I)^2=(I)\cdot(I)$) for the minor of $G$ where we keep rows $I$ and columns $J$ the radius of the sphere is simply 
\be \label{eq:radii}
	r^2_{I} = \frac{(I)^2}{(0I)^2} 
\ee
For example, the radius of the $m_1$-tadpole cut is
\be
 r^2_{1} = \det\left(-m_1^2\right) \Big/ \det\left(\begin{array}{c@{\hspace{4mm}}c} 0&1\\1&-m_1^2\end{array}\right) = m_1^2,
\ee
while the radius for a bubble cut with external momentum $p$ is
\begin{align}
 r^2_{12} &= 
 	\det\left(
		\begin{array}{c@{\hspace{4mm}}c} 
			-m_1^2 & -\frac{p^2+m_1^2+m_2^2}{2} 
		\\
  			-\frac{p^2+m_1^2+m_2^2}{2} & -m_2^2
		\end{array}\right) \Big/
  	\det\left(\begin{array}{c@{\hspace{4mm}}c@{\hspace{4mm}}c} 
		0&1&1 
		\\ 
		1& -m_1^2 & -\frac{p^2+m_1^2+m_2^2}{2} 
		\\ 
		1 & -\frac{p^2+m_1^2+m_2^2}{2} & -m_2^2
	\end{array}\right)
   \nn \\
   &= \frac{(p^2+(m_1+m_2)^2)(p^2+(m_1-m_2)^2)}{4p^2}.
\end{align}
When $r \to 0,\infty$ (or in fact whenever either determinant vanishes),
the corresponding dual form becomes reducible and are not generators of the cohomology.
These determinants are familiar from the Landau equations.

Thus massless tadpoles, bubbles with massless external momentum and bubbles with momentum at threshold are always reducible,
and are not included in the sum \eqref{eq:delta(sphere)}.
A triangle with two massless corners and no internal mass is similarly also reducible ($r=0$). 

For non-planar four-particle massless scattering, the basis is 6-dimensional (three bubbles and three boxes): \\
\begin{equation*}
\includegraphics[align=c,scale=.25]{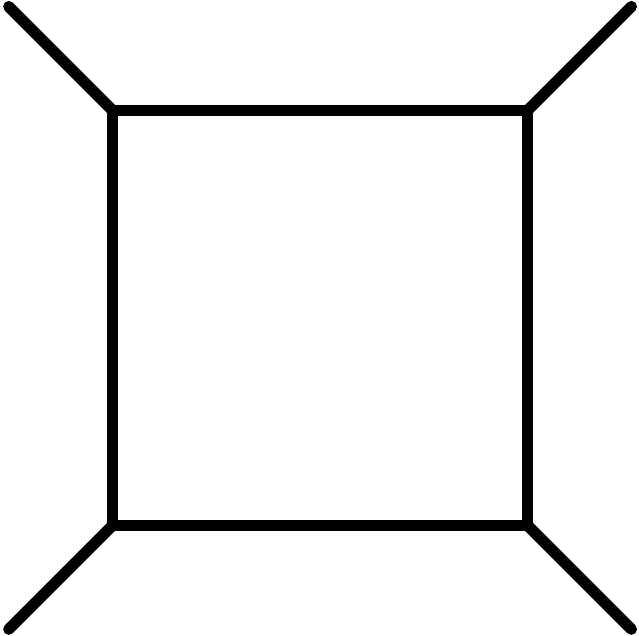}
\qquad
\includegraphics[align=c,scale=.25]{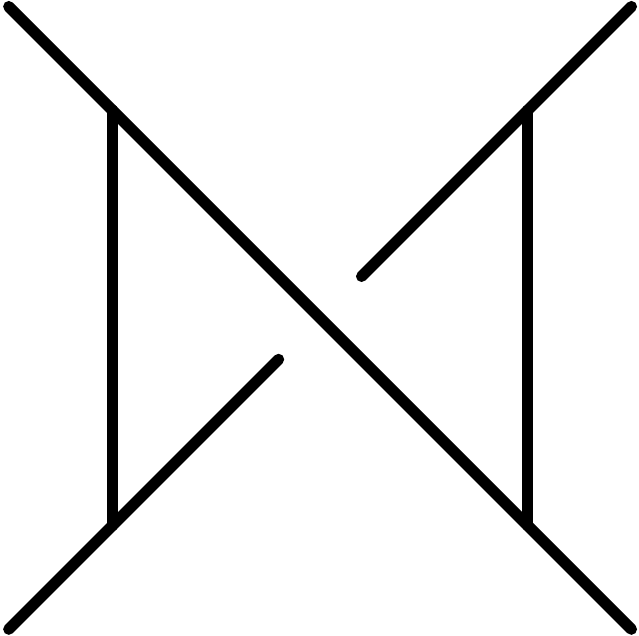}
\qquad
\includegraphics[align=c,scale=.25]{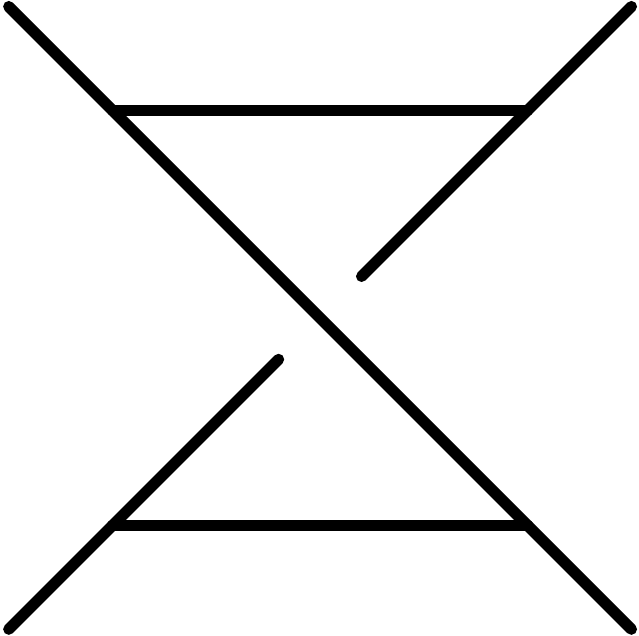}
\qquad
\includegraphics[align=c,scale=.25]{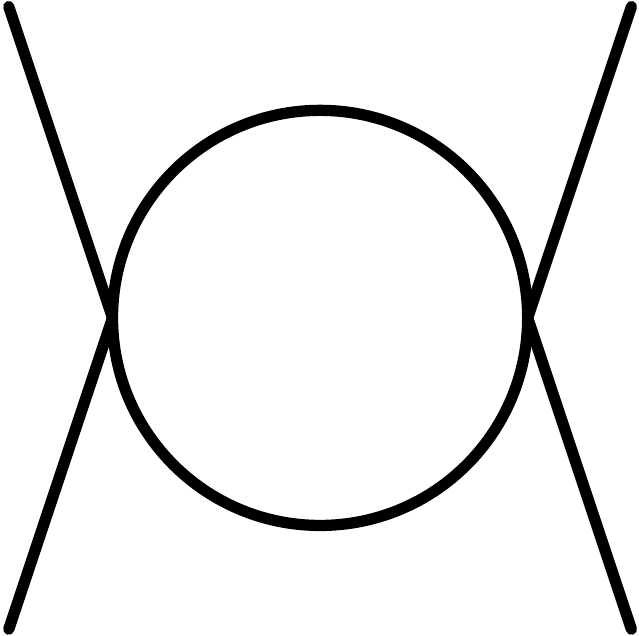}
\qquad
\includegraphics[align=c,scale=.25]{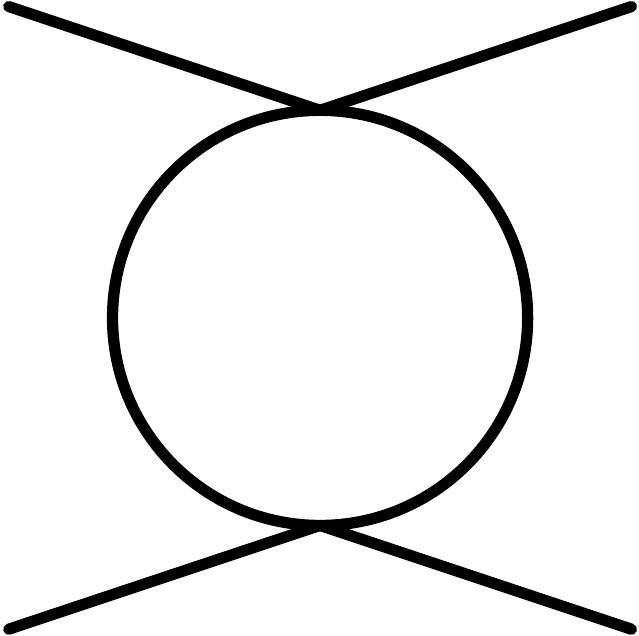}
\qquad
\includegraphics[align=c,scale=.25]{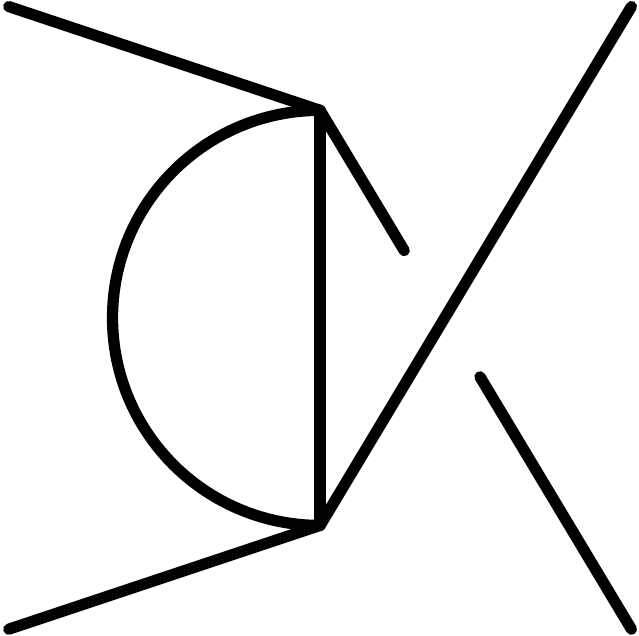}
\end{equation*}\\[0.2em]
while for planar five-particle massless scattering, the basis is 16-dimensional (67 in the non-planar case):\\
\begin{equation*}
5 \times \includegraphics[align=c,scale=.25]{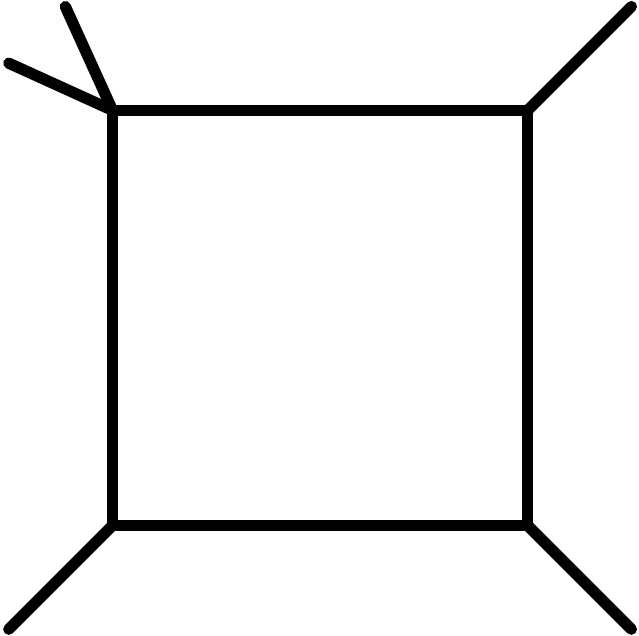}
\quad + \quad 
5 \times \includegraphics[align=c,scale=.25]{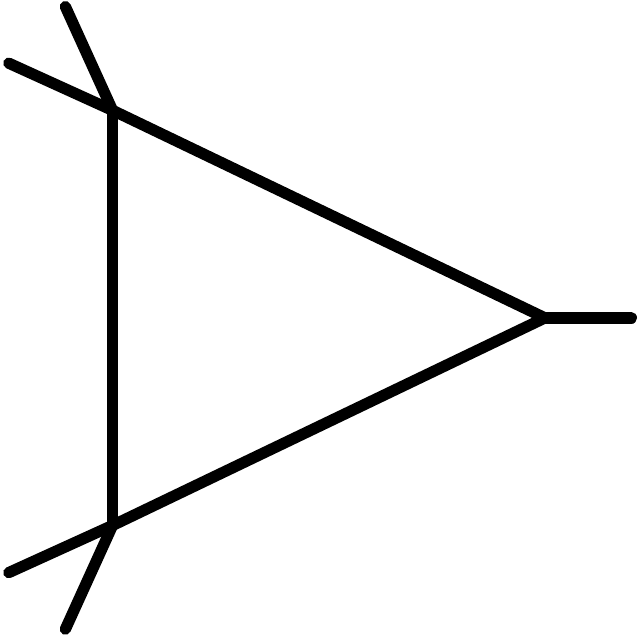}
\quad + \quad 
5 \times \includegraphics[align=c,scale=.25]{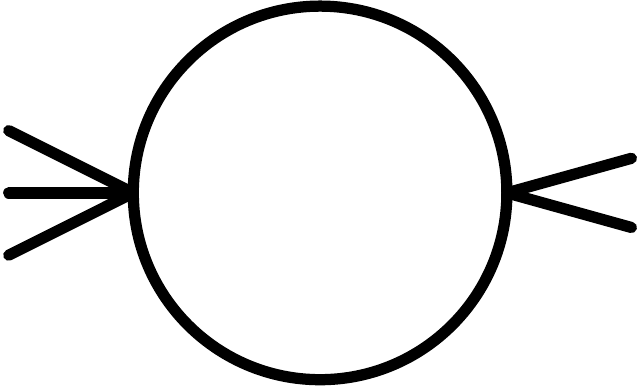}
\quad + \quad 
\includegraphics[align=c,scale=.25]{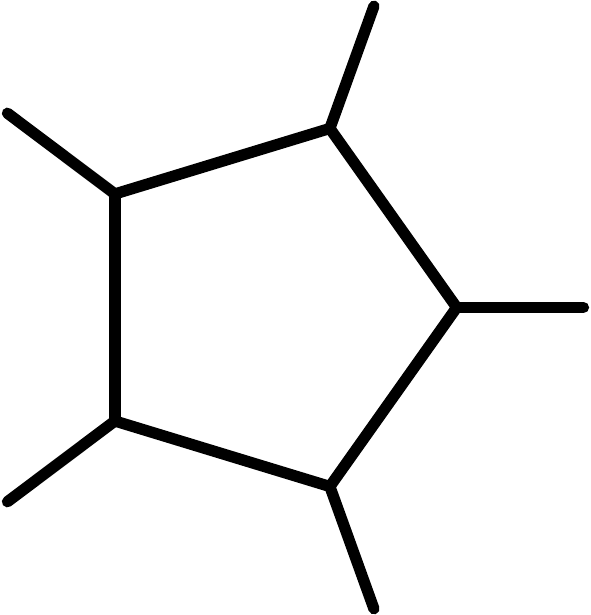}
\ .
\end{equation*}\\[0.2em]
In generic kinematics,
the dimension of the twisted cohomology group is given by the Euler characteristic of the underlying manifold \cite{Bitoun:2018afx, Mizera:2017rqa}. Since the Euler characteristic is given by the alternating sum $\chi = \sum_{p=0}^n (-1)^p\ \text{dim}H^p(\Space;\nabla_\omega)$ and generically only the middle dimensional cohomology is non-trivial in dimensional regularization, the sum collapses and $|\chi| = \text{dim} H^{n}(\Space;\nabla_\omega)$.
We caution the reader, however, that this rule does not always apply in degenerate limits, as exemplified in section \ref{sec:degenerate limits}.

\section{Differential equations for one-loop dual forms\label{sec:one-loop deqs}}

Ultimately, we are interested in integrated amplitudes.
While dual forms will allow us to project any given Feynman integral onto a chosen basis,
the basis integrals themselves must be integrated.
This is often not possible via brute force integration. Instead, it is often more convenient to derive differential equations for the basis integrals in terms of the kinematic parameters (Mandelstam invariants, masses, etc.) \cite{Kotikov:1990kg, Remiddi:1997ny, Gehrmann:1999as, Henn:2014qga, Dlapa:2020cwj, Chicherin:2018old}.
In this section we consider the question of deriving these differential equations, from the perspective of the dual forms.

The idea is that the system of
differential equations satisfied by integrals and their duals are just transpose of each other,
in a basis where the pairing is orthonormal.
Since dual forms are localized to cut surfaces and are pure numerators,
it appears easier to derive differential equations for them than
using standard IBP techniques for dual Feynman integrals. 

In subsection~\ref{sec:deqs gen}, we derive the relation between the FI differential equations and the corresponding dual differential equations. Then in  subsection~\ref{sec:2dDEqs} we compute the kinematic connection for dual forms with $\dint=2$ using standard IBP techniques. This simple example illustrates the main properties of dual forms without introducing additional formalism. Using the embedding space formalism, where the IBP vectors are particularly simple, these results are then generalized to arbitrary $d$. 

\subsection{Differential equations: generalities \label{sec:deqs gen}}

We denote the collection of kinematic parameters by $\{s_a\}$, which we call the kinematic space. The total space consists of kinematic space and internal loop space $\{\ell_a^\mu\}$. In previous sections, $d$ was the external derivative on the internal loop space. From now on, $d$ will mean the exterior derivative on the total space:
 $d = ds_a\ \partial_{s_a} + d\ell_a^\mu\ \partial_{\ell_a^\mu}$.  

\begin{figure}[t]
\begin{equation*}
\begin{minipage}{.4\textwidth}
    \centering
	\includegraphics[scale=.25]{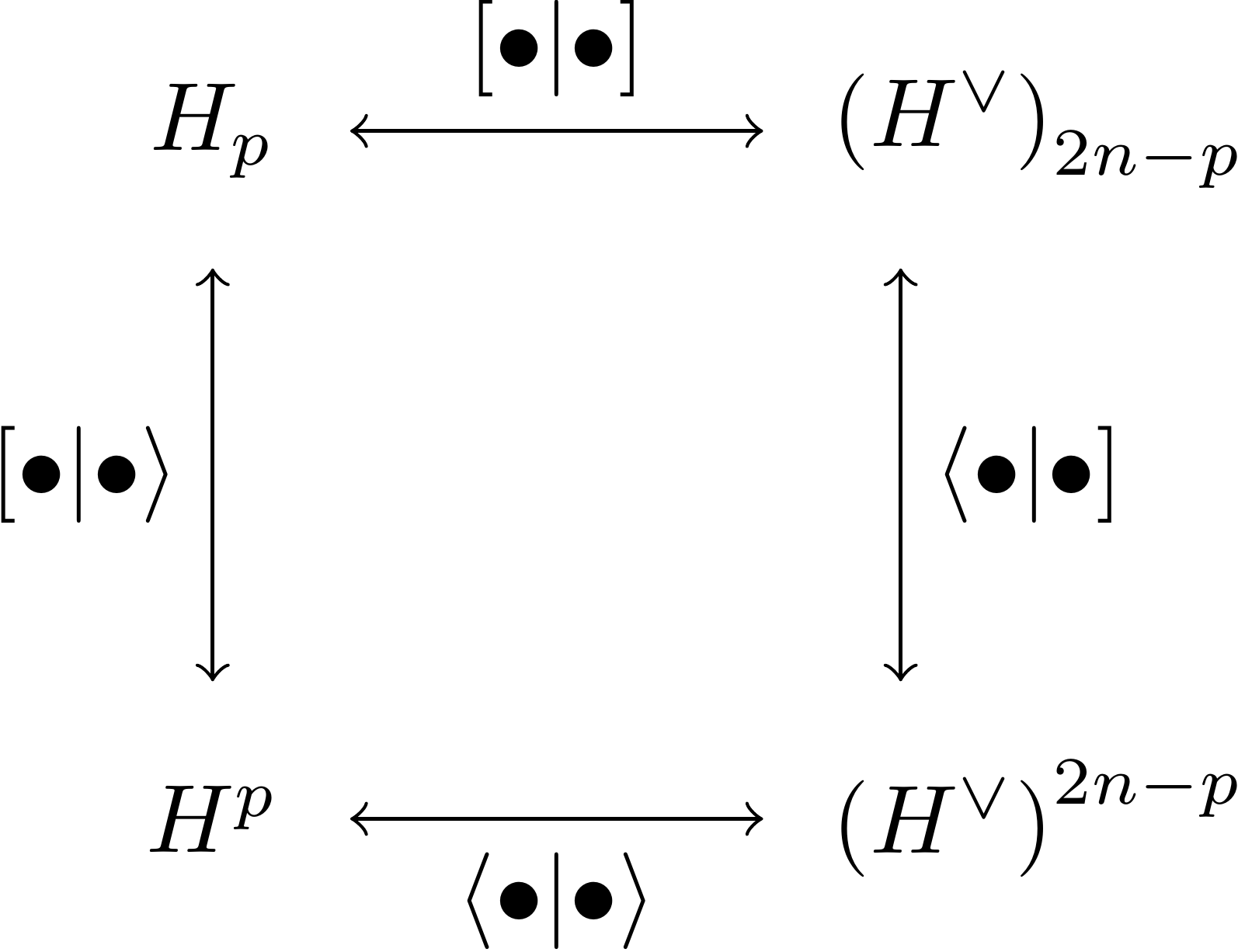}
\end{minipage}
\begin{minipage}{.5\textwidth}
    \centering
	\includegraphics[scale=.25]{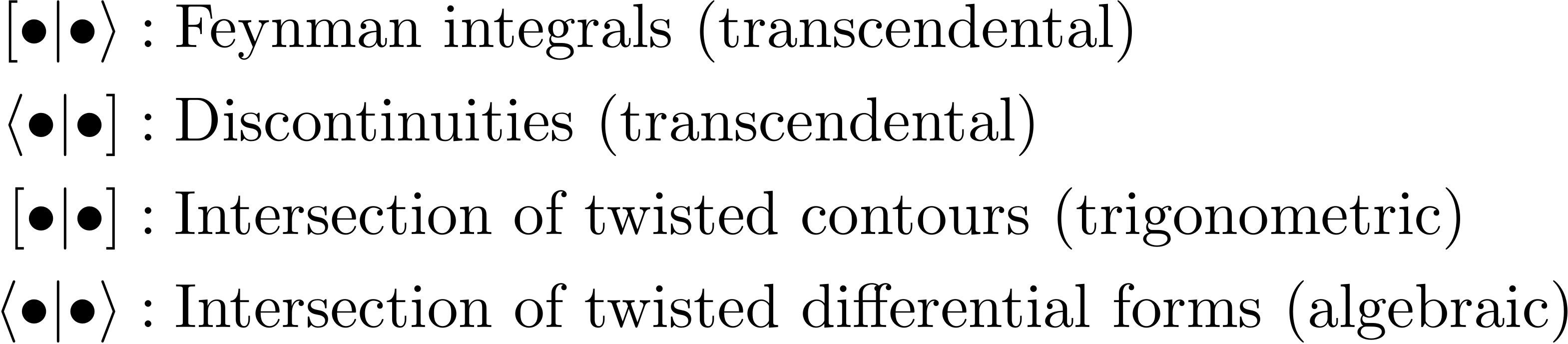}
\end{minipage}
\end{equation*}
\caption{Example pairings between various homologies (cycles $[\bullet|$), cohomologies (forms $|\bullet\rangle$), and their respective duals.
The vertical pairings are difficult and give transcendental functions that are basically the Feynman integrals and their discontinuities.
The horizontal pairings are simpler, and respectively give trigonometric or algebraic functions.
Composing two pairings produces isomorphisms along the diagonals, which are also difficult.
As emphasized in \cite{Mastrolia:2018uzb}, we use the bottom horizontal arrows as much as possible since this paring is always algebraic. \label{fig:cycles}}
\end{figure}

The integrals, technically, are pairings between integration contours and forms, denoted
$\ipsa{\gamma}{\vphi_a}$.   We will thus include the contour $[\gamma|$ in the present discussion,
although it will be a purely passive spectator since the integration-by-parts identities we are interested in are valid on any (closed) contour.
The cycles are part of a homology group $H_{\bullet}$, which describes the space of allowed integration cycles, see \cite{Mimachi:2004aa, aomoto2011theory, yoshida2013hypergeometric, Mizera:2017aa, Mastrolia:2018uzb, Matsumoto:2018aa, Casali:2019ihm}; physically, this space contains all possible analytic continuations and discontinuities (with respect to kinematic invariants) of the integral. That space has its own dual, with an intersection pairing between cycles.
Examples of pairings between cycles, forms and their duals are shown in fig.~\ref{fig:cycles}.

We aim to find a Gauss-Manin connection, $\Omega$, on the kinematic space of FIs \cite{aomoto1982, aomoto1983, aomoto1987gauss, aomoto2015a}.
Taking the total derivative of a FI yields
\be \label{eq:dFI}
	d \ipsa{\gamma}{\vphi_a} 
	= d \int_\gamma u\ \vphi_a 
	= \ipsa{\gamma}{\nabla_\omega \vphi_a}
	= \Omega_{ab} \ipsa{\gamma}{\vphi_b} 
\ee
where the covariant derivative $\nabla_\omega$ comes from $d$ acting on the twist (which is suppressed in the bra-ket notation).
Given a relative homology (dual) cycle $|\gamma]$, the dual connection would be defined analogously 
\be \label{eq:ddualFI}
	d \ipas{\vphi^\vee_a}{\gamma} 
	= \int_{\gamma^\vee} u^{-1}\ \vphi^\vee
	= \ipas{\nabla_{-\omega} \vphi^\vee_a}{\gamma^\vee}
	= \Omega^\vee_{ab} \ipas{\vphi^\vee_b}{\gamma^\vee}.
\ee
As mentioned, the cycles play a passive role and we can equivalently write
\be
|\nabla_\omega \vphi_a\rangle \simeq \Omega_{ab} |\vphi_b\rangle,
\qquad \langle\nabla_\omega \vphi_a| \simeq \Omega^\vee_{ab} \langle\vphi_b|
 \label{eq:dvphi}
\ee
modulo exact forms (with respect to internal $\nabla_\omega$),
and similarly for dual forms.
The kinematic connection $\Omega$ can then be formally computed by intersections:
insert the completeness relation 
\begin{align} \label{completeness relations}
	\mathds{1} = \ket{\vphi_a} (C^{-1})_{ab} \bra{\vphi_b^\dual}
\end{align}
into \eqref{eq:dvphi} to find:
\begin{align}
	\Omega_{ab} = C^{-1}_{bc} \la \vphi^\vee_c \vert \nabla_\omega \vphi_a \ra,
	\quad 
	\Omega_{ab}^\vee = \la  \nabla_{-\omega} \vphi^\vee_c \vert \vphi_c \ra C^{-1}_{cb},
\end{align} 
where $C_{ab} = \braket{\vphi_a^\dual}{\vphi_b}$ is the intersection matrix associated to our chosen basis.

The kinematic connections can also be computed directly from eq.~\eqref{eq:dvphi} using standard IBP techniques.
In this section, to familiarize ourselves with dual forms, we will compute $\Omega^\vee$ in this way using only IBP identities,
but we will test an interesting prediction of intersection theory.
Assuming an orthonormal basis $\ipaa{\vphi^\vee_a}{\vphi_b} = \delta_{ab}$,
it is simple to show that the dual connection is simply the minus-transpose of the kinematic connection:
\be \label{eq:OmegaVeeMinusOmegaT}
	d \ipaa{\vphi_a^\vee}{\vphi_b}
	= \left\la \vphi^\vee_a \right\vert \overset{\leftarrow}{\nabla} \left\vert \vphi_b \right\ra 
	+\left\la \vphi^\vee_a \right\vert \overset{\rightarrow}{\nabla} \left\vert \vphi_b \right\ra
	= \Omega^\vee_{ab} + \Omega_{ab}^T
	= 0 \,.
\ee

Since the dual forms are the $\delta$-image of lower dimensional forms, computing $\Omega^\vee$ appears simpler for dual forms.
For this reason, we advocate that dual forms should be thought of as more fundamental.
Once a nice (uniform transcendental) basis for dual forms is selected, a basis of FIs can be selected such that it is orthonormal to the basis of dual forms. 

We will see that for generic external momenta and internal masses, this approaches picks the standard ``scalar integral basis'', dimensionally shifted to their ``natural''
dimensions where they are pure: $d\simeq 2$ for tadpoles and bubbles, and $d\simeq 6$ for scalar pentagons.
Furthermore, in massless or singular limits, the dual form have smooth limits which automatically pick out special combinations of integrals
(such as the infrared-subtracted scalar boxes).

In the rest of this section we will derive the dual kinematic connection for a basis of dual forms near 2-dimensions, and, by passing to embedding space we generalize our construction to dual forms near any integer dimension (subsection~\ref{sec:4dDEqs}).

\subsection{Warm-up: differential equations near 2-dimensions\label{sec:2dDEqs}}

\begin{figure}[]
\centering
\includegraphics[align=c,width=.20\textwidth,]{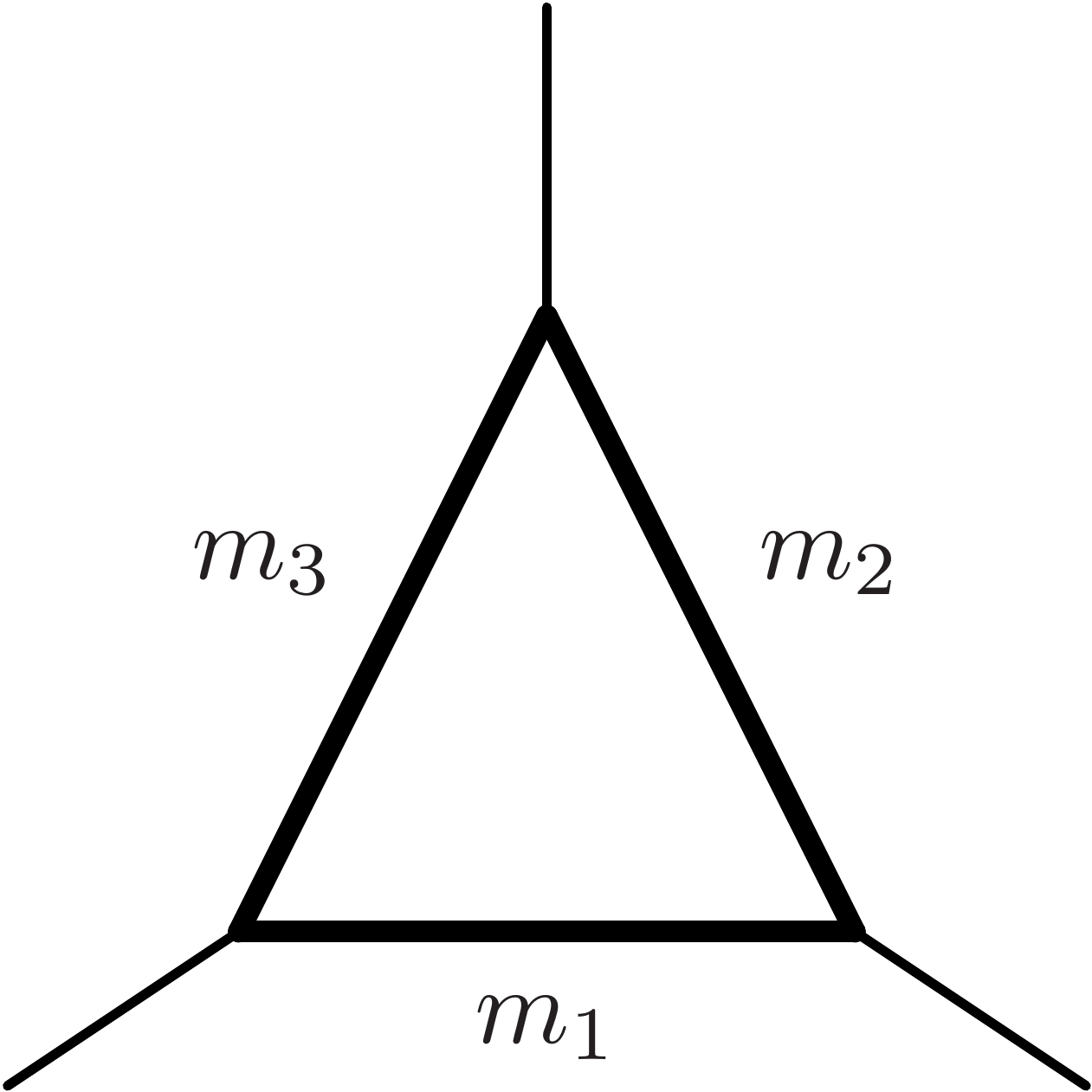}
\quad
\raisebox{0em}{\includegraphics[align=c,width=.20\textwidth]{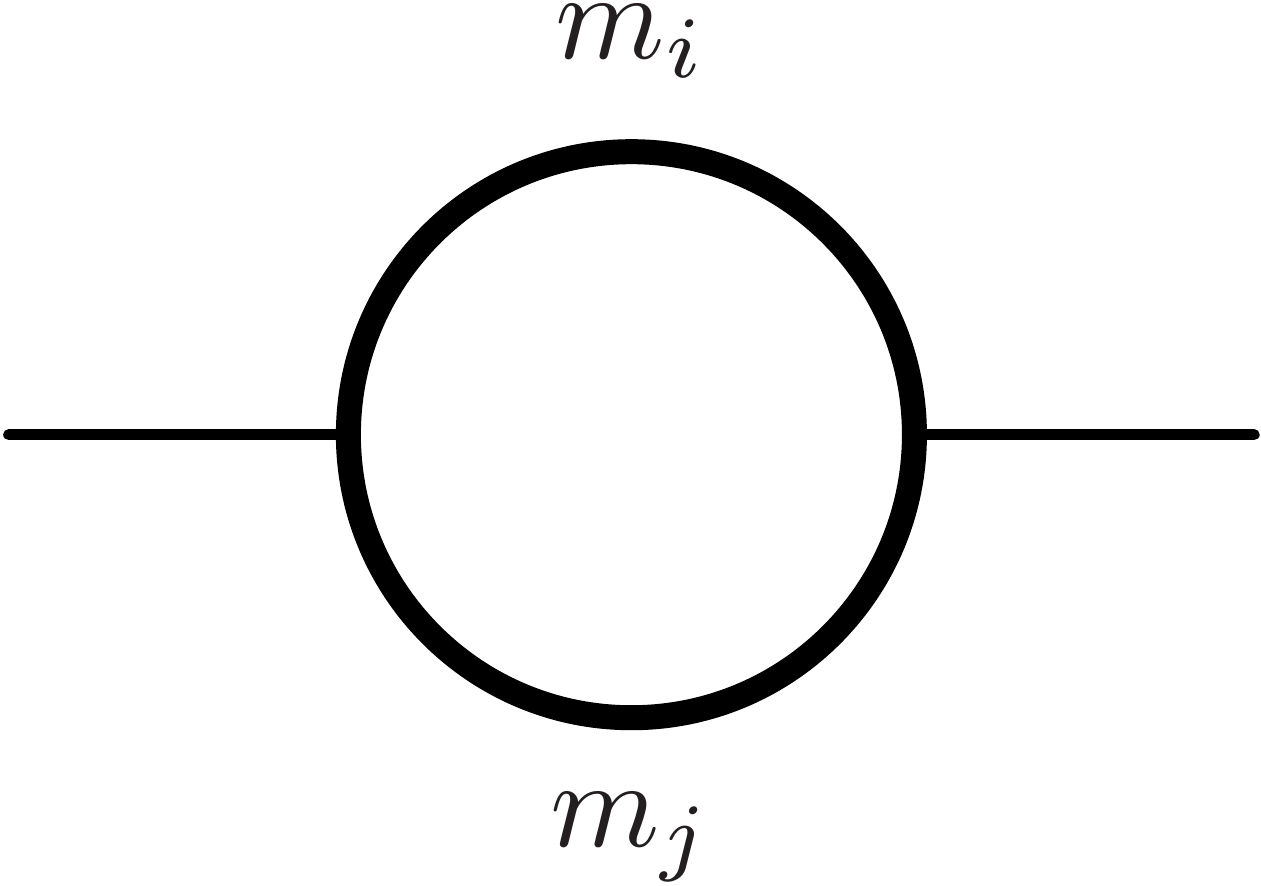}}
\quad
\raisebox{0em}{\includegraphics[align=c,width=.20\textwidth]{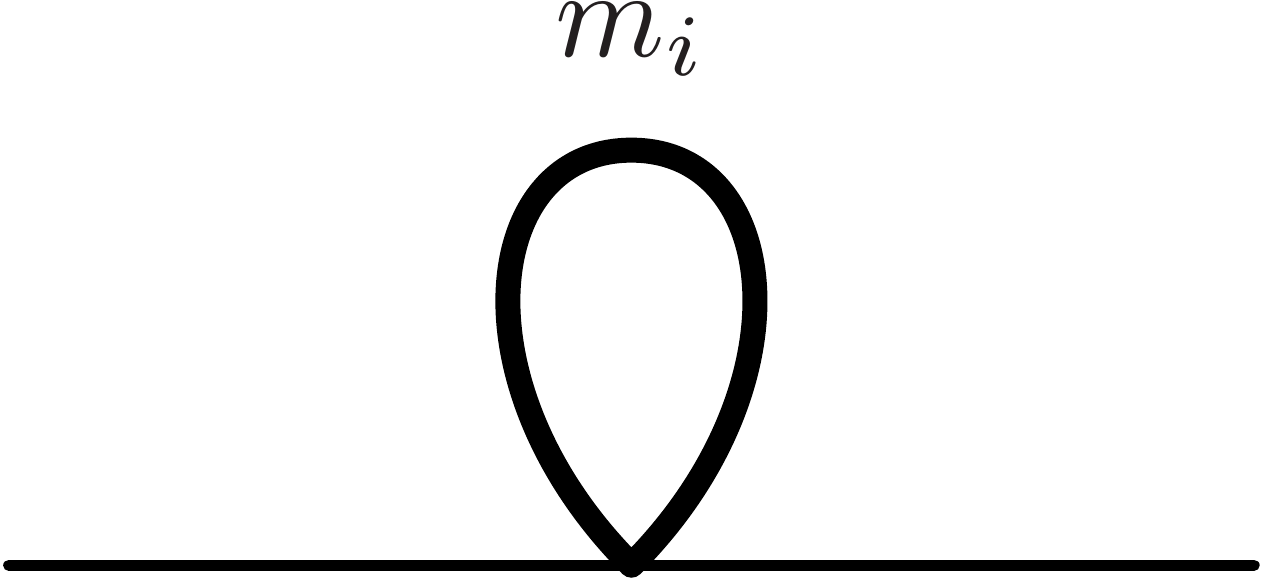}}
\caption{
\label{fig:2dim masters}
Master integrals for a simple one-loop 3-point process near 2-dimensions.
}
\end{figure}

Having set up the basic notation, we consider the basis of dual forms for a simple 3-point process near 2-dimensions (fig.~\ref{fig:2dim masters}),
with generic internal and external masses.
The basis consists of one triangle, three bubbles (different ways of pinching the triangle to a bubble) and three tadpoles (different ways of pinching to a tadpole).

Since we are near 2-dimensions we can assume that two legs of the triangle span the space of physical momenta without loss of generality. Let the propagators of the triangle be 
\begin{align}
	D_1 = \ell^2 + m_1^2, \quad
	D_2 = (\ell+p_1)^2 + m_2^2, \quad
	D_3 = (\ell^2+p_1+p_2) + m_3^2.
\end{align}
where each propagator, $D_i=(\ell+q_i)^2+m_i^2$, is a sphere of radius $m_i$ with centre at $-q_i$. Individually, the on-shell conditions place quadratic constraints on the loop momentum. However, once we take a single cut  (restrict $\ell_\perp$), the propagators are linearized and we are left with the hyperplanes defined by $D_{j}\vert_i = 2\ell \cdot (q_j - q_i) + q_j^2 - q_i^2 + m_j^2 - m_i^2$ where $j \in \{1,2,3\} \setminus \{i\}$.
The twist $u = (\ell_\perp^2)^{-\vep}$ has a spherical form with radius dictated by the first mass.
Further cuts only change the radius of the spherical twist, since the intersection of a sphere and a hyperplane is also a sphere.

Restricting to $\{D_i=0\}$, the only element of $H^2( D^{(i)} \setminus \{\G|_i=0\}, D_{(\{1,2,3\} \setminus \{i\})}|_i; \nabla_{-\omega})$ that is not the $\delta$-map of a lower-dimensional form, is the tadpole 
\begin{align}
	\vphi^\vee_{{\rm tad}_i} 
	= c_{1}\ \delta_i \left( \th\ \frac{ d^2k_i }{ r_i^2 + k^2_i } \right).
\end{align}
Here, $k_i$ is the physical component of the loop momentum perpendicular to the cut $D^{(i)}$ and $r_i^2 + k_i^2$ is the restriction of the twist to the cut surface. Likewise, the only element of $H^1( D^{(i,j)} \setminus \{\G|_{i,j}\}, D_{(\{1,2,3\}\setminus\{i,j\})}|_{i,j}; \nabla_{-\omega})$ that is not the $\delta$-map of a lower-dimensional form, is the bubble
\begin{align} \label{eq:2d bubble}
	\vphi^\vee_{{\rm bub}_{ij}} 
	= c_{0}\ \delta_{i,j} \left( \th\ \frac{ r_{ij}\ dk_{ij} }{ r_{ij}^2 + k^2_{ij} } \right),
\end{align}
where $k_{ij}$ is the physical component of the loop momentum perpendicular to the cut $D^{(i,j)}$ and $r_{ij}^2 + k_{ij}^2$ is the twist restricted to the cut surface. On the triple cut $D^{(1,2,3)}$, the cohomology $H^1( D^{(1,2,3)}; \nabla_{-\omega})$ consists of the single element, namely $1$. This corresponds to the triangle dual form
\be
	\vphi^\vee_\tri = c_{-1}\ \delta_{i,j,k}(d^0k_{ijk})
\ee
where $d^0k_{ijk}=d\ell_\perp^2\wedge d^2\ell/dD_i \wedge dD_j \wedge dD_k$ is a sign factor that preserves the orientation of the original measure and keeps the triangle independent of the order that the cuts are taken. 
Similarly, the anti-symmetry of $dk_{ij}$ compensates for the anti-symmetry of the $\delta_{i,j}$ so that the bubble is also independent of the order that cuts are taken.
The 3-tadpoles, 3-bubbles and single triangle constitute a basis for the relative cohomology for $\dint =2$.

\subsubsection{Derivative of tadpole-dual agrees with derivative of bubble-integral\label{sec:2dDEqs-tadpole}}

The algorithm for obtaining the kinematic connection for dual forms via integration-by-part identifies is similar to the standard procedure used on FIs -- the only difference being the presence of boundary terms. In this section, we focus on the tadpole components of the dual kinematic connection. Since the bubble and tadpole FIs are simple functions, the tadpole-tadpole and tadpole-bubble components are straightforward to verify explicitly. 

We further specialize to the tadpole defined by the single cut $D^{(1)}$ since the other tadpoles are related by symmetry
\be
	\vphi^\vee_\tad = c_1\ \delta_1 \left( \frac{\theta\; d^2k_1}{r_1^2 + k_1^2}\right).
\ee
On the $\tad$-cut, the twist is $u = (r_1^2 + k_1^2)^{-\vep}$ and we are working on $D^{(1)} \setminus \{\G|_{1}=0\}$ relative to $D_{(2,3)}|_{1}$ where
\begin{align}
	D_2\vert_{1} &= 2 k_1 \cdot p_1 + p_1^2 + m_2^2 - m_1^2, \\
	D_3\vert_{1} &= 2 k_1 \cdot (p_1 + p_2) + (p_1+p_2)^2 + m_3^2 - m_1^2.
\end{align}

Taking the covariant derivative of $\vphi_\tad^\vee$ yields
\begin{align} \label{eq:d(2d-tad) 1}
	\nabla\vphi_{1}^{\vee}
	&= c_1\ \delta_{1} \left( 
	\theta\ \frac{ 2 (1{-}\varepsilon) r_1\ dr_{1} \wedge d^{2}k_1 }{ (r_{1}^{2}+k_1^{2})^{2}} 
	\right)
	\nn	\\ & \qquad
	+ c_1\ \delta_{1,2} \left( -\theta\ \left. \frac{d^{2}k_1}{r_{1}^{2}+k_1^{2}} \right\vert_2 \right)
	+ c_1\ \delta_{1,3} \left( -\theta\ \left. \frac{d^{2}k_1}{r_{1}^{2}+k_1^{2}} \right\vert_3 \right)
	. 
\end{align}
Note that we have two different kinds of forms: one that is localized to the $\tad$-cut and two terms localized on bubble cuts ($\bub$ and $\bubbb$). None of the terms in \eqref{eq:d(2d-tad) 1} are in the desired form and must be reduced back to the original basis. This is done by adding total derivatives to \eqref{eq:d(2d-tad) 1} until each term is a kinematic 1-form (dual kinematic connection) wedged with a basis form. For $\text{tad}_1$, this requires the tracking of two feed downs: (1) $\text{tad}_1 \to \text{bub}_{12} \to \text{tri}$ and (2) $\text{tad}_1 \to \text{bub}_{13} \to \text{tri}$. 

To simplify the $\delta_1$ component of \eqref{eq:d(2d-tad) 1}, we subtract the covariant derivative of an IBP-form on the $\tad$-cut. Explicitly, 
\begin{align} \label{eq:d(2d-tad) 2}
	\nabla \vphi^\vee_\tad - \nabla \text{IBP}_1
	&= \vep\; d\log r_1^2 \wedge \vphi^\vee_\tad
	+ c_1\ \delta_{1,2}\left(
		\theta \left.
			\left( 
				\frac{ k_1^2\ dr_1 \wedge d\vartheta_{1} }{ r_1 (r_1^2 + k_1^2) }
				-\frac{ d^2k_1 }{ r_1^2 + k_1^2 }
			\right) \right\vert_2
	\right) 
	\nn \\ & \quad 
	+ c_1\ \delta_{1,3}\left(
		\theta \left.
		\left(
			\frac{ k_1^2\ dr_1 \wedge d\vartheta_{1} }{ r_1 (r_1^2 + k_1^2) } 
			-\frac{ d^2k_1 }{ r_1^2 + k_1^2 }
		\right) \right\vert_3
	\right) 
\end{align}
where 
\begin{align}
	\text{IBP}_1 
	&:= c_1\ \delta_1\left(
		\theta\; \frac{ k_1^2\ dr_1 \wedge d\vartheta_{1} }{ r_1 (r_1^2 + k_{1}^2) }
	\right)
\end{align}
and $\vartheta_{1}$ is the angle in the polar decomposition of $k_1$. Thus, the $\tad$-$\tad$ component of the dual kinematic connection is $\Omega_{1;1}^\vee = \vep\ d\log r_1^2$. More generally, the diagonal terms in the dual kinematic connection are always the $d\log$ of the associated radius (at one-loop). 

Next, the $\delta_{12}$ and $\delta_{13}$ components of \eqref{eq:d(2d-tad) 2} need to be reduced. To construct the necessary IBP-form, the restrictions in \eqref{eq:d(2d-tad) 2} must be taken (in particular, the restriction of $d^2k_1$ and $\vartheta_1$). Therefore, we parameterize $k_1$ in terms of the variables on the corresponding bubble cuts and the external kinematics. Since the procedure is symmetric, we illustrate it for the $\delta_{12}$ term only. On the $\bub$-cut, the vector $k_1$ naturally separates into a component that is fixed by the constraint from setting $D_2|_1=0$ and a component that lives on the $\bub$ cut. Once restricted, 
\be
	k_1^\mu = \alpha_{12}\ e_1^\mu + k_{12}\ e_2^\mu
\ee
where $\alpha_{12} = \sqrt{r_{12}^2 - r_{1}^2} = (p_1^2+m_2^2-m_1^2)/(2 \sqrt{p_1^2})$, $k_{12} = k_1 \cdot e_2$  and the $e_i^\mu$ are the orthonormal unit vectors
\begin{align}
	e_1^\mu = \frac{p_1^\mu}{\sqrt{p_1^2}},
	&\qquad 
	e_2^\mu = \sqrt{\frac{(1)^2}{(12)^2}} \left( p_2^\mu - \frac{(p_1 \cdot p_2)}{p_1^2} p_1^\mu \right).
\end{align}
Then, substituting into the measure $d^2k_1$ yields
\begin{align}
	d^2k_1\vert_2 = d\alpha_{12} \wedge dk_{12}
	+ (e_1 \cdot de_2) \wedge k_{12}\ dk_{12}. 
\end{align}
The dot product $e_1 \cdot de_2$ encodes how the hyperplanes, defined by orthongalization of the momenta running through $D_2$ and $D_3$, are changing relative to each other. Similarly, taking $\vartheta_1$ to be the angle $\vartheta_1= (e_1 \cdot k_1)/\sqrt{k_1^2}$ yields 
\begin{align}
	d\vartheta_1\vert_2 = \alpha_{12} \frac{dk_{12}}{(k_1^2)|_2}.
\end{align}
Thus, we find that the $\delta_{12}$ term of \eqref{eq:d(2d-tad) 2} simplifies to 
\begin{align}
	& c_1\ \delta_{1,2}\left(
		\theta\; \frac{\alpha_{12}\ dr_1 \wedge  dk_{12} }{ r_1 (r_{12}^2 + k_{12}^2) }
		-\theta\; \frac{d\alpha_{12} \wedge dk_{12}}{ r_{12}^2 + k_{12}^2 }
		-\theta\; \frac{(e_1 \cdot de_2) \wedge k_{12}\ dk_{12}}{ r_{12}^2 + k_{12}^2 }
	\right),
	\nn \\
	&\qquad = \frac{1}{2} \frac{c_1}{c_0} d\log
	\left(
		\frac{ r_{12} + \sqrt{r_{12}^2 - r_{1}^2} }{ r_{12} - \sqrt{r_{12}^2 - r_{1}^2} }
	\right)
	\wedge \vphi^\vee_\bub
	+ c_1\ \delta_{1,2}\left(
		-\theta\; \frac{(e_1 \cdot de_2) \wedge k_{12}\ dk_{12}}{ r_{12}^2 + k_{12}^2 }
	\right).
\end{align}
Note that the second term above is proportional to $\omega^\vee_{12}=\omega^\vee\vert_{12}$ and is therefore a total derivative on the $\bub$-cut
\begin{align}
	-\frac{(e_1 \cdot de_2) \wedge k_{12}\ dk_{12}}{ r_{12}^2 + k_{12}^2}
	= \frac{\omega_{12}^\vee \wedge (e_1 \cdot de_2) }{2\vep}
	= \nabla^\vee \left(\frac{e_1 \cdot de_2}{2\vep}\right).
\end{align}
This term can be removed by subtracting the IBP vector,
\begin{align}
	\text{IBP}_{12} &:=
	c_1\ \delta_{1,2} \left(\th\ \frac{e_1 \cdot de_2}{2\vep}\right)
\end{align} 
at the cost of introducing a term on the $\tri$ cut
\begin{align}
	\text{eq.~(\ref{eq:d(2d-tad) 2})} - \nabla \text{IBP}_{12}
	&= \vep\; d\log r_1^2 \wedge \vphi^\vee_\tad
	+\frac{1}{2} \frac{c_1}{c_0} d\log
	\left(
		\frac{ r_{12} + \sqrt{r_{12}^2 - r_{1}^2} }{ r_{12} - \sqrt{r_{12}^2 - r_{1}^2} }
	\right)	\wedge \vphi^\vee_\bub
	\nn \\
	&\quad 
	+ \frac{c_1 (e_1 \cdot de_2)}{2\vep c_0} \wedge \vphi^\vee_\tri
	+ \delta_{13}\text{-term} .
\end{align}
With the $\delta_{12}$ term now the desired form, we can read off the main result of this calculation: the
$\tad$-$\bub$ component of the dual kinematic connection
\begin{align}
	\Omega^\vee_{1;12} 
	= \vep d\log
	\left(
		\frac{ r_{12} + \sqrt{r_{12}^2 - r_{1}^2} }{ r_{12} - \sqrt{r_{12}^2 - r_{1}^2} }
	\right) \label{2d tadpole result}
\end{align}
where we set $c_1/c_0 = 2\vep$.  Let us now compare this result with the derivative of the familiar bubble Feynman integral.

We normalize the bubble as (see \cite{Spradlin:2011wp, Bourjaily:2019exo})
\begin{align}
	\mathscr{I}_{2(1{-}\vep)}[\vphi_{12}] 
	&\equiv\vep \int \frac{d^{2(1{-}\vep)}\ell}{\pi^{1{-}\vep}} \frac{2\sqrt{-(12)^2}}{(\ell^2{+}m_1^2)((\ell+p)^2{+}m_2^2)}
	\nn \\
	&= 2\vep\Gamma(1{+}\vep)\sqrt{-(12)^2} \int_{0}^1 \frac{dx}{\Delta^{1+\vep}}\,,
\end{align}
where $\Delta = x(1-x) p^2 + x m_2^2 + (1-x) m_1^2$, ${-}(12)^2 = [(p^2+m_1^2+m_2^2)^2-4m_1^2m_2^2]/ 4$ and $x$ is a standard Feynman parameter.
Similarly, we define the tadpole FI as
\begin{align} \label{tadpole norm}
	\mathscr{I}_{2(1-\vep)}[\vphi_{i}] 
	\equiv -\vep \int \frac{d^{2(1-\vep)}\ell}{\pi^{1-\vep}} \frac{\sqrt{-(0i)^2} }{(\ell^2+m_i^2)}
	= -\Gamma(1{+}\vep) \int_{0}^1 \frac{dx}{(m_i^2)^\vep}
	= -\Gamma(1{+}\vep) (m_i^2)^{-\vep}. 
\end{align}
Compared with the above references, we have included an extra factor of $\vep$ so that the integrals have pure transcendental weight zero.  The normalizations
are not independent: our tadpole coincides with a bubble where a propagator is taken to infinity
(replacing the label of the erased propagator with $0$ in the minor normalizing the bubble).\footnote{This is responsible for the slightly unusual sign of eq.~\eqref{tadpole norm}
since a propagator at infinity comes with $-1/2$, see eq.~\eqref{embedding YX}.
This convention leads to a cleaner differential equation.}
Using standard IBP techniques, it can be confirmed that these integrals satisfy a differential equation where the off-diagonal component
is precisely minus that in eq.~\eqref{2d tadpole result} (with diagonals minus $\vep d\log r^2$ as in eq.~\eqref{eq:d(2d-tad) 2}). By \eqref{eq:OmegaVeeMinusOmegaT}, this confirms the duality between our dual forms and Feynman forms. 
Here, we will examine the $\vep\to0$ limit of these integrals. In this limit, the bubble can be easily integrated
\begin{align}
	\mathscr{I}_{d=2}[\vphi_{12}] 
	&= \frac{2\vep \sqrt{-(12)^2}}{p^2 (R_+-R_-)} \int_{0}^1 d\log\left( \frac{x-R_-}{x-R_+} \right),
	\nn\\ 
	& = \vep\log\left( \frac{p^2+m_1^2+m_2^2+\sqrt{(p^2+m_1^2+m_2^2)^2-4m_1^2m_2^2}}{p^2+m_1^2+m_2^2-\sqrt{(p^2+m_1^2+m_2^2)^2-4m_1^2m_2^2}} \right)+O(\vep^2)
\end{align}
where $R_\pm = \frac{p^2-m_1^2+m_2^2\pm\sqrt{(p^2+m_1^2+m_2^2)^2-4m_1^2m_2^2}}{2p^2}$ are the $x$-roots of $\Delta$.
Taking the derivative of this function
we then find perfect agreement with eq.~\eqref{2d tadpole result}:
\begin{align}
	d\mathscr{I}_{d=2}[\vphi_{12}] 
	&= \vep d\log\left( \frac{p^2+m_1^2+m_2^2+\sqrt{(p^2+m_1^2+m_2^2)^2-4m_1^2m_2^2}}{p^2+m_1^2+m_2^2-\sqrt{(p^2+m_1^2+m_2^2)^2-4m_1^2m_2^2}} \right) +O(\vep^2)
	\nn \\
	&= -\Omega^\vee_{1;12} \mathscr{I}_{d=2}[\vphi_1] -\Omega^\vee_{2;12} \mathscr{I}_{d=2}[\vphi_2]
\end{align}
where $\mathscr{I}_{d=2}[\vphi_i] = -1+O(\vep)$, and the two terms come from the two tadpoles.
This confirms that FI satisfy precisely the (minus transposed) differential equation as the dual-forms, as predicted by \eqref{eq:OmegaVeeMinusOmegaT}.

Differential equations for other 2d integrals can be obtained in a similar fashion. For example,
repeating the above procedure for the $\delta_{13}$-term of \eqref{eq:d(2d-tad) 2},
one finds the $\tad$-$\bubb$ component of the dual kinematic connection as well as an additional  term localized to the $\tri$-cut. Fortunately, no integration-by-parts is needed on the $\tri$-cut and the corresponding $\tad$-$\tri$ component of the dual kinematic connection can immediately be read off 
\begin{align}
	\Omega^\vee_{(i),(ijk)} &= \vep \frac{i}{2} d\log\left( \frac{p_1 \cdot (p_1+p_2) - i \sqrt{(01)^2(0123)^2}}{p_1 \cdot (p_1+p_2) + i \sqrt{(01)^2(0123)^2}} \right)
\end{align}
where we used $c_1/c_{-1} = -2\vep^2$ from eq.~\eqref{eq:ci's}.

By symmetry, we generalize the $\tad$ components of the dual kinematic connection to arbitrary tadpoles
\begin{align}
	\label{eq:2d-tadpole-tadpole}
	\Omega^\vee_{i;i} &= \vep\ d\log r_i^2,
	\\
	\label{eq:2d-tadpole-bubble}
	\Omega^\vee_{i;ij} &= \vep\ d\log \left( \frac{\sqrt{(ij)^2(i0)^2}+(ij)\cdot(i0)}{\sqrt{(ij)^2(i0)^2}-(ij)\cdot(i0)} \right),
	\\
	\label{eq:2d-tadpole-triangle}
	\Omega^\vee_{i;ijk} &= \vep \frac{i}{2} d\log\left( \frac{(0ij)\cdot(0ik) + i \sqrt{(0i)^2(0ijk)^2}}{(0ij)\cdot(0ik) - i \sqrt{(0i)^2(0ijk)^2}} \right).
\end{align}
In order to make the pattern clear, we have expressed all external kinematics as minors of the Gram matrix \eqref{eq:gram}. 

The bubble and triangle components of the kinematic connection have simpler boundary structures
and will be presented as part of the general case.

\newpage

\subsection{Differential equations for one-loop dual forms in any dimension\label{sec:4dDEqs}}

In this section, we present the differential equation for a general dual form in any spacetime dimension.
Recall that dual forms near four-dimensions are 5-forms supported on cuts, leading to a basis
that (generically) contains tadpoles, bubbles, triangles, boxes and pentagons.
Near $d$ dimensions one can find up to a $(d+1)$-gon.

As will be clear from the calculation, the differential equations depends only the number of integration variables modulo 2.
Thus, for example, a tadpole-dual in $\dint= 4$, bubble-dual in $\dint= 3$ or
tadpoles, triangles, pentagons and heptagons in $\dint= 6$ are all basically the same problem.
The bubble-dual in $\dint= 4$ then covers all remaining cases.
For definiteness, we will work out in detail the tadpoles and bubbles in $\dint= 4$
and then spell out the generalization at the end.

We will use the embedding space formalism since it clarifies a number of points obscured by Minkowski momentum space, notably the appearance of Gram determinants.

Minkowski space in $d$-dimensions ($\eta_{_\text{Mink}} = \text{diag}(-1,1,1\dots)$) can be realized as a null cone in $d+2$-dimensional projective space ($\eta_{_\text{embed}} = \text{diag}(-1,1,1\dots,1,-1)$) \cite{Dirac:1936fq}. The loop momentum is mapped to an embedding space vector $\ell \to Y \equiv \lambda Y$ where $Y^2=0$. The propagators are similarly embedded, with the exception that massive propagators lie off the null-cone ($X_i^2 = -m_i^2$):
\be \label{embedding YX}
 Y \simeq (\ell^\mu, \tfrac{\ell^2-1}{2},\tfrac{\ell^2+1}{2}),\qquad
  X_i = (x_i^\mu, \tfrac{x_i^2+m_i^2-1}{2},\tfrac{x_i^2+m_i^2+1}{2})\,.
\ee
The symbol $\simeq$ reminds us that $Y$ is defined up to rescaling.
By introducing a null ``point at infinity'' $I=(0^\mu,-1,-1)$, propagators can be written in a homogeneous way:
$D_i = -2(YX_i)/(YI)$.

Cutting propagators in embedding space gives linear equations $Y{\cdot}X_i=0$:
cut integrals only know about other other momenta through their the orthogonal projection.
As a consequence, the Gram matrix \eqref{eq:radii} will arise very naturally in the embedding space formalism. 

The Minkowski space measure can be promoted to embedding space as follows
\be
	\int d^d\ell = \int \frac{d^{d+2}Y\ \delta(Y^2)}{\gl (IY)^d} \,.
\ee
The factor of $\gl$ in the denominator instructs us to divide by the volume of a $\gl$ orbit.
Essentially, it is the insertion of a gauge fixing delta function with the appropriate Faddeev-Popov determinant such that integrating over any $\text{gl}(1)$ orbit is unity \cite{Caron-Huot:2014lda}.\footnote{It is also common to define the projective measure using the Levi-Civita symbol $d^{d+2}Y\ \delta(Y^2) / \gl = Y_{\mu_1} \epsilon^{\mu_1 \mu_2 \cdots \mu_{d+2}} dY_{\mu_1} \wedge dY_{\mu_2} \wedge \cdots dY_{\mu_{d+2}}$ \cite{Herrmann:2019upk, Bourjaily:2019exo}.} A standard choice (which easily demonstrates the above) is $1/\gl \mapsto \delta((YI) +1)$, which has Fadeev-Popov determinant of unity. However, none of our manipulations will depend on a choice of gauge.

Thus, a generic ($\dint=4$) one-loop Feynman integrand in embedding space is 
\begin{align}
	\int \frac{d^d\ell}{D_1 \cdots D_n}
	&= \int \frac{d^{d+2}Y\ \delta(Y^2)}{\gl (IY)^d} \frac{(IY)^n}{(-2)^n (YX_1) \cdots (YX_n)} 
	\to \int \embedd\ u\ \vphi
	\\
	 \vphi 
	&= \frac{(IY)^{n+2-6}}{(-2)^n(YX_1) \cdots (YX_n)} \frac{d(Y_\perp^2) \wedge d^{6}Y}{Y_\perp^2}
	\label{eq:embedding space FI}
\end{align}
where the twist is 
\be
	u = \left( \frac{Y_\perp^2}{(IY)^2} \right)^{-\vep}
\ee
We have dropped the factor corresponding to the integration over the angles in the unphysical $-2\vep$-dimensions,
since we shall only be manipulating algebraic forms like $\vphi$.

On the other hand, the dual forms live on a cut indexed by $J$, where $Y{\cdot}X_j=0\ \forall\ j\in J$:
\be
 \vphi^\vee_J = \delta_J(\phi^\vee_J)\,. \label{eq:embed dual forms}
\ee
Let $N=d-|J|$ denote the number of unconstrained embedding coordinates.
A natural volume form on the $J$-cut is then written as
\be
 d^NY_J \equiv \frac{1}{\sqrt{(J)^2}} \langle X_{j_1}, \ldots, X_{j_{|J|}},\overset{N}{\overbrace{dY_J, \dots, dY_J }}\rangle/N!
\ee
where we use the angle brackets as a short hand for contraction with the totally anti-symmetric tensor in the
6-dimensional embedding space. The denominator is the Gram determinant of the vectors $X_{j_1},\ldots X_{j_{|J|}}$.

The measure is anti-symmetric under the exchange of any two indices in the set $J$.
The anti-symmetry in the volume form cancels the anti-symmetry in the $\delta$-map,
so that the dual forms in \eqref{eq:embed dual forms} are themselves symmetrical.

It will be convenient to introduce unit-normalized vectors: let $a^\mu=X_a^\mu/\sqrt{X_a^2}$
denote a unit vector in the direction of $X_a$. More generally, let $a_J^\mu$ denote the unit vector
along the projection of $a$ orthogonally to $J$. Thus, for example, when $J=\{b\}$ consists of a single index:
\be
 a_b^\mu = \frac{a^\mu - b^\mu (ab)}{\sqrt{1-(ab)^2}} = \frac{X_a^\mu X_b^2 - X_b^\mu (X_a X_b)}{\sqrt{X_a^2X_b^2-(X_aX_b)^2}}\,.
\ee
Similarly, $i_J$ denotes the unit vector along the projection of $I^\mu$ orthogonally to $J$
(this makes sense even though $I$ is a null vector, since its projection is generally not).
The volume form on the cut $J=\{a,b,c,\ldots\}$ is then written equivalently as:
\be
	d^NY_J = \langle a, b_a, c_{ab}, d_{abc}, \dots, \overset{N}{\overbrace{dY_J, \dots, dY_J }} \rangle / N!\,.
	\label{volume form embedding}
\ee

The specific form of the $\phi^\vee_J$ are fixed by homogeneity in $Y$ up to a factor of $(YI)^2/Y_\perp^2$.
We fix that freedom using the following simple principle: we require the dual forms to be either regular or have a single pole at infinity.
This determines the forms up to constants $c_i$:
\begin{eqnarray}
	\label{eq:tadpoles}
	&\text{Tadpoles:} \quad
	&\phi^\vee_{a} = c_3 \frac{dY_\perp^2 \wedge d^5Y_a}{(i_aY) (Y_\perp^2)^2} 
	\\
	\label{eq:bubbles}
	&\text{Bubbles:} \quad
	&\phi^\vee_{ab} = c_2 \frac{ dY_\perp^2 \wedge d^4Y_{ab} }{(Y_\perp^2)^2}
	\\
	\label{eq:triangles}
	&\text{Triangles:} \quad
	&\phi^\vee_{abc} = c_1 \frac{ dY_\perp^2 \wedge d^3Y_{abc} }{(i_{abc}Y) Y_\perp^2}
	\\
	\label{eq:boxes}
	&\text{Boxes:} \quad
	&\phi^\vee_{abcd} = c_0 \frac{ dY_\perp^2 \wedge d^2Y_{abcd} }{Y_\perp^2}
	\\
	\label{eq:pentagon}
	&\text{Pentagon:} \quad
	&\phi^\vee_{abcde} = c_{-1} \frac{ dY_\perp^2 \wedge d^1Y_{abcde} }{(i_{abcde}Y)} \,.
\end{eqnarray}
These are only meaningful for cuts that are non-degenerate --- as we saw above, the cohomology is trivial
when either $(J)^2=0$ or $(0J)^2=0$. We expect that by removing double poles at infinity, the dual forms will have uniform transcendentality. This is verified below by computing the dual kinematic connection. The coefficients $c_i$ are fixed by requiring the dual kinematic connection to be in canonical form and are given by \eqref{eq:ci's}. For general $\dint$, an $|J|$-gon is paired with the coefficient $c_{\dint-|J|}$.

Lastly, the above forms are equivalent to the spherical forms defined by equation \eqref{eq:sphereforms} once the embedding space null-cone and gauge fixing constraints are imposed. To gauge fix, we define $\tilde{Y}^\mu = Y^\mu - (i_J Y) i_J^\mu$, which trivializes the condition $(YI_J) = 1$ by setting $(Yi_J) = 1/|I_J| = r_J$. In terms of $\tilde{Y}$, a generic dual form becomes 
\begin{align} \label{eq:sphereforms2}
	\phi^\vee_J = \delta(Y_\perp^2 + Y^2) \delta( (YI_J) -1) 
	\frac{-c_{N-2}\ d(Yi_J) \wedge dY_\perp^2 \wedge d^{N-1}\tilde{Y}}{ (Yi_J)^{\delta_{|J|,\text{odd}}} (Y_\perp^2)^{\lfloor\frac{N}{2}\rfloor} } 
	= c_{N-2} \frac{r_J^{\delta_{N,\text{even}}} d^{N-1} \tilde{Y}}{(\tilde{Y}^2 + r_J^2)^{\lfloor\frac{N}{2}\rfloor}} .
\end{align}
Setting $\tilde{Y} = z$, we recover \eqref{eq:sphereforms} with $m=N-1=\dint+1-|J|$. 

\subsubsection{All the IBPs we will need}

The main advantage of dual forms is that the only allowed denominators on a cut
are $Y_\perp^2$ and $(YI)$, which preserve the spherical symmetry of a given cut.
The only purpose of integration-by-part identities is to symmetrically-integrate numerators.
We will need the explicit IBPs, however, not just the outcome of symmetrical integration,
in order to track boundary terms.

In embedding space, IBP identities are slightly more subtle than their momentum space counterparts
since they must preserve the null cone $Y^2_J+Y_\perp^2=0$ \cite{Caron-Huot:2014lda}.
Writing IBP identities in terms of a vector $V$,
\be
 0= \int d \left[ \left(V^\mu dY_\perp^2  \wedge d^{N-1}Y_{J\mu} + V^\perp d^NY_J\right) \frac{\delta(Y^2_\perp+Y^2)}{\gl}\right]\,,
\ee
the condition that the $\delta$-function does not get squared is $2 (YV) = V^\perp$.
Here we introduced the following shortcut, similar to eq.~\eqref{volume form embedding}, to convert from vectors to forms:
\be
 d^{N-1}Y_{J\mu} \equiv
 \langle a, b_a, c_{ab}, d_{abc}, \dots,\mu, \overset{N-1}{\overbrace{dY_J, \dots, dY_J }} \rangle / (N-1)!\,.
\ee

We will need two classes of IBPs. The first, ``dimension-preserving'' class, has $V^\perp=0$ and represent rotations
in the $Z_J-I$ plane for some vector $Z_J$:
\begin{align}
	V_J^{(m,n)} [Z_J]  &= \overset{V_J^\mu}{\overbrace{ Y_\nu Z_J^{[\nu} i_J^{\mu]} }}\ 
	\frac{dY_\perp^2 \wedge d^{N-1}Y_{J\mu}}{(Yi_J)^m (Y_\perp^2)^n}
	\label{eq:dimension_preserving_IBP}
\end{align}
where $Z_J^\mu$ is an embedding vector on the $J$-cut, that is, $(Z_J X_j)=0 \ \forall \ j \in J$.
The exponents $m,n$ are constrained so that the form has the correct homogeneity degree: $m+2n=N$.
\begin{align}
\nabla^\vee V_J^{(m,n)}[Z_J]
	&= (-1)^{|Z_J|}(m+2\vep)
		\left( \frac{(YZ_J)}{(Yi_J)} - (Z_J i_J)  \right) \wedge
		\frac{ d(Y_\perp^2) \wedge d^NY_J }{ (Yi_J)^{m} (Y_\perp^2)^n }
	\label{eq:V divergence}
\end{align}
We have included a sign $(-1)^{|Z_J|}$ and wedge symbol anticipating that in our applications the vector $Z^\mu_J$ will also be a one-form on the space of external momenta.\footnote{All components that are not top forms on the internal space are dropped since they are irrelevant for computing the kinematic connection.}

These IBPs are used to replace any $Y$-dependent ratio of the form $(YZ_J)/(Yi_J)$ with the $Y$-independent factor $(i_JZ_J)$.
Sometimes we will also need to shift the power of $Y_\perp^2$. This is achieved by the following IBP
\begin{align}
	U_J^{(m,n)} [Z_J] 
	&= \frac{Z^{\mu}_J \wedge dY_{\perp}^{2}\wedge d^{N-1}Y_{J\mu}+2 (Z_JY) \wedge d^NY_J}{(Yi_J)^{m} (Y_\perp^2)^n}\,,
	\label{eq:dimension_shifting_IBP}
\end{align}
which is effectively a rotation in the $Z{-}\!{\perp}$ plane, times $Y_\perp$ and satisfies 
\begin{align} \label{eq:U divergence}
	\nabla^\vee U_{J}^{(m,n)}[Z_J]
	&=\left[
			2(n - \vep)\frac{(YZ_J)}{Y_{\perp}^{2}}
			+ (-m - 2\vep)\frac{(i_{J}Z_J)}{(i_{J}Y)}
		\right]
	\wedge \frac{dY_{\perp}^{2}\wedge d^{N}Y_{J}}{\left(Yi_{J}\right)^{m}\left(Y_{\perp}^{2}\right)^{n}}.
\end{align}
The $U$-IBPs allow us to replace squared propagators with powers of the twist, much like \cite{Bosma:2018mtf}.
The IBP vectors $V$ and $U$, with suitable vectors $Z^\mu$,
generate all necessary identities to derive one-loop differential equations.

\subsubsection{Derivative of tadpole-dual \label{sec:4d-tadpole} }

\begin{figure}[]
\centering
\includegraphics[width=.3\textwidth]{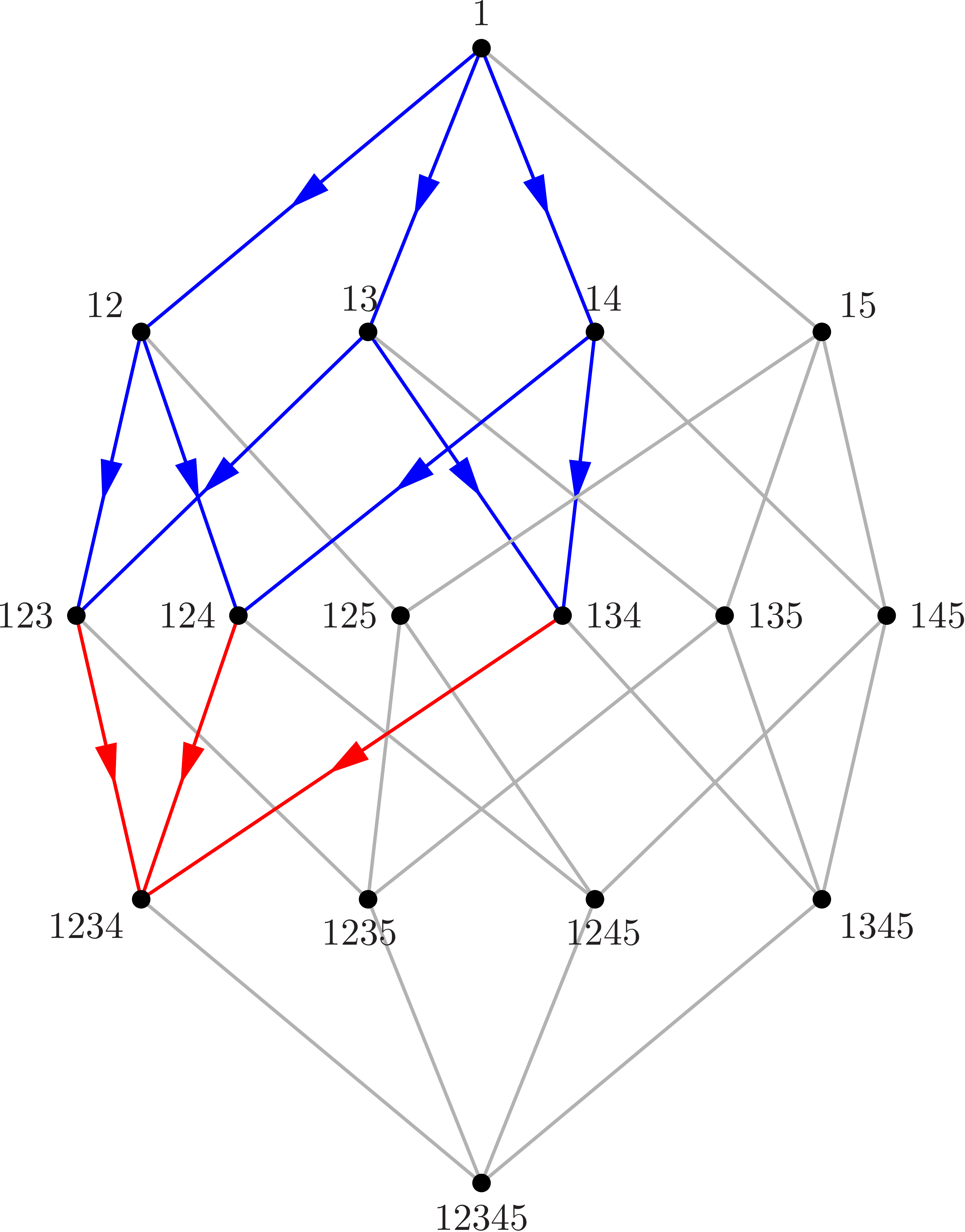} \
\includegraphics[width=.3\textwidth]{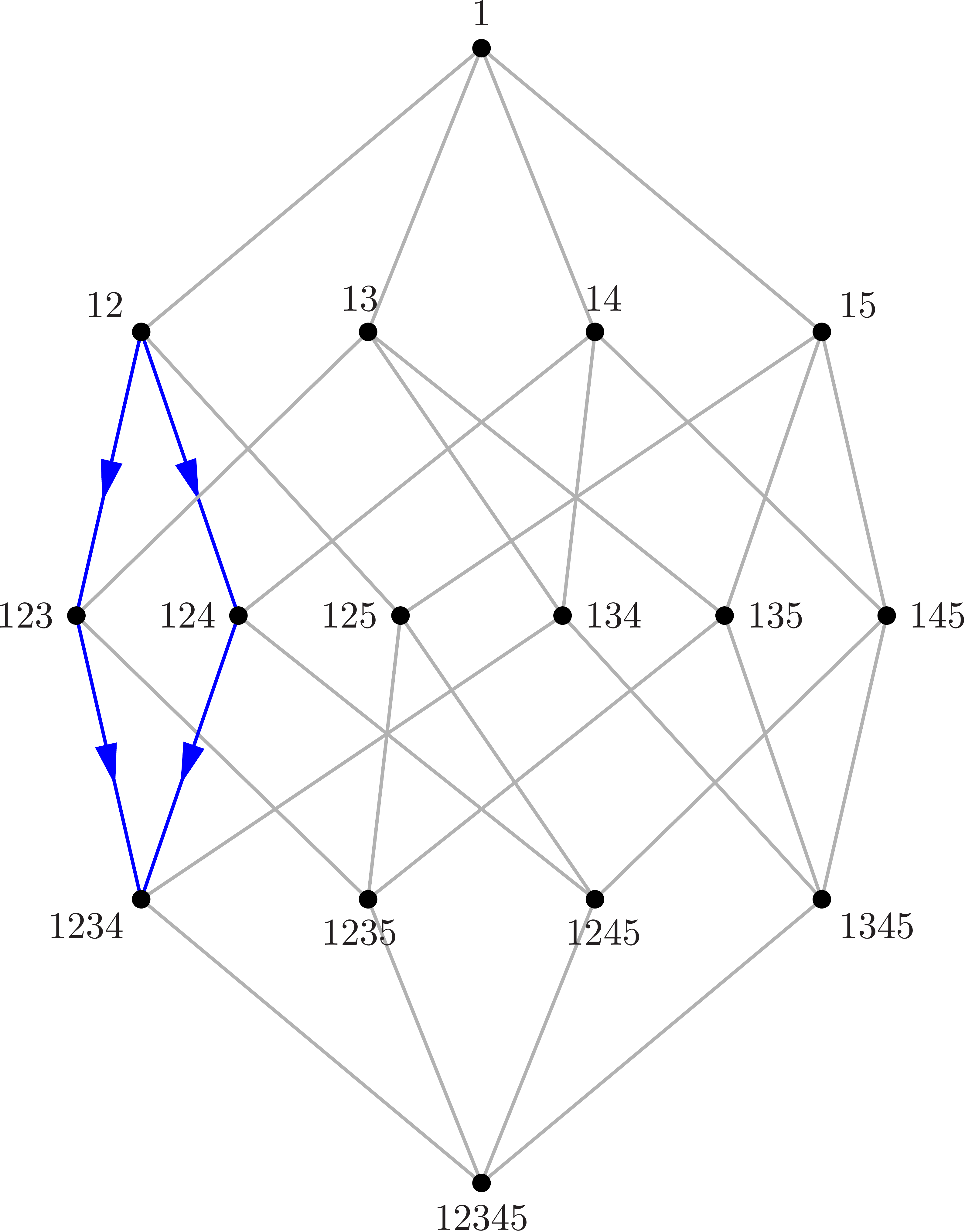} \
\includegraphics[width=.3\textwidth]{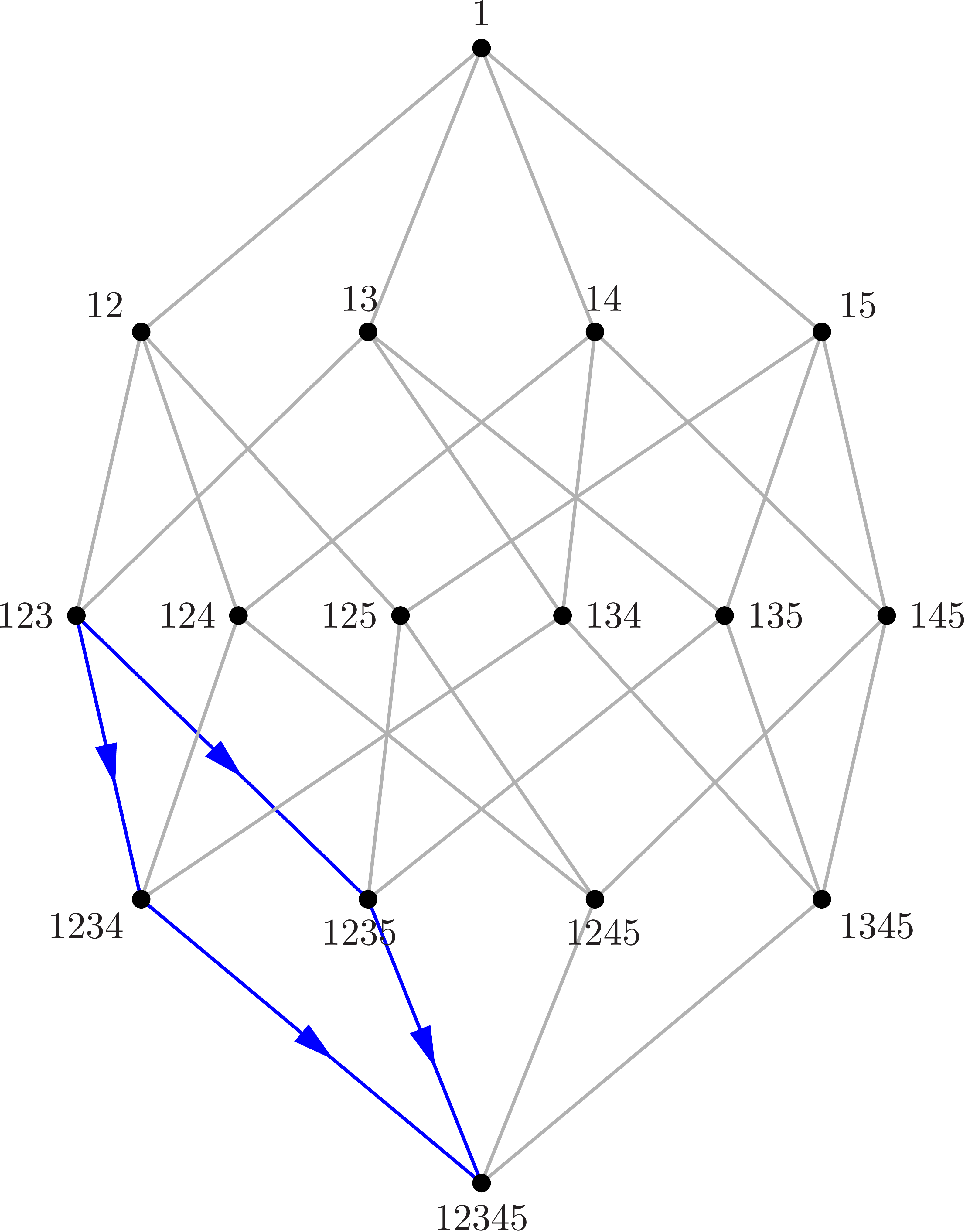} \

\caption{
\label{fig:boundary_paths}
The boundary structure of the tadpole (left), bubble (centre) and triangle (right) kinematic connections. 
Each node labels a cut surface and is associated to one of the basis integrals. 
Each edge represents the restriction to the corresponding boundary.
The nodes joined by blue arrows produce non-zero components of the kinematic connection. 
Nodes joined by red arrows signals that the feed-down from the starting node ends.
The boxes and pentagon have trivial boundary structure and are not pictured here.
}
\end{figure}

The first differential equation we'll compute is tadpole-dual near $d=4$ since the answer must putatively 
match the $\dint= 2$ analysis from  subsection~\ref{sec:2dDEqs-tadpole}.  As shown there, varying external kinematics and
integrating-by-parts on the tadpole cut gives boundary terms that account for
tadpole-bubble and tadpole-triangle terms in the differential equation.
In this section, we will show that the $\dint=2$ terms are complete: the tadpole-box vanishes!

Let us recall the tadpole-dual form:
\be \label{tadpole recall}
 \vphi_{a} = \delta_a( \phi_a^\vee), \qquad \phi_a^\vee= c_3 \frac{dY_\perp^2 \wedge d^5Y_a}{(i_aY) (Y_\perp^2)^2}\,.
\ee
Taking derivative (with respect to external variables), using the connection $\omega^\vee = -d\log u$, one readily finds:
\be
	\nabla^\vee \phi_a^\vee = \left[-\vep d\log |I_{a}|^2-(1+2\vep)\frac{(Ydi_a)}{(Yi_{a})}\right] \wedge \phi_a^\vee\,.
\label{tad der 1}
\ee
The $Y$ dependence in the second term can be brought back into our
basis by adding an integration-by-parts vector with $Z_a=di_a$:
\begin{align}
	\ibp_a = c_3 V_a^{(1,2)}[Z_a]
\end{align}
such that 
\begin{align} \label{eq:tad2}
 \nabla^\vee \ibp_a = -\left(1+2\epsilon\right)\left[\frac{\left(Z_{a}Y\right)}{\left(i_{a}Y\right)}-\cancel{\left(Z_{a}i_{a}\right)}^0\right]\wedge\phi_{a}^{\vee}.
\end{align}
The last term, equal to $(di_ai_{a})$, vanishes since $i_a^2=1$ is constant.
Then, subtracting \eqref{eq:tad2} from the left hand side fo \eqref{tad der 1}, we find
\begin{align}
\nabla^\vee \vphi_a - \nabla^\vee \delta_a(\ibp_a)
	&= - \delta_a( \nabla^\vee \phi^\vee_a - \nabla^\vee \ibp_a ) 
		- \sum_{b\neq a} \delta_{a,b}(\phi^\vee_a\vert_b - \ibp_a \vert_b) \nn \\
	&=  \Omega^\vee_{a;a} \wedge \vphi_a^\vee - \sum_{b\neq a} \delta_{a,b}(\phi^\vee_a\vert_b - \ibp_a \vert_b)
\label{eq:dtad1}
\end{align}
where  
\begin{align} \label{eq:O(a,a)}
	\Omega^\vee_{a;a} = -\vep\ d\log|I_a|^2 = \vep\ d\log r_a^2  
\end{align}
is the tadpole-tadpole component of the kinematic connection.
In general, the diagonal part of any kinematic connection is given by the simple $d\log$ form $\Omega_{\bullet;\bullet}^\vee = -\vep d\log |I_\bullet|^2$.


While the result eq.~\eqref{eq:O(a,a)} could have been easily guessed by symmetrical integration of eq.~\eqref{tad der 1}
(which just replaces $Y$ by $i_a$ in the numerator),
the explicit IBP vector is needed to  predict the tadpole-bubble feed-down.
Restriction of the form $\phi^\vee_{a}$ to the $ab$-bubble cut is given by 
\begin{align}
	\phi^\vee_{a}\vert_{b}
	&= c_3 \left.\frac{ dY_\perp^2 \wedge d^5Y_{a} }{ (i_{a} Y) (Y_\perp^2)^2 }\right\vert_{b}
	= c_3 \frac{ |I_{a}| }{ |I_{ab}| } \frac{ (Ydb_a) }{ (i_{ab}Y) (Y_\perp^2)^2 } 
		\wedge dY_\perp^2 \wedge d^4Y_{ab}
\end{align}
where  $(Yb_a)=0 \implies (b_a dY)=-(Ydb_a)$ on the $ab$-cut,
and the factor of $|I_a|/|I_{ab}|$ comes from the restriction of $(Yi_a)$.
For the restriction of the IBP-form $V_a[Z_a]$, we are only interested in the part that is a top-form
on the $ab$-cut, which is obtained by replacing $Z_a^\mu \to (Z_a b_{a}) b_a^\mu$ in the definition of $V_a[Z_a]$:
\begin{align} \label{eq:IBPa|b}
	\ibp_{a}\vert_{b}
	= c_3 V_a^{(1,2)}[Z_a]\vert_{b}
	= c_3 \frac{|I_{a}|}{|I_{ab}|} Y_\mu Z_{a}^{[\mu} i_{a}^{\nu]} b_{a\nu}
		\wedge \frac{dY_{\perp}^{2}\wedge d^{4}Y_{ab}}{(Yi_{ab}) (Y_\perp^2)^2}\,.
\end{align} 
Combining the above, we find the total feed-down to the $ab$-bubble cut
\begin{align}
	\phi^\vee_a\vert_b - \ibp_a \vert_b
	= c_3 (YZ_{ab})\wedge\frac{dY_{\perp}^{2}\wedge d^{4}Y_{ab}}{(i_{ab}Y)(Y_{\perp}^{2})^{2}}
	= \frac{c_3}{c_2} \frac{(YZ_{ab})}{(Yi_{ab})}\wedge\phi_{ab}^{\vee}  \label{bubble feed}
\end{align}
where 
\begin{align}
	Z_{ab}^\mu 
	= \frac{|I_{a}|}{|I_{ab}|} \left(db_a^\mu + i_{a}^{[\mu}di_{a}^{\nu]}b_{a\nu}\right).
\end{align}
While eq.~\eqref{bubble feed}
is proportional to $\phi_{ab}^\vee$, it's coefficient still contains some $Y$-dependence, which must be removed via integration-by-parts. 
Before doing so, it is convenient to rewrite $Z_{ab}^\mu$ in terms of geometrical quantities on the $ab$-cut.
Using the identity:
\be
 b_{ai}^\mu = \frac{b_a^\mu - i_a^\mu (i_ab_a)}{\sqrt{1-(i_ab_a)^2}} =
 \frac{|I_{a}|}{|I_{ab}|} \left(b_a^\mu - i_a^\mu (b_a i_a)\right)\,,
\ee
a bit of algebra shows that
\be
 Z_{ab}^\mu = \frac{i_{ab}^\mu dx}{1-x_{ab}^2}+ db_{ai}^\mu - i_{ab}^\mu (i_{ab}db_{ai})\,. \label{Zab identity}
\ee
with $x_{ab}=(i_ab_a)$. The vector $i_{ab}$ is the point at infinity projected to the $ab$-cut, while $b_{ai}$ is $X_b$ projected orthogonally from the point at infinity. We then remove the unwanted $Y$-dependence using the IBP-form
\be
 \ibp_{ab} = -\frac{c_3}{2\vep} V_{ab}^{(0,2)}[Z_{ab}], \label{ibp ab}
\ee
which gives
\be
\mbox{eq.~\eqref{bubble feed}} - \nabla^\vee\ibp_{ab} = \frac{c_3}{c_2}\left(i_{ab}Z_{ab}\right)\wedge\phi_{ab}^{\vee}.
\ee
Summarizing, we have computed the derivative of tadpole-dual up to triple cuts:
\begin{align} 
	&\nabla^\vee \Big( \vphi_{a}^\vee - \delta_{a} (\ibp_{a}) + \sum_{b\neq a} \delta_{a,b} (\ibp_{ab}) \Big) \nn \\
	&\qquad = \Omega_{a;a}^{\vee} \wedge \vphi_{a}^{\vee}
	+ \sum_{b\neq a} \Omega_{a;ab}^{\vee} \wedge \vphi_{ab}^{\vee}
	+ \sum_{b\neq a} \sum_{c>b} \delta_{a,b,c} ( \ibp_{ab}\vert_{c} - \ibp_{ac}\vert_{b} )
\label{eq:dtad2}
\end{align}
where
\begin{align} \label{eq:O(a,ab)}
	\Omega^\vee_{a;ab} 
	= -\frac{c_3}{c_2}(i_{ab}Z_{ab}) 	
	= +\frac{c_3}{2c_2} d\log\left(\frac{1{-}x_{ab}}{1{+}x_{ab}}\right)
	= +\vep\frac{1}{2}\ d\log\left(\frac{\sqrt{(ab)^{2}(a0)^{2}}-(a0)\cdot(ab)}{\sqrt{(ab)^{2}(a0)^{2}}+(a0)\cdot(ab)}\right)
\end{align}
with $\frac{c_3}{c_2} = 2\vep$. Up to overall rational functions of $\vep$, equation \eqref{eq:O(a,ab)} is equivalent to its 2-dimensional counterpart \eqref{eq:2d-tadpole-bubble}. 


It will be important that the triple-cut contribution in eq.~\eqref{eq:dtad2} originates from two pathways:
one coming from the $a \to ab \to abc$ feed-down and the other from  $a \to ac \to acb$ (see Fig.~\ref{fig:boundary_paths}). 
The first restriction, comes from the IBP in eq.~\eqref{ibp ab}:
\begin{align}
	\ibp_{ab}\vert_{c} 
	= -\frac{c_3}{2\vep} Y_\mu \ db^{[\mu}_{ai} i^{\nu]}_{ab}\ c_{ab\nu}  \frac{dY_\perp^2 \wedge d^3Y_{abc}}{Y_\perp^2}. 
\end{align}
The second restriction is similar but with $b$ and $c$ exchanged. Since $d^3Y_{acb} = -d^3Y_{abc}$, both terms contribute with the same sign:
\be
	\ibp_{ab}\vert_c - \ibp_{ac}\vert_b
	= -\frac{c_3}{2\vep c_1} \frac{(YZ_{abc})}{Y_\perp^2}(Yi_{abc}) \wedge  \phi^\vee_{abc}
\label{step abc}
\ee
where 
\be
	Z_{abc}^\mu = db^{[\mu}_{ai} i^{\nu]}_{ab} c_{ab\nu}+dc^{[\mu}_{ai} i^{\nu]}_{ac} b_{ac\nu}\ .
\ee
This is consistent with the fact that the dual form $\vphi_{abc}^\vee$ is symmetric under exchange of $b$ and $c$. Again, we must add a suitable IBP-form to remove the $Y$-dependent pre-factor in front of $\phi_{abc}^\vee$. Because of the additional power of $Y_\perp^2$, we must use the \emph{dimension shifting} IBP-form \eqref{eq:dimension_shifting_IBP}
\begin{align}
	\ibp_{abc} = -\frac{c_3}{4\vep(1 - \vep)} U_{abc}^{(0,1)}[Z_{abc}],
\end{align}
whose covariant derivative gives 
\be
\mbox{eq.~\eqref{step abc}} - \nabla^\vee \ibp_{abc} = -\Omega^\vee_{a;abc} \wedge \phi_{abc}^\vee,
\quad
\Omega^\vee_{a;abc}\equiv + \frac{c_3}{2(1 - \vep)c_1} (i_{abc}Z_{abc}) \label{Omega a;abc}
\ee
Before we simplify this, let us explain why the system will terminate at the triangle.
This is due to the especially simple form of the $U$-IBP form, whose restriction to the box cut will give:
\be
 \ibp_{abc}\vert_d \propto
 \left(Z_{abc}d_{abc}\right)\wedge \frac{dY_{\perp}^{2}\wedge d^{2}Y_{abcd}}{Y_\perp^2}\,.
\ee
The vector $Z_{abc}$ contains a term $db_{ai}^\mu$, which will reach a given box cut through two pathways,
$ab\to abc\to abcd$ and $ab\to abd\to abcd$:
\be
 \left(Z_{abc}d_{abc}\right)
+\left(Z_{abd}c_{abd}\right)   \supset   (db_{ai}^{[\mu}i^{\nu]}_{ab}) (d_{abc\mu} c_{ab\nu}+ c_{abd\mu} d_{ab\nu}) = 0
\label{no box}
\ee
where we used antisymmetry in $\mu,\nu$.  Geometrically, the two bivectors in the parentheses represent the
$cd$ and $dc$ planes, which are equal up to opposite orientations. Similar identities cancel the $bc$ and $bd$ terms, and the sum of all triangle IBPs restrict to zero identically on any box cut:
\be  \ibp_{abc}\vert_d -  \ibp_{abd}\vert_c +  \ibp_{acd}\vert_b = 0\,.
\ee
We have thus shown that, adding suitable IBP vectors, the derivative of tadpole-dual only contains three sorts of terms: 
\begin{align} 
\nabla^\vee \vphi_a^\vee \simeq
	&\nabla^\vee 
	\Big( 
		\vphi_{a}^\vee 
		- \delta_{a} (\ibp_{a}) 
		+ \sum_{b\neq a} \delta_{a,b} (\ibp_{ab}) 
		+ \sum_{b\neq a} \sum_{c>b} \delta_{a,b,c} (\ibp_{abc}) 
	\Big) 
	\nn \\
	&\quad = \Omega_{a;a}^{\vee} \wedge \vphi_{a}^{\vee}
	+ \sum_{b\neq a} \Omega_{a;ab}^{\vee} \wedge \vphi_{ab}^{\vee}
	+ \sum_{b\neq a, c>b} \Omega_{a;abc}^{\vee}\wedge \vphi^\vee_{abc}\,.
\label{eq:dtad3}
\end{align}
It remains only to simplify the last connection, given in eq.~\eqref{Omega a;abc}.
By an identity similar to eq.~\eqref{no box}, 
\begin{align} \label{eq:iabcZabc}
 (i_{abc}Z_{abc}) &= db^{[\mu}_{ai} i^{\nu]}_{ab}\ i_{abc\mu} c_{ab\nu} + (b\leftrightarrow c) \\
&=  db^{[\mu}_{ai} i^{\nu]}_{ab}\ i_{ab\mu} c_{abi\nu} + (b\leftrightarrow c)
  = -(db_{ai} c_{abi})-(dc_{ai} b_{aci})
\end{align}
where in the last step we used $(i_{ab}c_{abi})=0$.
Further, noting that $c_{abi}^\mu = c_{aib}^\mu = \frac{c_{ai}^\mu - b_{ai}^\mu x_{abc}}{\sqrt{1-x^2_{abc}}}$ with
$x_{abc}\equiv (b_{ai}c_{ai})$,  and that $(b_{ai}db_{ai})=0$, this reduces to
\be
 (i_{abc}Z_{abc}) = \frac{-dx_{abc}}{\sqrt{1-x_{abc}^2}}= d\cos^{-1}(x_{abc})\,.
\ee
Thus the tadpole-triangle kinematic connection is 
\begin{align} \label{eq:O(a,abc)}
	\Omega^\vee_{a;abc}  
	= \frac{c_3}{2(1 - \vep)c_1} d\cos^{-1}(b_{ai}c_{ai}) 
	= \vep \frac{i}{2} d\log\left( \frac{ \sqrt{(a0)^2(abc0)^2} + i(ab0){\cdot}(ac0) }
	{ \sqrt{(a0)^2(abc0)^2} - i(ab0){\cdot}(ac0)}\right)
\end{align}
where $\frac{c_3}{2(1 - \vep)c_1}=\vep$. The above can be generalized to any odd-dimensional cut simply by replacing the index $a$ with a multi-index $J$.

Since the cut surfaces are effectively hyperplanes in embedding space, the off-diagonal connections can only depend on the angles between these hyperplanes (i.e., $x_{ab}$ and $x_{abc}$). In particular, the tadpole-box component must vanish since there are no viable invariants. 

\subsubsection{Derivative of bubble-dual \label{sec:4d-bubble}}

As explained at the beginning of the section, the derivative of any dual form is either similar to that
of tadpole-dual or bubble-dual, depending on whether the dimensionality of the cut is odd or even.
We will find that the bubble-dual is somewhat simpler, and terminates on the box cut for a simple reason:
no IBPs will be needed at the second step.

Let's start by computing the covariant derivative of the $ab$ bubble, defined in eq.~\eqref{eq:bubbles}:
\begin{align} \label{eq:dbub1}
	\nabla^\vee \phi^\vee_{ab} = -2\vep \frac{(YZ_{ab})}{(Yi_{ab})} \wedge \phi^\vee_{ab}, \qquad
	Z_{ab}^\mu = d\log|I_{ab}| i_{ab}^\mu + di_{ab}^\mu.
\end{align}
In perfect analogy to the diagonal tadpole-tadpole component, the $Y$-dependent factor can be removed
by using a suitable IBP-from:
\begin{align}
	\ibp_{ab} = c_2 V_{ab}^{(0,2)}[Z_{ab}].
\end{align}
This gives:
\begin{align} \label{eq:dbub2}
	\nabla^\vee \vphi_{ab}^{\vee} - \nabla^\vee\delta_{ab}(\ibp_{ab})
	= \Omega^\vee_{ab;ab} \wedge \vphi^\vee_{ab} + \sum_{c \neq a,b} \delta_{a,b,c}(\phi^\vee_{ab}\vert_{c} - \ibp_{ab}\vert_{c})
\end{align}
where 
\begin{align} \label{eq:O(ab,ab)}
	\Omega^\vee_{ab;ab} 
	= -2\vep (Z_{ab}i_{ab}) 
	= -\vep\ d\log|I_{ab}|^2 
	= \vep\ d\log r_{ab}^2 
\end{align}
is the bubble-bubble component of the kinematic connection. 

The boundary term in \eqref{eq:dbub2} is proportional to a $abc$-triangle form up to a $Y$-dependent factor.
Evaluating the restrictions as in eq.~\eqref{eq:IBPa|b} above, we get
\be
	\phi_{ab}^\vee\vert_c - \ibp_{ab}\vert_c 
	= \frac{c_2}{c_1} \frac{(YZ_{abc})}{ Y_\perp^2 }(i_{abc}Y)  \wedge  \phi_{abc}^\vee\,,
\label{bub on tri}
\ee
where
\begin{align}
	Z_{abc}^\mu  &= dc_{ab}^\mu + i_{ab}^{[\mu} di_{ab}^{\nu]} c_{ab\nu} \label{Zabc first form}
	\\ &= \frac{i_{abc}^\mu dx}{\sqrt{1-x^2}} + \sqrt{1-x^2}\left( dc_{iab}^\mu - i_{abc}^\mu (i_{abc}dc_{iab})\right)
\end{align}
where on the second line we used the same identity as in eq.~\eqref{Zab identity} above, with $x=(i_{ab}c_{ab})$. 

The unwanted $Y$ and $Y_\perp^2$ dependence in eq.~\eqref{bub on tri}
can be removed at once using the dimension shifting IBP-form:
\be
	\ibp_{abc} = \frac{c_2}{2(1 - \vep)} U_{abc}^{(0,1)}[Z_{abc}],  
\ee
such that eq.~\eqref{eq:dbub2} becomes:
\begin{align} \label{eq:dbub3}
	&\nabla^\vee \vphi^\vee_{ab} - \nabla^\vee \delta_{a,b} (\ibp_{ab}) + \sum_{c\neq a,b} \nabla^\vee \delta_{a,b,c}(\ibp_{abc})
	\nn \\ & \qquad
	= \Omega^\vee_{ab;ab} \wedge \vphi^\vee_{ab} 
		+ \sum_{c\neq a,b} \Omega^\vee_{ab;abc} \wedge \vphi_{a,b,c}^\vee 
		+ \sum_{d>c} \sum_{c \neq a,b} \delta_{a,b,c,d} ( \ibp_{abc}\vert_{d} - \ibp_{abd}\vert_{c} )
\end{align}
where the bubble-triangle component of the kinematic connection is:
\begin{align} \label{eq:O(ab,abc)}
	\Omega^\vee_{ab;abc} 
	&= - \frac{c_2}{c_1} \frac{\vep}{1 - \vep} (i_{abc} Z_{abc})
	= - \frac{c_2}{c_1} \frac{\vep}{1 - \vep} \frac{dx}{\sqrt{1-x^2}}
	\nn \\ 
	&= 2\vep \frac{i}{2} 
	d\log \left(\frac{ \sqrt{ (ab)^2 (abc0)^2 } + i (ab0)\cdot(abc) }{ \sqrt{ (ab)^2 (abc0)^2 } - i (ab0)\cdot(abc) }\right)
\end{align}
where $\frac{c_2}{c_1} \frac{\vep}{1 - \vep} = \vep$. The $d\log$ form of the above becomes obvious once the top line is written using $\frac{dx}{\sqrt{1-x^2}}=-d\cos^{-1}(x)$ and $x=(ab0)\cdot(abc)/\sqrt{(ab0)^2(abc)^2}$.

There remains to compute the last sum in eq.~\eqref{eq:dbub3}: the bubble-box connection.
The nature of the $U$ IBP-form makes this particularly simple: there is no $Y$ dependence to remove!
This is why the differential equation terminates here. Proceeding as in eq.~\eqref{step abc}, the restriction evaluates to:
\begin{align}
\ibp_{abc}\vert_{d} - \ibp_{abd}\vert_{c} &= \frac{ c_2 }{ 2(1 - \vep) c_0 }
		[ (Z_{abc}d_{abc}) + (Z_{abd}c_{abd}) ] \wedge \phi^\vee_{abcd} \nn \\
	&\equiv -\Omega^\vee_{ab;abcd} \wedge \phi^\vee_{abcd}\,.
\end{align}
The square bracket is evaluated most easily using the first form in eq.~\eqref{Zabc first form}:
\begin{align}
(Z_{abc}d_{abc}) +(c\leftrightarrow d) &= (dc_{ab}d_{abc}) + i_{ab}^{[\mu} di_{ab}^{\nu]} ( d_{abc\mu} c_{ab\nu})+(c\leftrightarrow d)
	\nn \\ &=(dc_{ab}d_{abc}) + (c\leftrightarrow d) = \frac{dx}{\sqrt{1-x^2}}
\end{align}
where $x=(c_{ab}d_{ab})$. On the first step we have used that the last term is proportional to the volume element $c_{ab}\wedge d_{ab}$ which cancels when symmetrizing in $c\leftrightarrow d$, and in the second step we used
$d_{abc}^\mu =\frac{d_{ab}^\mu - xc_{ab}^\mu}{\sqrt{1-x^2}}$ together with the fact that $(c_{ab}dc_{ab})=0$.
(Similar identities were used in the tadpole-triangle calculation in eq.~\eqref{eq:iabcZabc}.)
Thus:
\be \label{eq:O(ab,abcd)}
	\Omega^\vee_{ab;abcd} \wedge \phi^\vee_{abcd}=
	+\vep \frac{i}{2} d\log\left(\frac{ \sqrt{(abcd)^2 (ab)^2} + i (abd) \cdot (abc) }{ \sqrt{(abcd)^2 (ab)^2} - i (abd) \cdot (abc) }\right)
\ee
where $\frac{ c_2 }{ 2(1 - \vep)  c_0 }=\vep$.

\subsection{Summary: one-loop differential equations in any dimension}
\label{ssec:summary}

By replacing the indices $a$ and $ab$ in the tadpole and bubble examples by a generic multi-index $J$,
we obtain the general form of the differential equation around $(\dint+2\vep)$-dimensions (recall that dual integrals come with $+\vep$ rather than $-\vep$):
\be
 \nabla^\vee \vphi^\vee_{J} \simeq \Omega_{J;J}^\vee \wedge \vphi^\vee_{J} 
		+ \sum_{a\notin J} \Omega^\vee_{J;Ja} \wedge \vphi_{Ja}^\vee 
		+ \sum_{b>a} \sum_{a \notin J} \Omega^\vee_{J;Jab} \wedge \vphi_{Jab}^\vee \,.
\ee
The explicit form of the connection depends on $\dint-|J|$,
the number of integration variables on the cut at the stated integer dimension.
If it is odd (as for the tadpole in four dimensions), we get from eqs.~\eqref{eq:O(a,a)},\eqref{eq:O(a,ab)},\eqref{eq:O(a,abc)}:
\begin{align} \label{eq:oddOmega}
\left.\begin{array}{ll} \displaystyle
	\Omega_{J;J}^\vee 	&= \vep\ d\log (J)^2/(J0)^2
	\\[0.1em]\displaystyle
	\Omega_{J;Ja}^\vee &=
		2\vep\ d\tanh^{-1}(i_J a_J)
	\\[0.1em]\displaystyle
	\Omega_{J;Jab}^\vee  &= \vep\ d\cos^{-1}(a_{J0}b_{J0})
\end{array}
\quad\right\}\quad	\text{for } \dint-|J| \text{ odd}\,.
\end{align}
When $\dint-|J|$ is even (as for the bubble in four dimensions), we get from eqs.~\eqref{eq:O(ab,ab)},\eqref{eq:O(ab,abc)},\eqref{eq:O(ab,abcd)}:
\begin{align}
\left.\begin{array}{ll} \displaystyle
	\Omega_{J;J}^\vee 
		&=\vep\ d\log(J)^2/(J0)^2
	\\[0.1em]\displaystyle
	\Omega_{J;Ja}^\vee 
		&= \vep\ d\cos^{-1}(i_{J}a_{J})
	\\[0.1em]\displaystyle
	\Omega_{J;Jab}^\vee
	&=\vep\ d\cos^{-1}(a_{J}b_{J})
\end{array}
\quad\right\}\quad \text{for } \dint-|J| \text{ even}\,.	\label{eq:evenOmega}
\end{align}
In terms of Gram determinants defined in eq.~\eqref{eq:gram},
the dot products are explicitly:
$(i_Ja_J) = \frac{(J0){\cdot}(Ja)}{\sqrt{(J0)^2(Ja)^2}}$, 
and $(a_{J0}b_{J0})=\frac{(J0a){\cdot}(J0b)}{\sqrt{(J0a)^2(J0b)^2}}$.

This gives the differential equation for dual-forms in any spacetime dimension, and, 
thus, according to eq.~\eqref{eq:OmegaVeeMinusOmegaT}, the minus-transposed of that of Feynman forms in a suitable normalization.
We can thus compare with the extensively known results for one-loop integrals,
for example \cite{Spradlin:2011wp, Arkani-Hamed:2017ahv, Bourjaily:2019exo} in integer spacetime dimensions. We find perfect agreement: The components $\Omega_{J;Jab}^\vee$ (where the number of propagators shifts by 2) and in fact all $d\cos^{-1}$ terms are familiar from the Schl\"afi
differential equation in integer dimensions.
The 0's in the formulas, representing the point at infinity $I=X_0$
are simply explained by viewing {\it ie.} a triangle in four-dimension as a box with a point at infinity.
The diagonal terms $\Omega_{J;J}^\vee$ are also easy to understand and reflect $r^{2\epsilon}$ factors in the leading singularities,
with the radius defined in eq.~\eqref{eq:radii}.
The main novel component of our general formula,
not obvious from integer dimensions, is $d\tanh^{-1}$.

The agreement of differential equations means that
our dual forms, normalized here by the criterion that they satisfy a differential equation in canonical form, must be orthonormal to canonically-normalized Feynman integrals:
\be
 \langle \vphi^\vee_J|\vphi_{J'} \rangle = \delta_{J,J'} \times
 \mbox{constant}
\ee
where the constant is independent of $|J|$ and on kinematics.
In the next section we confirm this (for the diagonal terms),
deducing the corresponding normalizations of the FIs in \eqref{normalized FI}.


\section{Relation to Feynman integrals \label{sec:FI and degenerate limits}}

In this section, we provide evidence for the duality between dual forms and regular Feynman integrands. 
Assuming minimal knowledge of intersection theory, in sec.~\ref{sec:normalization} we determine the normalization of Feynman integrands dual to the dual forms of sec.~\ref{sec:one-loop deqs}, \emph{assuming} that the intersection is diagonal (a guess which will be confirmed by agreement of the differential equations).
The normalized Feynman integrals are shown to correspond to the familiar uniform transcendental one-loop integrals. 
In sec.~\ref{sec:degenerate limits}, 
we analyze several degenerate but well studied kinematic limits and see that the agreement continues to hold, although 
the dual forms are now dual to special IR-finite combinations of FIs!

\subsection{Normalizing Feynman integrals from diagonal intersections \label{sec:normalization}}

The normalization of the Feynman integrands dual to the dual forms can be predicted by computing the diagonal $JJ$-components of the intersection matrix ignoring boundaries/propagators not contained in the $J$-cut. 

More concretely, this corresponds to the intersection of dual forms localized to the $J$-cut ($\vphi^\vee=\delta_J(\phi_J^\vee)$ with $\phi^\vee_J \in H^{\dint+1-|J|}(D^{(J)}\setminus \{\G=0\},\nabla^\vee)$) and Feynman integrands $\vphi_J$ whose $J$-cut residue is non-vanishing and only has singularities on the twisted boundaries ($\phi_J = \res_J[\vphi_J] \in  H^{\dint+1-|J|}(D^{(J)}\setminus \{\G=0\},\nabla)$). Since additional boundaries/propagators are absent in these cohomology groups, it is easy to use integration-by-parts and replace all forms by a cohomologous sphere-form with logarithmic singularities. 

We will use the embedding space formalism here since it is slightly easier to take residues on cuts (see subsection~\ref{sec:4dDEqs} for a review). 
Starting from \eqref{eq:embedding space FI}, let $\mathscr{I}_J = \int \vphi_J$ be our ansatz for the normalized Feynman integral where
\begin{align} \label{eq:phi ansatz}
	\vphi_J &= 
	\frac{a_{\dint-|J|+1} \sqrt{-(J0)^2}}{(-2)^{|J|}(YX_{j_1}) \cdots (YX_{j_{|J|}})}
\frac{r_J^{\delta_{\dint-|J|,\text{even}}}}{(IY)^{\delta_{d_{\rm int}-|J|,\rm odd}}}
		\frac{d(Y_\perp^2) \wedge d^{\dint+2}Y}{(Y_\perp^2)^{\lfloor\frac{\dint{-}|J|+2}{2}\rfloor}
	}.
\end{align}
and $a_{\dint-|J|}$ is the undetermined normalization. Note that the power of $Y_\perp^2$ is the same for both form and dual form (see \eqref{eq:sphereforms} and \eqref{eq:sphereforms2}). Also, recall that there is an implicit factor of $\delta(Y_\perp^2+Y^2) \delta((YI)-1)$ multiplying all forms in embedding space (assuming the standard $\gl$ gauge fixing condition $(YI)=1$). Lastly, the power of $(YI)$ in \eqref{eq:phi ansatz} is fixed by requiring that $\vphi_J$ is invariant under rescaling $Y\to \lambda Y$.

While we will work primarily in embedding space, it is important to connect \eqref{eq:phi ansatz} to integrands written in momentum space. 
The corresponding momentum space integrand is 
\begin{align}  \label{normalized FI}
	\vphi_J 
	&= 
	a_{\dint-|J|+1}\ \sqrt{-(J0)^2}\ r_J^{\delta_{\dint-|J|,\text{even}}} 
	\frac{
		d\ell_\perp^2 \wedge d^{\dint}\ell
	}{
		D_1 \cdots D_{|J|} (\ell_\perp^2)^{\lfloor\frac{\dint{-}|J|+2}{2}\rfloor}
	} .
\end{align}
where the map to embedding space sends propagators $D_i$ to $-\frac12 (YX_i)/(YI)$.
We will now determine the numerical constants $a_m$ by requiring that intersections
with dual forms be independent of $m$.  According to eq.~\eqref{eq:OmegaVeeMinusOmegaT}
this will ensure that the integrals $\int \vphi_J$ satisfy the (minus transpose)
of the differential equation in eqs.~\eqref{eq:oddOmega}-\eqref{eq:evenOmega}.

Let us thus compute the diagonal intersection number $\la \vphi^\vee_J \vert \vphi_J \ra$. From equations \eqref{eq:action of delta in intersection numbers} and \eqref{eq:sphereforms2},
\begin{align} 
	\label{eq:diag int num}
	\la \vphi^\vee_J \vert \vphi_J \ra 
	= \la \delta_{J}(\phi^\vee_J) \vert \vphi_J \ra
	= \la \phi^\vee_J \vert \res_J[\vphi_J] \ra
	\equiv \la \phi^\vee_J \vert \phi_J \ra
\end{align}
where we have defined $\phi_J = \res_J[\vphi_J]$.
The residue step is a standard calculation in the context of generalized unitarity.
We describe it in the embedding space formalism. Defining a basis of vectors for the $J$-cut
\begin{align}
	e_1 = j_1,
	\quad 
	e_2 = (j_2)_{j_1},
	\quad \cdots \quad
	e_{|J|} = (j_{|J|})_{j_1 \cdots j_{|J|-1}},
	\quad
	e_{|J|+1} = i_{J},
\end{align}
we split $Y$ into components parallel and perpendicular to the $J$-cut
\begin{align}
	Y = \sum_{i=1}^{|J|+1} Y_i e_i + \tilde{Y},
	\quad 
	Y_i = (Ye_i),
	\quad 
	(\tilde{Y} X_i) = 0.
\end{align}
In these coordinates, the dual form $\phi^\vee_J$ is given by \eqref{eq:sphereforms2} and $\vphi_J$ is 
\begin{align}
	\vphi_J 
	&=
	\frac{ dY_1 \cdots dY_{|J|} }{ Y_{j_1} \cdots Y_{j_{|J|}} }\
	\delta\left(|I_J| Y_{|J|+1} - 1 \right)  dY_{|J|+1}\
	\delta\left(Y_\perp^2+\sum_{i=1}^{|J|} Y_i^2 + Y_{|J|+1}^2 + \tilde{Y}^2\right) dY_\perp^2  
	\nn\\&\qquad
	\times \frac{
	 	a_{m-1}\ r_J^{\delta_{m,\text{odd}}}\ d^{m}\tilde{Y}
	}{ 
		(-2)^{|J|}  (Y_\perp^2)^n  
	}  	
\end{align}
where $m=\dint-|J|+1$, $n=\lfloor\frac{m+1}{2}\rfloor$ and $d^{m}\tilde{Y} = \la e_1 \cdots e_{|J|+1}\ d\tilde{Y} \cdots d\tilde{Y} \ra / m!$. Taking the residue and integrating over $dY_{|J|+1}$ and $dY_\perp^2$, the top line above becomes unity and 
\begin{align}
	\phi_J	
	= \res_J[\vphi_J]
	= \frac{ (-1)^{|J|+n} a_{m-1} r_J^{\delta_{m,\text{odd}}}}{ 2^{|J|} }  \frac{ d^{m}\tilde{Y} }{ (r_J^2 + \tilde{Y}^2)^{n} }.
\end{align}
On the $J$-cut, $\phi^\vee_J$ and $\phi_J$ are elements of the twisted cohomologies with the respective connections $\pm\vep \ d\log(r_J^2+z^2)$.
Since these connections have a single critical point
the corresponding twisted cohomologies are 1-dimensional (we are ignoring any other boundaries). Thus, $\phi^\vee_J$ and $\phi_J$ can be reduced to a cohomologous sphere-form with logarithmic singularities using integration-by-parts
\begin{align}
	\frac{ d^{m}\tilde{Y} }{ (r_J^2 + \tilde{Y}^2)^n }
	&\simeq b^{(m,n)}_{\alpha} \left(\frac{1}{r_J^2}\right)^{n-1}
	\frac{ d^{m}\tilde{Y} }{ r_J^2 + \tilde{Y}^2 },
	\\ 
	b^{(m,n)}_{\alpha} &= (-1)^{n-1} \left(\alpha\right)_{1-n}
		\left(-\alpha-\frac{m}{2}+1\right)_{n-1}.
\end{align}
While we have $n=\lfloor\frac{m+1}{2}\rfloor$ in mind, the above formula is valid for any integer $n$.
Thus our intersection number is proportional to 
\begin{align} \label{eq:log sphere-form}
	\la \vphi^\vee_J \vert \vphi_J \ra  \propto \bigg\la
	\frac{ d^{m}\tilde{Y} }{ r_J^2 + \tilde{Y}^2 } \bigg\vert \frac{ d^{m}\tilde{Y} }{ r_J^2 + \tilde{Y}^2 } \bigg\ra,
\end{align}
which involves logarithmic forms whose intersections can be computed by repeated application of a one-dimensional formula
(see appendix \ref{app:log sphere int}).  Explicitly, 
\begin{align} 
	\la \vphi^\vee_J \vert \vphi_J \ra 
	&= \frac{ (-1)^{|J|-\lfloor|J|/2\rfloor} }{ 2^{\dint+1} } 
	\frac{a_{m-1}\ c_{m-1}}{ \left(-\vep\right)^{\delta_{m,\text{even}}} 
	\left(\vep {-} \left\lfloor\frac{m{-}1}{2}\right\rfloor \right)_{m-\delta_{m,\rm even}}
	} .
	\label{eq:diag int} 
\end{align}
Setting $\la \vphi^\vee_J \vert \vphi_J \ra = - \vep^{\lfloor\dint/2\rfloor}/2^{\dint}$ fixes the normalization $a_{m-1}$ (see table \ref{tab:norm}). 
The power of $\vep$ is chosen to match the powers of $\pi$ in the volume of a $\dint$-sphere and  fact that the intersections do not depend on any kinematic variables reflects the fact that we are dealing with pure transcendental functions.


\begin{table}[h]
	\centering
	\bgroup
	\def\arraystretch{1.5}
	\begin{tabular}{c|ccccc}
		\multicolumn{6}{c}{$\dint=4 \implies |J|=1,\dots,5$} \\
		\hline 
		$|J|$ & 1 & 2 & 3 & 4 & 5
		\\
		\hline
		$a_{4-|J|}$ 
			& $\vep(\vep+1)$ 
			& $2\vep(\vep+1)$ 
			& $2\vep^2$ 
			& $4\vep^2$
			& $4\vep^2$
	\end{tabular}
	\egroup 
	\caption{ \label{tab:norm}
		The normalization factors for the Feynman integrals ``dual'' to the one-loop uniform transcendental dual forms.
		The factors of $(\vep+1)$ in the tadpole and bubble normalization are related to
	our version of dimension shifting
		(multiplying or dividing by $Y_\perp^2$) versus the usual notion of integrating in a different dimension.  The pentagon (effectively shifted to $d=6$ due to
		a $Y_\perp^2$ numerator, see eq.~\eqref{normalized FI})
		integrates to $O(\vep^3)$ and decouples as $d\to 4$.
	}
\end{table}

As discussed at the end of the preceding section,
Feynman integrals normalized as in eq.~\eqref{normalized FI} will automatically
satisfy the (minus-transpose of the) canonical differential equation \eqref{eq:oddOmega}-\eqref{eq:evenOmega}.
In particular, just independence on kinematic variables forced us to use the combinations:
\be
	\sqrt{-(J0)^2}\ r_J^{\delta_{m,\text{odd}}} 
	= \begin{cases}
		\sqrt{-(J0)^2} & m=\dint-|J|-1 \text{ even}
		\\
		\sqrt{-(J)^2} & m=\dint-|J|-1 \text{ odd}
	\end{cases}
\ee
which are the familiar normalizations which remove the leading singularity 
of $2n$-gons in $2n$-dimensions \cite{Spradlin:2011wp, Bourjaily:2019exo}.
The square root is positive for real Euclidean momenta and positive masses, but can become imaginary in Minkowski signature (explaining factors of $i$ which will appear below).

Non-diagonal intersections (for example between bubble-dual and box-form)
will be described in a companion paper and confirmed directly
to vanish, which will then be applied to extract integral coefficients in specific scattering amplitudes. 
In non-generic kinematics, we will now see that the pairing is diagonal not in the standard scalar basis of FIs, but in a modified, infrared-finite basis.

\subsection{Some degenerate limits \label{sec:degenerate limits}}

To illustrate the formulas of subsections~\ref{ssec:summary} and \ref{sec:normalization},
we consider a four-point amplitude with massless external legs. In the massless external limit ($p_i^2\to0$), all bubbles with massless incoming/outgoing external momentum ($\vphi^\vee_{12}, \vphi^\vee_{14}, \vphi^\vee_{23}, \vphi^\vee_{34}$) become exact on their respective cuts and our basis shrinks from 15 to 11 elements.  Applying integration-by-parts
to remove these ``bad'' bubble-duals we find a sum of triangle-duals,
for example
\be
\vphi_{12} \vert_{p_1^2=0} \simeq i \left( \vphi^\vee_{124} -\vphi^\vee_{123} \right) \vert_{p_1^2=0}\,.
\ee
This has an interesting effect on the differential equation.  Inspection of eqs.~\eqref{eq:oddOmega}-\eqref{eq:evenOmega} reveals that divergences in matrix elements neatly cancel, for example:
\begin{align}
	\Omega_{1;123}^\vee 
	&\underset{p_1^2\to0}{\longrightarrow} \left(\Omega^\vee_{1;123} - i \Omega^\vee_{1,12} \right)_{p_i^2=0}
	\nn \\
	&= \frac{\vep}{2} d\log \frac{ s{+}m_1^2{-}m_3^2 {+} \sqrt{s^2{+}(m_1^2{-}m_3^2)^2{-}2s(m_1^2{+}m_3^2)} }{ s{+}m_1^2{-}m_3^2 {-} \sqrt{s^2{+}(m_1^2{-}m_3^2)^2{-}2s(m_1^2{+}m_3^2)} }\,.
\end{align}
The two elements on the first line separately diverge, but their combination is finite.
Thanks to this mechanism, one finds an unambiguous differential equation in the smaller basis.

An interesting subtlety happens in the equal internal mass limit $m_i\to m$. Naively, the size of the basis does not change in this limit (the tadpole-duals look like valid forms in the algebraic cohomology). However, it turns out this would overcount (on the FI side, one could also show that equal-mass tadpoles are cohomologous).

As mentioned near \eqref{longexact}, there are two ways that we can overcount. The second of these applies here: the tadpole forms cannot be uplifted from their cut space to the full space. While this is not obvious using the algebraic techniques introduced so far, the obstruction is clear in the compactly supported world. As emphasized in section \ref{ssec:duals}, the compactly supported cohomologies are really the main objects of interest and we only use the algebraic description when it streamlines calculation.

For concreteness, we focus on the obstruction for $\vphi^\vee_1$.
Ignoring the other three propagators ($D_{2,3,4}$), a compactly supported tadpole form can be constructed $[\tilde{\vphi}^\vee_1]_c \in H^4_c( \mathbb{C}^4 \setminus \{\G=0\})$. However, one finds that its restriction to massless-bubble cuts fails to be exact (see eq.~\ref{dphi example 1}),
an obstruction which is captured by a nontrivial cohomology class in $H_c^{4}$ of the bubble cut.
Naively, the restriction of a four-form to the bubble-cut (a three-dimensional complex manifold)
would naively vanish, however, the point is that the compact-supported form contains an anti-holomorphic component. This obstruction is detailed in appendix 
\ref{app:equal mass dual tadpole} where we show furthermore that  it cancels pairwise between two tadpole-dual forms which share the same bubble cut.  Therefore, for all massless-external momenta and equal-mass internal (light-by-light scattering kinematics), only the sum of all four tadpole-duals is a valid form.
This phenomenon could also have been deduced from the differential equation, where the derivatives of tadpole-duals individually diverge but sum up to a finite equation.

Therefore, in the equal internal mass limit $m_i\to m$, the basis shrinks from 11 elements to 8, as known ie. from \cite{Caron-Huot:2014lda} (if we don't identify triangles related by symmetry).
Thus, in the limit $p_i^2\to0, m_i\to m$, the basis consists of one tadpole, two bubbles, four triangles, and one box:
\begin{align}
	\bs{\vphi}^\vee 
	&= \begin{pmatrix}
		\vphi^\vee_{\text{tad}}
		& \vphi^\vee_{\text{bub}_s} 
		& \vphi^\vee_{\text{bub}_t}
		& \vphi^\vee_{\text{tri}_{s,1}}
		& \vphi^\vee_{\text{tri}_{s,2}}
		& \vphi^\vee_{\text{tri}_{t,1}}
		& \vphi^\vee_{\text{tri}_{t,2}}
		& \vphi^\vee_{\text{box}}
	\end{pmatrix},
	\nn \\
	&\equiv \begin{pmatrix}
		\sum_{i=1}^4 \vphi^\vee_{i}
		& \vphi^\vee_{13} 
		& \vphi^\vee_{24}
		& \vphi^\vee_{123}
		& \vphi^\vee_{134}
		& \vphi^\vee_{124}
		& \vphi^\vee_{234}
		& \vphi^\vee_{1234}
	\end{pmatrix}\bigg\vert_{p_i^2=0,m_i=m}. 
\end{align}
The corresponding connection, obtained from
\eqref{eq:oddOmega}-\eqref{eq:evenOmega} by
summing the tadpole components of $\Omega^\vee_{\bullet;\bullet}\vert_{p^2\to0}$:
\begin{align}
	\nabla^\vee \vphi_i^\vee = \sum_i \vep\ d\left(\delta_{ij} \log(m^2) + \tilde{\Omega}^\vee_{ij})\right) \vphi_j^\vee,
\end{align}
where $i,j \in \{\text{tad}, \text{bub}_s, \text{bub}_t, \text{tri}_{s,1}, \text{tri}_{s,2}, \text{tri}_{t,1}, \text{tri}_{t,2}, \text{box} \}$ and 
\small
\begin{align}
	\tilde{\mat{\Omega}}^\vee &=  
	\begin{pmatrix}
		0 
		& \log \frac{\beta_u{-}1}{\beta_u{+}1}
		& \log \frac{\beta_v{-}1}{\beta_v{+}1}
		& 0
		& 0
		& 0
		& 0
		& 0
	\\[0.2em]
		0 
		& -\log \frac{u}{1+u}
		& 0
		& i \log \frac{\beta_u-1}{\beta_u+1}
		& i \log \frac{\beta_u-1}{\beta_u+1}
		& 0
		& 0
		& i\log \frac{\beta_u-\beta_{uv}}{\beta_u+\beta_{uv}}
	\\[0.2em]
		0 
		& 0
		& -\log \frac{v}{1+v}
		& 0
		& 0
		& i \log \frac{\beta_v-1}{\beta_v+1}
		& i \log \frac{\beta_v-1}{\beta_v+1}
		& i\log \frac{\beta_v-\beta_{uv}}{\beta_v+\beta_{uv}}
	\\[0.2em]
		0 
		& 0
		& 0
		& 0
		& 0
		& 0
		& 0
		& \frac12 \log \frac{\beta_{uv}-1}{\beta_{uv}+1}
	\\[0.2em]
		0 
		& 0
		& 0
		& 0
		& 0
		& 0
		& 0
		& \frac12 \log \frac{\beta_{uv}-1}{\beta_{uv}+1}
	\\[0.2em]
		0 
		& 0
		& 0
		& 0
		& 0
		& 0
		& 0
		& \frac12 \log \frac{\beta_{uv}-1}{\beta_{uv}+1}
	\\[0.2em]
		0 
		& 0
		& 0
		& 0
		& 0
		& 0
		& 0
		& \frac12 \log \frac{\beta_{uv}-1}{\beta_{uv}+1}
	\\[0.2em]
		0 
		& 0
		& 0
		& 0
		& 0
		& 0
		& 0
		& \log \frac{1+u+v}{u+v}
	\end{pmatrix}.
\label{eq:Oeqm}
\end{align}
\normalsize
Here, we have adopted the notation of \cite{Caron-Huot:2014lda}, where $u=-4m^2/s$, $v=-4m^2/t$, $\beta_u=\sqrt{1+u}$, $\beta_v=\sqrt{1+v}$ and $\beta_{uv}=\sqrt{1+u+v}$ for ease of comparison. 
Rescaling the $\vphi^\vee_i$ by $\mat{B}=\text{diag}(-1, 1, 1, i,i,-i)$ and taking the minus transpose (see eq.~\eqref{eq:OmegaVeeMinusOmegaT}) of the resulting connection ($\mat{B}\cdot\tilde{\mat{\Omega}}\cdot\mat{B}^{-1}$) we recover the connection in \cite{Caron-Huot:2014lda}. This confirms that our basis of dual forms is indeed dual (after constant rescaling) to the uniform transcendental basis of equal mass FIs!

Next, we take the completely massless limit $m\to0$. This time, the tadpoles and triangles become exact on their respective cuts since $r^2_\text{tad} = r^2_{\text{tri}_{s,t}} = m^2 \to 0$. Moreover, the tadpoles are exact on all boundaries and therefore can be removed. Since the triangles are only exact on their respective cuts, they can be reduced to topologies with more cuts (i.e., the box). Integrating the triangles by parts,
\begin{align} \label{eq:tri to box}
	\vphi^\vee_{\text{tri}_{s,t}} \vert_{m\to0} \simeq \frac{1}{2\vep} \frac{c_1}{c_0} \vphi^\vee_{\text{box}}\vert_{m\to0} = \vphi^\vee_{\text{box}}\vert_{m\to0}
\end{align}
and
\begin{align}
	\nabla^\vee \vphi^\vee_{\rm{bub}_s} \vert_{m_i=0}
	&= \Omega^\vee_{\text{bub}_s;\text{bub}_s}\vert_{m_i=0} \wedge  \vphi^\vee_{\rm{bub}_s}\vert_{m_i=0} 
	\nn \\
	&\qquad +\left( 
		\Omega^\vee_{\text{bub}_s;\text{box}}\vert_{m_i=0} 
		+ \sum_{j=1}^2 \Omega^\vee_{\text{bub}_s;\text{tri}_{s,j}}\vert_{m_i=0} 
	\right) \wedge \vphi^\vee_{\rm{box}}\vert_{m_i=0}.
\end{align}
The $t$-channel bubble connection is obtained by exchanging $s$ and $t$. This leaves only the box-box component of the connection, which is finite in the completely massless limit. Explicitly, the kinematic connection becomes
\begin{align} \label{eq:massless 4pt DEQ}
	\nabla^\vee \begin{pmatrix} \vphi^\vee_{\text{bub}_s} \\ \vphi^\vee_{\text{bub}_t} \\ \vphi^\vee_{\text{box}} \end{pmatrix}
	= \vep 
	\begin{pmatrix}
		d\log(s) & 0 & i\, d\log(s/t)  \\
		0 & d\log(t) & i\, d\log(t/s) \\
		0 & 0 & d\log(st/u)
	\end{pmatrix}
	\wedge \begin{pmatrix} \vphi^\vee_{\text{bub}_s} \\ \vphi^\vee_{\text{bub}_t} \\ \vphi^\vee_{\text{box}} \end{pmatrix}
\end{align}
in the completely massless limit. The lack of $u=-(s+t)$ poles in the off-diagonal terms is related to the fact that our box-dual is dual to an IR-finite box!

To see this more explicitly, consider the massless box differential equation of \cite{Henn:2014qga}
\begin{align} \label{eq:4ptHenn}
	\nabla\begin{pmatrix}\vphi_{\text{bub}_s} \\ \vphi_{\text{bub}_t} \\ \vphi_\text{box}\end{pmatrix}
	= -\vep 
	\begin{pmatrix}
		d\log(s) & 0 & 0 \\
		0 & d\log(t) & 0 \\
		d\log(s^2/u^2) & d\log(t^2/u^2) & d\log(st/u)
	\end{pmatrix}
	\wedge \begin{pmatrix}\vphi_{\text{bub}_s} \\ \vphi_{\text{bub}_t} \\ \vphi_\text{box}\end{pmatrix}
\end{align}
where $\vphi_{\rm{bub}_s} = \vep s\ G_{1,0,2,0}$, $\vphi_{\rm{bub}_t} = \vep t\ G_{0,1,0,2}$, $\vphi_{\rm{box}} = \vep^2 st\ G_{1,1,1,1}$. Here, $G_{\nu_1,\nu_2,\nu_3,\nu_4}$ is the Feynman integral with propagator $D_i$ to the power $\nu_i$. Squaring a propagator in the definition of the $s,t$-bubbles effectively reduces the $4d$-bubble to the $2d$-bubble. This reflects analogous powers of $1/\ell_\perp^2$ in our bubble duals (dimension shifting will be further discussed in \cite{schap}). 

We can now find which combinations of FI are orthogonal to our $\vphi^\vee$ basis by comparing the differential equations.  It turns out we can guess the result by trying combinations with nice properties,
namely infrared safety.
From the integrated form in the standard basis:
\begin{align}
	\mathscr{I}_{{\rm bub}_{x=s,t}} 
	&= - \vep \frac{ (-x)^{-\vep} \Gamma^2(1-\vep)\Gamma(1+\vep) }{\Gamma(1-2\vep)},
	\\
	\mathscr{I}_{\rm box} 
	&= e^{-\vep\gamma_E} \left[ 
		4 
		+ \vep \left( - 2 \log (st) \right) 
		+ \vep^2 \left( -\frac{4\pi^2}{3} + 2\log(-s)\log(-t) \right) 
		+ \O(\vep^3)
	\right],
\end{align}
we see that the combination
\be
\mathscr{I}_{\rm box}^{\rm finite}=
\mathscr{I}_{\rm box} + 2 \mathscr{I}_{{\rm bub}_s} + 2 \mathscr{I}_{{\rm bub}_t} = -\vep^2 (\pi^2 + \log^2(st)) + \O(\vep^3)
\ee
is nicely IR finite (ie. the soft-collinear double pole after dividing by the factor of $\vep^2$ from the normalization of $\mathscr{I}_{\rm box} = \int \vphi_{\rm box}$ canceled out).
From here, it is simple to guess the transformation to an IR-finite basis:
\begin{align}
	\bs{\vphi}_\text{finite} = \mat{B} \cdot \bs{\vphi} 
	= \begin{pmatrix}
		2\, \vphi_{\text{bub}_t} \\
		2\, \vphi_{\text{bub}_s} \\
		-i\left( \vphi_\text{box} + 2 \vphi_{\text{bub}_s} + 2 \vphi_{\text{bub}_t} \right)
	\end{pmatrix},
	\qquad 
	\underline{\bs{B}} = \begin{pmatrix}
		2 & 0 & 0 \\
		0 & 2 & 0 \\
		-2i & -2i & -i 
	\end{pmatrix}.
\end{align}
Implementing this change of basis, the kinematic connection becomes
\begin{align}
	\underline{\bs{\Omega}}_\text{finite} = \mat{B}\cdot\mat{\Omega}\cdot\mat{B}^{-1}
	=  -\vep 
	\begin{pmatrix}
		d\log(s) & 0 & 0 \\
		0 & d\log(t) & 0 \\
		i\, d\log(s/t) & i\, d\log(t/s) & d\log(st/u)
	\end{pmatrix},
\end{align}
which is precisely the minus transpose of \eqref{eq:massless 4pt DEQ} (see eq.~\eqref{eq:OmegaVeeMinusOmegaT})!  This indicates that the integrals orthonormal
to our dual forms are the $\vphi_{j, \rm finite}$, that is:
\be
 \langle \vphi^\vee_i | \vphi_{j, \rm finite}\rangle = \delta_{i,j}\,,
\ee
which will be confirmed explicitly in our subsequent paper.
With very little input (we simply required our dual forms to have at most simple poles in $(YI)$), we thus produced a uniform transcendental basis dual to IR finite Feynman integrals.
It would very interesting if a similar property were to hold at higher loops as well.

\subsection{A simple two-loop (elliptic) example \label{sec:elliptic}}

Having performed a general treatment of one-loop dual forms,
we will analyze the simplest two-loop example: the sunrise diagram, which has three massive propagators.
We study the equal mass limit where the problem is especially simple; there is a single scale $x=m^2/p^2$. While there are only three independent functions due to symmetries,
the cohomology is 7-dimensional (see for example \cite{Mizera:2019vvs})
(each double cut has a 1-dimensional cohomology while the max-cut has a 4-dimensional cohomology).

Since there is a single external momentum, we can use eq.~\eqref{L loops from Gram}
working around $d_{\rm int}=1$
(or equivalently use the Baikov parametrization, see appendix A of \cite{Mastrolia:2018uzb})
to determine the integration measure on the un-cut space.
We choose variables $z_{i=1,2}=\ell_i^2/p^2$, $z_3=\ell_1\cdot\ell_2/p^2$ and
$z_{3+i}=\ell_{i}\cdot p/p^2$ so that we can trivialize the dependence on $p^2$.
Ignoring numerical constants, the measure is
\begin{align}
	\prod_{i=1}^2 \frac{d^d\ell_j}{\pi^{d/2}} \propto (p^2)^4\ d^5z\  u,
\end{align}
where the twist $u=(p^4\B)^{(d-4)/2}$ is expressed in terms of the rescaled Gram determinant
\begin{align}
	\B = \det\mat{G}(\ell_1,\ell_2,p)/p^6 =z_3^2-z_1 z_2  +z_2 z_4^2 - 2 z_3 z_4 z_5 + z_1 z_5^2\,.
\end{align} 
The propagators (boundaries) are
\begin{align}
	D_1/p^2  = x + z_1\,,	\quad
	D_2/p^2  = x + z_2\,,	\quad
	D_3/p^2  = 1 + x + z_1 + z_2 + 2(z_3+z_4+z_5)\,.
\end{align} 
For dual forms we write explicitly all factors except for the twist $u^\vee=u^{-1}$.
On the double cuts, we choose the basis
\begin{align}
	\vphi^\vee_1 = \delta_{1,2} \left(\frac{d^3z}{\B\vert_{12}} \right),
	\quad
	\vphi^\vee_2 = \delta_{1,3} \left(\frac{d^3z}{\B\vert_{13}} \right),
	\quad
	\vphi^\vee_3 = \delta_{2,3} \left(-\frac{d^3z}{\B\vert_{23}} \right),
\end{align}
 where $d^3z = dz_3 \wedge dz_4 \wedge dz_5$ and $z_{i<3}$ have been fixed by the cut conditions. On the maximal 3-cut, the twist is
\begin{align}
	\B\vert_{123} =  -\frac14 x (1+x)^2 -2 L_1 L_2 L_3
\end{align}
where 
\begin{align}
L_1 = \frac12(1 - x) + z_4,
\quad 
L_2 = \frac12(1 - x) + z_5,
\quad 
L_3 = \frac12(-1 - x) - z_4 - z_5,
\end{align}
are the asymptotes of $\B\vert_{123}$, which permute into one another under
$S_3$ permutations. On the max cut, a basis of dual forms is:
\begin{align} \label{elliptic vee}
	\vphi^\vee_4 = \delta_{1,2,3} \left( \frac{d^2z}{\B \vert_{123}} \right),
	\quad&
	\vphi^\vee_5 = 
		\delta_{1,2,3} \left(\frac{L_1L_2+L_2L_3+L_3L_1}{x(1+x)} \frac{d^2z}{\B\vert_{123}} \right),
	\\
	\vphi^\vee_6 = \delta_{1,2,3} \left( (1+3z_4) \frac{d^2z}{\B\vert_{123}} \right),
	\quad&
	\vphi^\vee_7 = \delta_{1,2,3} \left( (1+3z_5) \frac{d^2z}{\B\vert_{123}} \right),
\end{align}
where $d^2z = dz_4 \wedge dz_5$. The dual forms $\vphi^\vee_{6,7}$ would vanish
upon symmetrizing  propagators and are thus dual to FIs which integrate to zero by permutation symmetry.

We obtained differential equations for these dual forms using the same procedure as in
section \ref{sec:one-loop deqs}: we take the covariant derivative of $\vphi^\vee_i$ and reduce using integration-by-parts identities. The identities are somewhat distinct from the usual ones for FI because the only possible denominator is $\B$ (even on two-particle cuts) so effectively we are working with polynomials; however we must retain boundary terms that mix the two- and three-particle cuts.
We then obtain the following differential equation:
\begin{align}
	\nabla^\vee \bs{\vphi}^\vee = \left(2\vep d\log p^2 + \mat{\Omega}^\vee dx\right) \cdot \bs{\vphi}^\vee,
	\qquad 	\nabla^\vee = (d+ d\log u^\vee \wedge\bullet)
\end{align}
where $\vep=\frac{4-d}{2}$ and $d$ is the exterior derivative
on the \emph{total} space (including kinematic derivatives in $p^2$ and $x$)
and 
\begin{align}
	\mat{\Omega}^\vee =
	\begin{pmatrix}
		\frac{2\vep}{x} & 0 & 0 & -\frac{1}{6x} & 0 & -\frac{1}{6x} & -\frac{1}{6x} 
		\\
		0 & \frac{2\vep}{x} & 0 & -\frac{1}{6x} & 0 & 0 & \frac{1}{6x}
		\\
		0 & 0 & \frac{2\vep}{x} & -\frac{1}{6x} & 0 & \frac{1}{6x} & 0 
		\\
		0 & 0 & 0 & \frac{\vep}{x} + \frac{2\vep}{1+x}-\frac{6}{1+9x} 
			& -\frac{4(1+3\vep)}{1+9x} & 0 & 0
		\\
		0 & 0 & 0 & -\frac{3}{4x} + \frac{9}{1+9x} 
			& {-}\frac{1}{x} -\frac{1}{1+x}+\frac{6(1+3\vep)}{1+9x} & 0 & 0
		\\
		0 & 0 & 0 & 0 & 0 & \frac{\vep}{x} +\frac{2\vep}{1+x} & 0 
		\\
		0 & 0 & 0 & 0 & 0 & 0 & \frac{\vep}{x} +\frac{2\vep}{1+x}
	\end{pmatrix}\ . \label{dual diff}
\end{align}
Note that this is linear in $\vep$ and all entries take d-$\log$ forms; this is a consequence of the normalization
choices in eq.~\eqref{elliptic vee}, although there were not many other
options given the symmetries of this problem.  This is however not in $\vep$-canonical form (not proportional to $\vep$).
The $\vep=0$ part can be removed in principle by solving the equations with $\vep=0$, at the cost of introducing elliptic functions coming from the $2\times 2$ block at the center of the matrix
(see \cite{Laporta:2004rb,Tarasov:2006nk,Broedel:2017siw,Bogner:2019lfa} for further explanations
and references).

By itself, even without knowledge of which FIs our forms are dual to,
the dual differential equation \eqref{dual diff} determines which transcendental functions
can appear in this topology. Our purpose here is only to confirm this.
To compare with the standard result for that family of integrals
(see for example \cite{Laporta:2004rb,Mizera:2019vvs}),
we looked for a change of basis connecting the two distinct forms of the differential equations.
This change of basis is in principle precisely the intersection numbers, but we did not use that,
and simply observed
that the following basis of 7 FIs satisfies precisely the minus-transpose of eq.~\eqref{dual diff}:
\be
 \vphi_i=\left\{\begin{array}{l}
 (d-2)^2G_{110}\,,\qquad (d-2)^2G_{101}\,, \qquad (d-2)^2G_{011}\,, \\[2pt]
(1+x)(8-3d)m^2G_{111}-\frac{(1+x)(1+9x)}{3x}m^4(G_{211}+G_{121}+G_{112}),\\[2pt]
\frac23(8-3d)m^2G_{111}\,,\qquad
\frac23\frac{1+x}{x}m^4(G_{112}-G_{211}),\qquad
\frac23\frac{1+x}{x}m^4(G_{112}-G_{121})
\end{array}\right\} \,,
\ee
where we use the standard notation $G_{ijk}=\int \frac{1}{D_1^iD_2^jD_3^k}$,
here for integrals in $2{-}2\vep$ dimensions
(which are linear combinations of integral in $4{-}2\vep$ due to dimension shifting identities
and thus span the same basis of functions).
The agreement indicates that in this basis $\langle \vphi^\vee_i |\vphi_j\rangle\propto \delta_{ij}$.

It is an important problem in general to find bases of
forms (either dual or standard), if possible, in which differential equations are linear in $\vep$;
the mechanism by which this turned out to be the case for eqs.~\eqref{elliptic vee}
remains to be elucidated.
Since the mechanics of dual forms are somewhat reversed (higher-dimensional dual forms being conjugate to simpler FIs) we hope that combining the two viewpoints will lead to progress on the question.

\section{Conclusion \label{sec:conclusion}}

In this work, we introduced dual forms that are Poincar\'e dual to Feynman integrands. Formally, dual forms are elements of the twisted relative cohomology. Perhaps more intuitively, they are Feynman integrands in ($\dint+2\vep$)-dimensions that vanish (instead of having poles) near the zero locus of the propagators ($\{D_i=0\}$).
The product of a dual form with an ordinary integrand can be meaningfully integrated over $\mathbb{C}^n$, 
yielding an algebraic invariant known as the intersection number.

Intersection numbers have been recently proposed as an efficient tool to reduce complicated integrands to a minimal basis of master integrals \cite{Mastrolia:2018uzb, Frellesvig:2019kgj, Frellesvig:2019uqt,  Mizera:2019vvs, Mizera:2020wdt, Frellesvig:2020qot},
bypassing the generation integration-by-parts identities.
Previous treatments required to deform Feynman integrals by raising propagators to non-integer powers.
The novelty of the dual forms introduced in this paper is that they deal directly with undeformed integrands (although we still require non-integral spacetime dimension $d=4{-}2\vep$).
This causes them to have support only on cut sub-manifolds,
making them significantly simpler than their Feynman counterparts (see section \ref{sec:dual form localization}).
This echoes a key insight from the method of generalized unitarity: loop corrections can be determined from on-shell information.

Intuitively, the space of dual forms is closely related to the space of complex integration contours in momentum space. 
In practice, the forms are much simpler, because they are purely algebraic and don't pick factors of $e^{i\pi \epsilon}$ under monodromy. Their pairing with Feynman forms is also algebraic, which 
contrasts with the integration pairing of contours with forms, which typically yields transcendental numbers (periods).


This paper, the first of a series of two, focuses on intrinsic properties of dual forms: their transcendentality properties and the differential equations they satisfy.
In a basis in which intersections are orthonormal, one expects dual forms to satisfy equivalent
(minus transpose) differential equations as the Feynman forms they are dual to.
We confirmed this by explicitly computing the differential equations satisfied by a basis of one-loop dual forms,
reproducing known examples.  Since dual forms live on cuts, the necessary integration-by-part identities are very symmetrical and considerably simpler than
for Feynman forms. This allowed us to obtain (for the first time to our knowledge) differential equations for arbitrary one-loop integrals in
non-integer spacetime dimensions (see equations (\ref{eq:oddOmega}-\ref{eq:evenOmega})).

For generic kinematics, this provides a natural basis of one-loop
dual forms of pure transcendental weight, which turns out to be dual to a standard basis of scalar integrals.  Surprisingly, when taking degenerate limits (massless, or equal internal masses), we found that they are no longer dual to simple scalar integrals, but to special IR-finite combinations!
We also discussed phenomena that occur in degenerate limits: new relations may appear, or certain forms may cease to be allowed (and should be discarded), as discussed in sections \ref{sec:dual form localization} and \ref{sec:degenerate limits}.  Both phenomena could be explained by non-middle-dimensional cohomology groups.

While we focused on applications to one-loop integrals and only considered the simplest of
two-loop examples,
we expect many aspects to straightforwardly extend over to higher loops, such as the localization to cuts
and the comparatively simple form of dual integration-by-parts vectors.
Even so, there are still many open questions for both one-loop and multi-loop integrals.
For example, are degenerate limits of dual forms always naturally dual to IR-finite combinations of Feynman integrals? If so, the dual perspective would be an important step towards working directly with 4-dimensional integrals.  Also, how does the $\vep\to0$ limit relates to integer-dimensional cohomology group, as studied initially
for example in \cite{FOTIADI1965159, Hwa:102287, Lascoux:1968bor}?
We also expect our dual forms to help extend the diagrammatic coproduct,
since they could be an effective replacement for integration contours (mod $i\pi$)
\cite{Panzer:2016snt, Schnetz:2017bko, Abreu:2017enx, Abreu:2018sat, Abreu:2018nzy, Abreu:2019wzk, Abreu:2019eyg, Brown:2019jng}.

\section{Acknowledgements}
The authors would like to thank Sebastian Mizera, Mathieu Giroux, Brent Pym, Kale Coville and Samuel Abreu for helpful discussions at various stages of this work, and the participants of the 2019 Padova workshop ``Intersection theory \& Feynman Integrals''. 
This work (S.C.-H.) was supported by the National Science and Engineering Council of Canada, the Canada Research Chair program,
the Fonds de Recherche du Qu\'ebec - Nature et Technologies, and the Sloan Foundation.
A.P. is grateful for support provided by the National Science and Engineering Council of Canada and the Fonds de Recherche du Qu\'ebec - Nature et Technologies.

\appendix

\section{Interpretation of relative $\theta$ and $\delta$ as distributions\label{app:delta}}

In this appendix we construct the smooth-compact counterpart to the combinatorial boundaries
in eq.~\eqref{delta map}.  The analogous construction in homology is known as the Leray coboundary.
It is useful to define a step function which vanishes in some neighborhood $\mathcal{U}_i=\{y\in Y: |D_i(y)|< \epsilon\}$ of a boundary:
\be \label{eq:theta def}
 \theta(D_i) \equiv \theta( |D_i|>\epsilon) = \left\{\begin{array}{ll} 1,\qquad & |D_i|\geq \epsilon\ ,\\ 0,& |D_i|< \epsilon\ .\end{array}\right.
\ee
(One could make $\theta$ as smooth as desired without affecting the discussion.)
We describe the map in a special case; for the general case see \cite{pham2011singularities}.
\begin{theorem}
Suppose $D\subset Y$ is a compact codimension-1 sub-manifold, and let $\psi \in H^p(D)$ be an element of its cohomology.
Then the following defines a representative of $\delta_D(\psi)\in H^{p+1}(Y,D)$ which vanishes within a neighborhood $\mathcal{U}$ of $D$:
\be
\delta_D(\psi) \simeq \delta_D^{(\mathcal{U})}(\psi) \equiv \theta\ d\theta(D) \wedge u \pi^*(\psi/u) \label{smooth delta}
\ee
where $\pi*$ is the pull-back of a projection $\pi: \mathcal{U} \mapsto D$ onto the sub-manifold from $\mathcal{U}$.
\end{theorem}

\begin{figure}
\centering
\includegraphics[scale=.4]{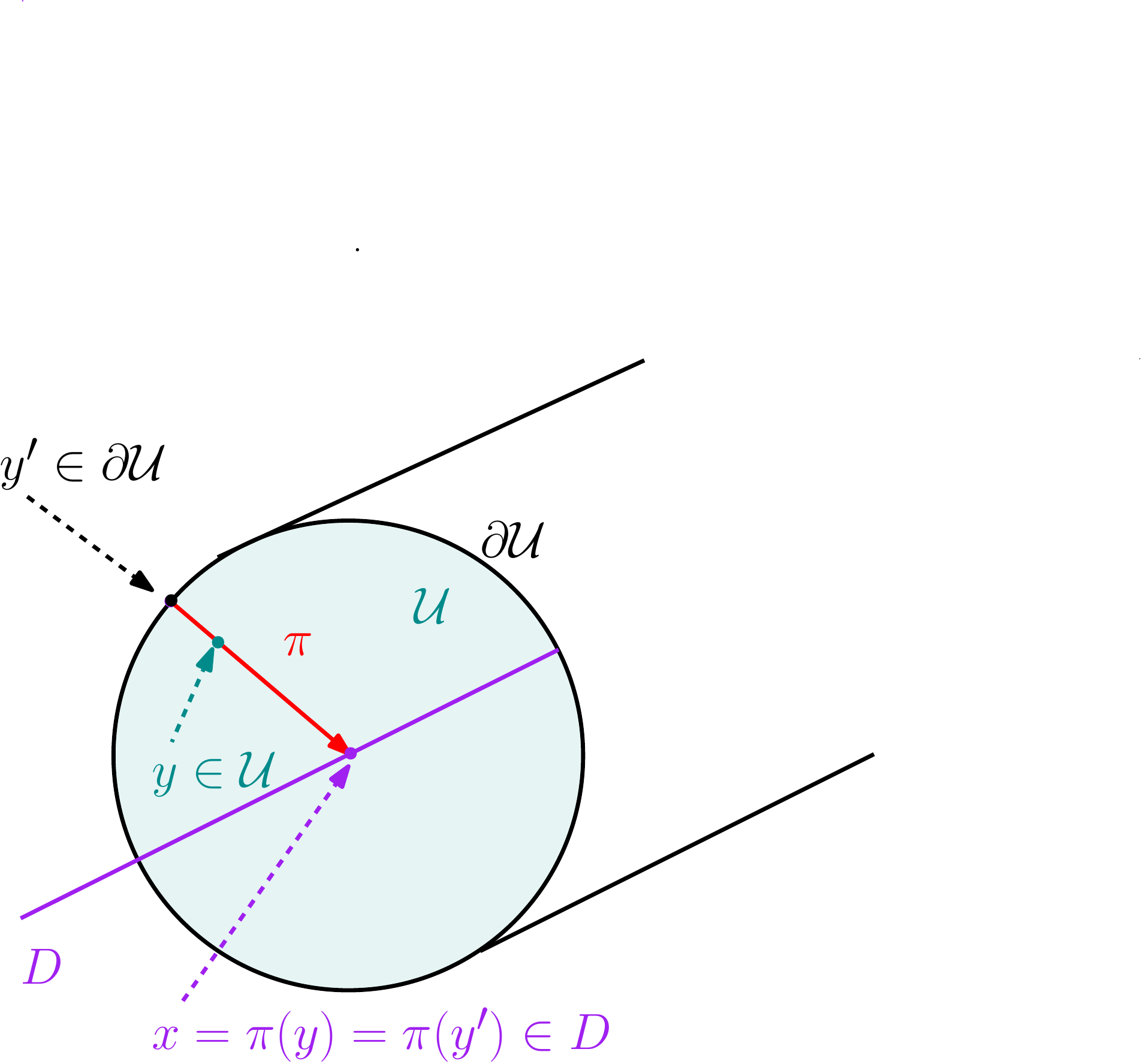}
\caption{
\label{fig:tubular_neigh}
A cross section of the tubular neighbourhood $\mathcal{U}$ is shown in green along with its boundary $\partial \mathcal{U}$ in black. The sub-manifold $D \subset Y$ is shown in purple and the projection map $\pi$ is red. 
}
\end{figure}

The construction is illustrated in fig.~\ref{fig:tubular_neigh}. Thanks to the $d\theta(D)$ factor,
the form $\delta^{(\mathcal{U})}(\psi)$ is evidently supported on the circle $\partial \mathcal{U}$.
The second $\delta_{D}$ is the combinatorial symbol introduced in eq.~\eqref{delta notation}.
The theorem allows us to view $\theta$ and $\delta$, which were introduced in eq.~\eqref{delta notation} as purely combinatorial symbols,
as literal step- and delta- functions: $\theta\equiv \prod_i \theta(D_i)$, $\delta_i(\bullet)\equiv d\theta(D_i)\wedge u \pi^*(\bullet/u)$.

The factors of the twist $u$ in eq.~\eqref{smooth delta} require some explanation.  As will be clear from the proof, they can be interpreted as parallel transport
from a point $y$ in the ambient space to the point $\pi(y)$ on its boundary. This serves to convert from the boundary covariant derivative $\nabla^\vee_D$ into the ambient one $\nabla^\vee$.

The theorem is proven by explicitly constructing the requisite total derivative:
\be \label{eq:delta-map def}
	\delta_{D}^{(\mathcal{U})}(\psi) -\delta_{D}(\psi) 
	= \nabla^\vee \Bigg( \theta\ (\theta(D)-1) u \pi^*(\psi/u)\Bigg) \,.
\ee
The right-hand-side can be computed using the combinatorial rule in eq.~\eqref{boundary term}:
\begin{align}
	\nabla^\vee\Bigg(\theta\ \big(\theta(D)-1\big) u \pi^*(\psi/u)\Bigg) 
	&=\theta\ d\theta(D)\wedge u\pi^*(\psi/u) 
	+ \theta\ \theta(D) \nabla^\vee\big(u\pi^*(\psi/u)\big)  
	\nn\\ 
	&\phantom{=} + \delta_D\big(\big[(\theta(D)-1) u \pi^*(\psi/u)\big]\big|_D\big)
	\nn\\ 
	&= \delta_{D}^{(\mathcal{U})}(\psi)
		+ \theta\ \theta(D)u\pi^*(\nabla^\vee \psi/u) 
		-\delta_D(\psi)\,.
	\nn\\ 
	&= \delta_{D}^{(\mathcal{U})}(\psi)
		-\delta_D(\psi)\,,
\end{align}
since $\psi$ is assumed to be an element of the cohomology $\in H^p(D)$. 

To summarize, the compactly supported Leray forms given by 
\begin{align} \label{leray summary}
	\delta_{i}(\psi) &=  \frac{u}{u\vert_i} d\theta_i \wedge \psi,
	\\
	\delta_{i_1,\cdots,i_{|I|}}(\psi) &= \frac{u}{u\vert_I}
	d\theta_{i_{1}} \wedge \cdots \wedge d\theta_{i_{|I|}} \wedge \psi,
\end{align}
are cohomologous to their combinatorial counterparts!
Notice that all anti-holomorphic dependence is concentrated inside $d\theta$, which in practice
multiply holomorphic forms $\psi$.

\section{Calculation of diagonal intersections \label{app:log sphere int} }

While the technology to calculate intersections will be detailed in the subsequent paper in this series, 
we present a derivation of eq.~\eqref{eq:log sphere-form} for completeness and as a particularly simple example.
We will need two main ingredients: (1) fibration (2) the intersection number of logarithmic forms. 

Fibration is a method to compute multi-variate intersection numbers one variable at a time while keeping all other variables fixed. Schematically, 
\begin{align}
	\la \hat{\vphi}^\vee\ dz_1 {\wedge} {\cdots} {\wedge} dz_m \vert \hat{\vphi}\ dz_1 {\wedge} {\cdots} {\wedge} dz_m \ra
	= \la dz_2 {\wedge} {\cdots} {\wedge} dz_m \vert 
		\la \hat{\vphi}^\vee \vert dz_1 \hat{\vphi}\ dz_1 \ra\
		dz_2 {\wedge} {\cdots} {\wedge} dz_m \ra.
\end{align} 
The intersection number $\la \hat{\vphi}^\vee  dz_1 \vert \hat{\vphi}\ dz_1 \ra$ is well defined since the connection on the single variable $z_1$-cohomology is known; it is the $dz_1$ component of the full space connection $\omega^{(\vee)}\vert_{dz_{i>1}=0}$. 
However, a calculation is required to find the connection to use after $z_1$ is integrated out.

To obtain the connection on the base space (all $z_{i>1}$), a basis for the single variable (fibre) $z_1$-cohomology must be chosen. Then, the connection is obtained by examining how the covariant derivative commutes past the basis elements of the fibre cohomology. In the case of \eqref{eq:log sphere-form}, each fibre cohomology and base cohomology are 1-dimensional making the procedure particularly straightforward.

In order to set up the fibration and recursion more clearly, we define the quadrics
\begin{align}
	Q^{(i)} &= r_J^2 + \sum_{k=i+1}^m z_k^2
	= r_J^2 + z_{i+1}^2 + Q^{(i+1)},
\end{align}
where $z=\tilde{Y}$. Choosing
\begin{align}
	f^{(i)} = Q^{(i)} \frac{d^{m-(i-1)}z_{i}}{Q^{(i-1)}}
\end{align}
as the basis for the $i^\text{th}$ dual fibre cohomology, yields the dual-connection  
\begin{align}
	\omega^{(i)} = \alpha_i\ d\log Q^{(i)},
	\quad
	\alpha_i = \alpha_{i-1}+1/2 = \alpha_0+i/2,
	\quad
	\alpha_0 = \vep,
\end{align}
on the $i^\text{th}$ base.\footnote{
In this simple case, the twist on the $i^\text{th}$ base can be obtained by setting the $i^\text{th}$ fibre variable to zero $Q^{(i)}=Q^{(i-1)}\vert_{z_i\to0}$. However, this will not work in general. For generic quadrics, the twist $Q^{(i)}$ is the discriminant of $Q^{(i-1)}$: $Q^{(i)}=-\frac14 \disc_{z_i}Q^{(i-1)}$. It is just coincidence that the discriminant of an $(m-i+1)$-sphere is a $(m-i)$-sphere.  
}
Here, the connections are obtained by acting on the basis of the $i^\text{th}$ fibre cohomology with the covariant derivative of the $(i-1)^\text{th}$ base $\nabla^{(i-1)} = d + \omega^{(i-1)} \wedge$. Explicitly,
\begin{align}
	\nabla^{(i-1)} f^{(i)} \simeq -f^{(i)} \wedge \omega^{(i)}.
\end{align}
Note that on each successive base the twisting ($\alpha_i$) increases by $1/2$. To keep track of this change in twist, we display the corresponding $\alpha_i$'s on the bra's and ket's. 

With these definitions, \eqref{eq:log sphere-form} becomes
\begin{align}
	\tensor[_{\alpha_0}]{\bigg\la}{}
			\frac{d^{m}z}{Q^{(0)}}
		\bigg\vert 
			\frac{d^{m}z}{Q^{(0)}}
		\bigg\ra_{-\alpha_0}.
\end{align}
Taking the intersection over the first fibre yields
\begin{align}
	\tensor[_{\alpha_1}]{\bigg\la}{}
		\frac{d^{m-1}z}{Q^{(1)}}
	\bigg\vert\
		\tensor[_{\alpha_0}]{\bigg\la}{} f^{(1)} 
		\bigg\vert\frac{dz_1}{Q^{(0)}}\bigg\ra_{\alpha_0}
	\bigg\vert 
		d^{m-1}z
	\bigg\ra_{-\alpha_1} 
	=-\frac{1}{2\alpha_0}\ 
	\tensor[_{\alpha_1}]{\bigg\la}{}
		\frac{d^{m-1}z}{Q^{(1)}}
	\bigg\vert
		d^{m-1}z
	\bigg\ra_{-\alpha_1} .
	\label{eq:vol}
\end{align}
where 
\begin{align}
	\tensor[_{\alpha_i}]{\bigg\la}{}
		f^{(i)}
	\bigg\vert 
		\frac{dz_{i+1}}{Q^{(i)}}
	\bigg\ra_{-\alpha_i}
	= \tensor[_{\alpha_i}]{\bigg\la}{}
		\frac{Q^{(i+1)}\ dz_{i+1}}{Q^{(i)}}
	\bigg\vert 
		\frac{dz_{i+1}}{Q^{(i)}}
	\bigg\ra_{-\alpha_i}
	= -\frac{1}{2\alpha_i}.
	\label{eq:dlog intersection}
\end{align}
Here, we have used the well-known fact that the intersection number of $d\log$ forms localizes on the critical points of the connection \cite{Mizera:2017rqa, Mastrolia:2018uzb, Mizera:2019vvs} (note that $dz_{i+1}/Q^{(i)} \propto d\log (z_{i+1}+\sqrt{Q^{(i+1)}})/(z_{i+1}-\sqrt{Q^{(i+1)}})$ is indeed a $d\log$ form).

The volume form in \eqref{eq:vol} can be converted to a $d\log$ form using integration-by-parts on the $i^\text{th}$ base 
\begin{align}
	\bigg\vert d^{m-i}z \bigg\ra_{-\alpha_i}
	\simeq \frac{-2\alpha_i\ r_J^2}{-2\alpha_i+(m-i)}\ 
	\bigg\vert\frac{d^{m-i}z}{Q^{(i)}}\bigg\ra_{-\alpha_i}. 
\end{align}
Substituting the above into \eqref{eq:vol}, yields
\begin{align}
	\tensor[_{\alpha_0}]{\bigg\la}{}
			\frac{d^{m}z}{Q^{(0)}}
		\bigg\vert 
			\frac{d^{m}z}{Q^{(0)}}
		\bigg\ra_{-\alpha_0}
	&=-\frac{1}{2\alpha_0}
	\frac{-2\alpha_1\ r_J^2}{m{-}2\alpha_1{-}1}\
	\tensor[_{\alpha_1}]{\bigg\la}{}
		\frac{d^{m-1}z}{Q^{(1)}}
	\bigg\vert
		\frac{d^{m-1}z}{Q^{(1)}}
	\bigg\ra_{-\alpha_1} 
	\nn\\
	&=-\frac{1}{2\alpha_0}
	\frac{-2\alpha_1\ r_J^2}{m{-}2\alpha_1{-}1}\
	\tensor[_{\alpha_2}]{\bigg\la}{}
		\frac{d^{m-2}z}{Q^{(2)}}
	\bigg\vert\
		\tensor[_{\alpha_1}]{\bigg\la}{}
			f^{(2)}
		\bigg\vert
			\frac{dz_2}{Q^{(1)}}
		\bigg\ra_{-\alpha_1}\
	\bigg\vert
		d^{m-2}z
	\bigg\ra_{-\alpha_2}
	\nn\\
	&=-\frac{1}{2\alpha_0}
	\frac{r_J^2}{m{-}2\alpha_1{-}1}\
	\tensor[_{\alpha_2}]{\bigg\la}{}
		\frac{d^{m-2}z}{Q^{(2)}}
	\bigg\vert
		d^{m-2}z
	\bigg\ra_{-\alpha_2}.
\end{align}
After repeating this procedure for each remaining fibre, we find
\begin{align}
	\tensor[_{\alpha_0}]{\bigg\la}{}
			\frac{d^{m}z}{Q^{(0)}}
		\bigg\vert 
			\frac{d^{m}z}{Q^{(0)}}
		\bigg\ra_{-\alpha_0 }
	&=-\frac{1}{2\alpha_0\ r_J^2}
	\prod_{i=1}^{m-1}\frac{r_J^2}{m{-}2\alpha_i{-}i}
	\nn\\&= -\frac{(r_J^2)^{m-2}}{2^m \vep} \left(-\vep +\frac{m}{2}\right)_{1-m}.
\end{align}

\section{Dual tadpoles with equal masses but translated by a null momentum\label{app:equal mass dual tadpole} }

In this section, we elaborate on why there is only one dual-tadpole form in the equal mass limit. We explicitly apply the $c$-map to an individual tadpole dual form and find an obstruction. The $c$-map is split into two steps: (1) obtaining a cohomologous form that is compactly supported about the twisted boundaries (2) making it also compactly supported about the untwisted boundaries. After demonstrating the obstruction, which lies in step (2), we provide the proper tadpole dual form that can be made compactly supported without obstruction.

Since the argument is easily generalized to higher dimensional cases, we will specialize to $(2-2\vep)$-dimensional integrals for which the tadpole duals are 2-forms and the top topology is a triangle:
\begin{align}
	D_1 = \ell_\perp^2 + \ell_\parallel^2 + m^2,
	\quad 
	D_2 = \ell_\perp^2 + (\ell_\parallel+p_1)^2 + m^2,
	\quad 
	D_3 = \ell_\perp^2 + (\ell_\parallel+p_1+p_2)^2 + m^2.
\end{align}

The $D_1$ tadpole dual form is 
\begin{align}
	\vphi^\vee_1 = \delta_1(\phi^\vee_1),
	\qquad 
	\phi^\vee_1 = \frac{dz_1 \wedge dz_2}{m^2/s - z_1^2 + z_2^2}
\end{align}
where we have set $\ell_\parallel = z_1 e_1 + z_2 e_2$, $e_1 = p_1 + p_2$, $e_2 = p_1 - p_2$, $p_i^2=0$ and $p_1\cdot p_2 = -s/2$. In order to streamline the $c$-map (compactly supported isomorphism), we change to light-cone-like coordinates $z_1 = a-b$,$z_2 = a+b$. In these coordinates, the boundaries and twist on the 1-cut become
\begin{align}
	u^\vee = s^\vep \left(\frac{m^2}{s}- 4ab\right)^\vep,
	\quad 
	D_2 = -2s a,
	\quad 
	D_3 = -s \left( a + b + 1 \right),
\end{align} 
while the tadpole dual form is
\begin{align}
	\phi^\vee_1 &= 2 \frac{da \wedge db}{m^2/s - 4ab}.
\end{align}
In the light-cone-like coordinates, there exists of a pair of almost global primitives 
\begin{eqnarray}
	&\psi_a = \frac{1}{2\vep} \frac{da}{a},
	\qquad 
	&\nabla^\vee \psi_a = \phi^\vee_1 \text{ for } a \neq 0,
	\\
	&\psi_b = -\frac{1}{2\vep} \frac{db}{b},
	\qquad
	&\nabla^\vee \psi_b = \phi^\vee_1 \text{ for } b \neq 0,
\end{eqnarray}
which makes this $c$-map particularly simple. 

\begin{figure}
\centering
\includegraphics[width=\textwidth]{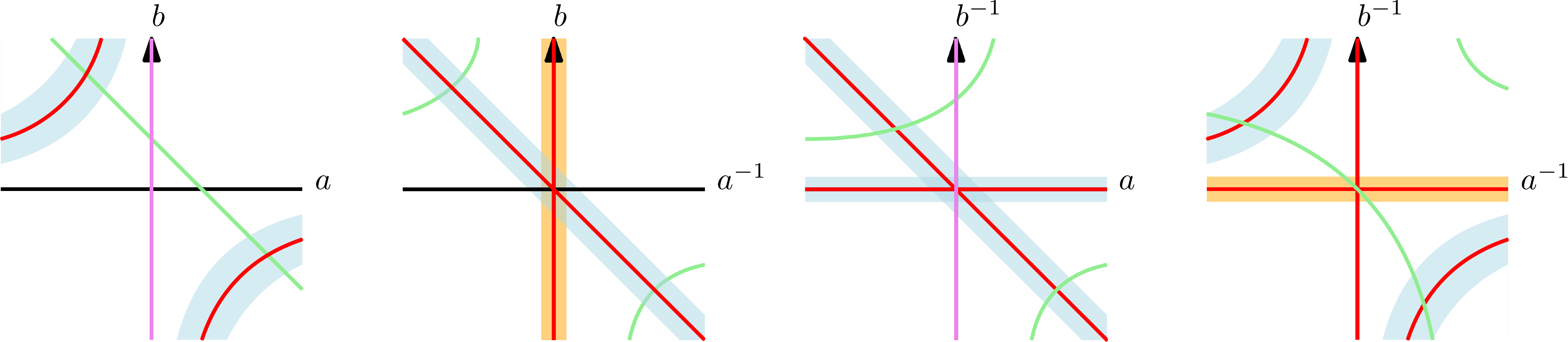}
\caption{ \label{fig:tadpole geometry}
	Real sections of $\mathbb{C}\times\mathbb{C}\setminus\{m^2/s-4ab=0\}$. 
	The twisted boundaries are red while the untwisted boundaries $D_2$ and $D_3$
	are violet and green respectively. The shaded regions represent the
	$d\theta$'s that remove support about the twisted boundaries. In the 
	light-blue regions we use the primitive $\psi_b$, while in the orange 
	regions we use $\psi_a$. 
}
\end{figure}

To make $\phi^\vee_1$ compactly supported about the twisted boundaries (singularities),
we will have to use the above primitives in small tubes around these singularities.
A systematic method is to first make one compact-support form by slapping step functions,
and then we iteratively correct that guess by adding delta-function terms to make it closed.
For the zeroth order guess, we set
\begin{align}
	[\phi^\vee_1]_c^{(0)} 
	= \theta_u \theta_{1/a} \theta_{1/b} \phi^\vee_1 
	=\left( 
		1 - \bar{\theta}_u  - \bar{\theta}_{1/a} - \bar{\theta}_{1/b} 
		+ \bar{\theta}_u \bar{\theta}_{1/a} 
		+ \bar{\theta}_u \bar{\theta}_{1/b}
		+ \bar{\theta}_{1/a} \bar{\theta}_{1/b}
	\right) \phi^\vee_1 
\end{align}
where  $\theta_i$ is given by equation \eqref{eq:theta def}
and $\bar{\theta}_i = 1-\theta_i$ represents a step function \emph{inside} a small circle.
Thanks to the step functions, $[\phi^\vee_1]_c^{(0)}$ clearly has compact support about the twisted boundaries. However, $[\phi^\vee_1]_c^{(0)}$ is not closed
\begin{align}
	\nabla^\vee [\phi^\vee_1]_c^{(0)}
	&= -\left( 
		d\bar{\theta}_u  + d\bar{\theta}_{1/a} + d\bar{\theta}_{1/b} 
	\right) \wedge \phi^\vee_1 
	\\ &\qquad 
	+ \left( 
		\bar{\theta}_{1/a} d\bar{\theta}_u 
		+ \bar{\theta}_u d\bar{\theta}_{1/a} 
		+ \bar{\theta}_{1/b} d\bar{\theta}_u 
		+ \bar{\theta}_u d\bar{\theta}_{1/b}
		+ \bar{\theta}_{1/b} d\bar{\theta}_{1/a} 
		+ \bar{\theta}_{1/a} d\bar{\theta}_{1/b}
	\right) \wedge \phi^\vee_1 .
	\nn
\end{align}
This can be patched-up by adding delta-functions times primitives:
\begin{align}
	[\phi^\vee_1]_c^{(1)} 
	&= [\phi^\vee_1]_c^{(0)} 
		+ \left(1 - \bar{\theta}_{1/a} - \bar{\theta}_{1/b} \right) 
			d\bar{\theta}_u \wedge \psi_b
		+ \left(1 - \bar{\theta}_u - \bar{\theta}_{1/b} \right) 
			d\bar{\theta}_{1/a} \wedge \psi_a 
	\nn \\ & \qquad  
		+ \left( 1 - \bar{\theta}_u - \bar{\theta}_{1/a} \right) 
			d\bar{\theta}_{1/b} \wedge \psi_b.
\end{align}
Note that we have discarded terms with empty support like
$\bar{\theta}_{1/b}\bar{\theta}_{1/a}d\bar{\theta}_u$, since the three curves don't have a triple-intersection
(see fig.~\ref{fig:tadpole geometry}).
We also used that in this specific example the same primitive $\psi_b$ works for both lines $b$ and $u$ (but it is important the the region where we use
$\psi_b$ does \emph{not} include $b=0$).
However, $[\phi^\vee_1]_c^{(1)}$ is still not closed near double intersections
$(a,b) = (\infty,0), (\infty,\infty)$:
\begin{align}
	\nabla^\vee [\phi^\vee_1]_c^{(1)} 
	= d\bar{\theta}_{1/a} \wedge d\bar{\theta}_u \wedge (\psi_b-\psi_a )
	+ d\bar{\theta}_{1/a} \wedge d\bar{\theta}_{1/b} \wedge (\psi_b - \psi_a).
\end{align}
This may be finally patched-up by constructing local 0-form primitives ($\psi_{(\infty,0)}$ and $\psi_{(\infty,\infty)}$) for $\psi_b-\psi_a$ near these points
(the primitives are simple functions of $x=ab$ which we do not write down explicitly):
\begin{align}
	[\phi^\vee_1]_c = [\phi^\vee_1]_c^{(1)} 
	+ d\bar{\theta}_{1/a} \wedge d\bar{\theta}_u \, \psi_{(\infty,0)}
	+ d\bar{\theta}_{1/a} \wedge d\bar{\theta}_{1/b} \, \psi_{(\infty,\infty)}.
\end{align}
Lastly, it is easy to check that $[\phi^\vee_1]_c$ and $\phi^\vee_1$ are indeed cohomologous, essentially they differ by terms of the form $\nabla^\vee(\bar{\theta}\psi \cdots)$.
Thus, we have succeeded in finding a form that is cohomologous to $\phi^\vee_1$ and compactly supported about the twisted boundaries.

We now turn to step (2) of our procedure: to remove support near \emph{untwisted} boundaries $D_2$ and $D_3$.
This is much simpler and the trick is to consider the restriction of $[\phi^\vee_1]_c$ to these boundaries.
If we can find a primitive for them, then the Leray construction of appendix \ref{app:delta} effectively solves our problem. For example, let $\psi_3$ be a (compact-support) primitive for $[\phi^\vee_1]_c\vert_{D_3=0}$.
Then, schematically (in a notation similar to eq.~\eqref{leray summary}), the following is cohomologous
to $[\phi^\vee_1]_c$ above:
\be
 [\phi^\vee_1]_c \simeq \theta_{D_3}[\phi^\vee_1]_c + \frac{u}{u\vert_{D_3=0}}
 d\theta_{D_3}\wedge \pi_*\psi_3\,\quad\mbox{if}\quad \nabla \psi_3 = [\phi^\vee_1]_c\vert_{D_3=0}\,,
\label{leray pullback}
\ee
which is now readily verified to be closed, and to vanish in a neighborhood of $D_3=0$.
The map is well-defined since $[\phi^\vee_1]_c$ and $\psi_3$ already have compact support:
since they vanish near all twist singularities, the pullback takes place within a regular neighborhood of $D_3=0$.
The procdure can then be repeated one boundary at a time.

Generally, a primitive like $\psi_3$ is guaranteed to
exist if $[\phi^\vee_1]_c$ is closed in relative cohomology,
since we recall that on a manifold with boundary the restriction of a closed form to any boundary is exact
(see eq.~\eqref{dphi example 1}). Hence the $c$-map can only fail if the original form $\phi^\vee_1$
was not legitimately closed.

For the 12-cut, we find indeed an obstruction to the $c$-map:
primitives for $[\phi^\vee_1]_c\vert_{D_2=0}$ do not exist near all boundaries. 
This is related to the fact that the twist disappears when $D_2=0$: $u\vert_{D_2=0} = (m^2)^\vep$.
Let us compute the restriction of $[\phi^\vee_1]_c$ to $D_2$:
\begin{align}
	[\phi^\vee_1]_c\vert_{D_2=0}
	= - d(\theta_{1/b} \bar\theta_u) \wedge \psi_b
	\propto \frac{db}{b}d\theta_{1/b} \text{ near } (a,b) = (0,\infty). 
\end{align}
The issue is that this 2-form is not exact. (If $b=\infty$ were a twisted point with exponent $\epsilon$,
a primitive of the form $\frac{d\theta_{1/b}}{\epsilon}$ would exist, but without a twist, $db/b$
does not admit any primitive throughout the circle $d\theta_{1/b}$.
Relaxing $d\theta\mapsto \theta$ or $\bar\theta$ also does not work since this would either
introduce support near $b=\infty$ or a naked singularity.)
We conclude that the compactly supported tadpole dual form does not exist on its own!

To better understand the root of this obstruction, we examine the compactly supported cohomology of the double boundary $D^{(12)}$. We can view it as the complex $b-$plane.
As mentioned, the twist is trivial on this boundary, so the relevant space is:
\be
 H_c^\bullet(\mathbb{C},D_3\vert_{12})\,.
\ee
The reason the space is $\mathbb{C}$ and \emph{not} the Riemann sphere is that $b=\infty$
is a twisted line \emph{in a neighborhood} of the cut, see fig.~\ref{fig:tadpole geometry}.
The construction in eq.~\eqref{leray pullback} requires the primitive $\psi$ to vanish
in a neighborhood of infinity, which is what $H_c^\bullet(\mathbb{C})$ requires.
The crucial observation is that $H_c^2(\mathbb{C},D_3\vert_{12})$ is non-empty.

To understand why $H_c^2(\mathbb{C},D_3\vert_{12})$ is non-empty, we start by considering a simpler example: the compactly supported cohomology of the complex plane $H^\bullet_c(\mathbb{C})$.
Since the closed 0-forms are constant functions and the only constant compactly supported function is zero, $H^0_c(\mathbb{C})=0$. Moreover, $H^1_c(\mathbb{C})=0$ because all 1-forms on $\mathbb{C}$ have a 0-form primitive (the integral). Only $H^2_c(\mathbb{C})$ is non-trivial. In particular, a basis for $H^2_c(\mathbb{C})$ is simply $\delta^2(z-a)$ for any $a\in\mathbb{C}$\footnote{Note that $\delta^2(z-b)$ is cohomologous to $\delta^2(z-a)$, as the difference is proportional to
$d\left(\frac{1}{z-a}-\frac{1}{z-b}\right)$ by the holomorphic anomaly.}, which is co-homologous to
$d\theta_{1/b}\frac{db}{b}$ by adding $d(\theta_{1/b}/b)$ and using the holomorphic anomaly
(for an algebraic treatment, see appendix \ref{app:algH2}).
Adding $D_3\vert_{12}$ as a relative point only changes the $H^1$, which in this case contains one element supported on the point $D_3\vert_{12}$.

Since the obstruction is one-dimensional, it can potentially cancel among different tadpoles.
Indeed, considering the restriction to the $12$ cut from the relevant dual tadpoles we find that
they have precisely opposite sign and cancel when we sum them:
\begin{align}
	[\vphi^\vee_\text{tad\ 1}+ \vphi^\vee_\text{tad\ 2}]_c\vert_{12}
	\sim \frac{db}{b} - \frac{db^\prime}{b^\prime} \simeq 0
\end{align}
where $b^\prime = b-c$ and $c$ is some constant.  This confirms that the sum over all tadpole-duals is a valid form, as used in section \ref{sec:FI and degenerate limits} of the main text.

\section{Equivalence of algebraic and distributional $H^2(\mathbb{P}^1)$ \label{app:algH2}}

In this appendix we elaborate on the algebraic understanding of forms like $H^2(\CP^1)$,
which we represented above using
antiholomorphic components $d\bar z$ hidden inside delta-functions.
This would not seem make sense for more generic fields (i.e., $\mathbb{R},\mathbb{Q},\mathbb{Z}$ etc., instead of $\mathbb{C}$).
What does a representative of $H^2(\mathbb{P}^1)$ look like if we are not allowed to use a $d\bar z$?
Algebraic geometry offers a rather simple answer but it does involve the introduction of some additional mathematical structure -- the \v{C}ech-de Rham complex (see \cite{bott1995} for a good introduction). 

Simply put, the \v{C}ech-de Rham complex is a double complex that has two differentials: $d$ (the normal exterior derivative) and $\delta$ (a ``differential'' that acts on the data of a good chart). For $\mathbb{P}^1$, we can cover it with two charts $U_0={z}$ and $U_1={w}$ where $w=1/z$ on $U_{01}=U_0 \cap U_1$. The relevant double complex is 
\begin{equation}
\begin{tikzcd}
	0 & 0 & {}
\\
	{\Omega^0(U_{01})} \arrow[r, "d"] \arrow[u,"\delta"] 
	& {\Omega^1(U_{01})} \arrow[r, "d"] \arrow[u,"\delta"]                
	& 0                   
\\
	{\Omega^0(U_0) \oplus \Omega^0(U_1)} \arrow[r, "d"] \arrow[u,"\delta"] 
	& {\Omega^1(U_0) \oplus \Omega^1(U_1)} \arrow[r, "d"] \arrow[u,"\delta"]                
	& 0                  
\end{tikzcd}
\end{equation}
where $\delta: \Omega^p(U_0) \oplus \Omega^p(U_1) \to \Omega^p(U_{01})$ by $(\alpha,\beta) \mapsto \alpha - \beta$. 

A rather surprising but powerful fact is that the ``diagonal''-cohomology of \v{C}ech-de Rham complex is equivalent to de Rham cohomology \cite{bott1995} (there is an equivalence between the combinatorics of a good covering and differential forms). More precisely, if we define
\begin{align}
	K^0 = \Omega^0(U_0) \oplus \Omega^0(U_1), \quad
	K^1 = \Omega^0(U_{01}) \oplus \Omega^1(U_0) \oplus \Omega^1(U_1), \quad
	K^2 = \Omega^1(U_{01}),
\end{align}
then 
\begin{align}
	H^0 = \ker(\mathcal{D}: K^0 \to K^1),\quad 
	H^1 = \frac{\ker(\mathcal{D}: K^1 \to K^2)}{\text{im}(\mathcal{D}:K^0 \to K^1)}, \quad
	H^2 = \frac{K^2}{\text{im}(\mathcal{D}:K^1 \to K^2)}
\end{align}
are equivalent to de Rham cohomology. Here, $\mathcal{D}$ is the differential given by
\begin{align}
	\mathcal{D}(\vphi \in K^a) = (-1)^\text{deg} d\vphi + \delta \vphi \in K^{a+1}
\end{align}
where $\text{deg}$ is the degree of the form in a given component of $\vphi$ (one needs to be careful with this definition since the elements of the \v{C}ech-de Rham complex have components with varying degrees). 

Now, lets construct $H^2(\mathbb{P})$. A general element of $K^2$ is a 1-form on the overlap $U_{01}$:
\begin{align}
	\mathbb{C}[z]\; dz
	\quad \text{or} \quad
	\mathbb{C}[w]\; dw.
\end{align}
We can always find a primitive for such 1-forms except for $dz/z$. For example, $z dz \in K^2$ has the primitive $ \frac{1}{2} z^2 \oplus 0 \in K^1$ (forms in $K^2$ do not have primitives when integration gives non-algebraic functions or when the rational form does not have a decomposition via partial fractions). Thus, we are forced to conclude that 
\begin{align}
	H^2(\mathbb{P}^1) = \left[ \frac{dz}{z} \right]_{01}. 
\end{align}

While this result may seem strange at first, there is an intuitive picture. Since $H^2(U_i)$ on its own is trivial, the only non-trivial contribution to $H^2(\mathbb{P})$ can come from the way that the two
charts are glued together -- the equator of the sphere. This 1-dimensional space supports the 1-form $dz/z$, which parameterizes the non-trivial structure of $\mathbb{P}^1$. 
The smooth-compact two-form $d\theta_z\wedge \frac{dz}{z}$
that arose above is obviously closely related, with the
overlap $U_{01}$ effectively replaced by the equator $d\theta_z$.

\bibliographystyle{JHEP}
\bibliography{physics}

\end{document}